\documentclass[fleqn,usenatbib]{mnras}

\usepackage{newtxtext,newtxmath}

\usepackage[T1]{fontenc}
\usepackage{xfrac}

\DeclareRobustCommand{\VAN}[3]{#2}
\let\VANthebibliography\thebibliography
\def\thebibliography{\DeclareRobustCommand{\VAN}[3]{##3}\VANthebibliography}

\usepackage{graphicx}	
\usepackage{amsmath}	
\usepackage{amssymb}	
\usepackage{makecell}
\usepackage{multirow}
\usepackage{ulem}
\usepackage{xcolor}
\usepackage{float}

\newcommand{\kms}{km s$^{-1}$}
\newcommand{\dego}{$^\circ$}
\newcommand{\msun}{M$_\odot$}

\title[Morpho-kinematics of the wind of L$_2$ Pup]{Morpho-kinematics of the wind of AGB star L$_2$ Pup}

\author[D.T. Hoai et al.]{D.T. Hoai \thanks{E-mail: dthoai@vnsc.org.vn},
  P.T. Nhung \thanks{E-mail: pttnhung@vnsc.org.vn},
  P. Darriulat,
  P.N. Diep,
  N.B. Ngoc,
  T.T. Thai
  and P. Tuan-Anh
\\
Department of Astrophysics, Vietnam National Space Center, Vietnam Academy of Science and Technology, \\
18, Hoang Quoc Viet, Nghia Do, Cau Giay, Ha Noi, Vietnam
}

\date{Accepted XXX. Received YYY; in original form ZZZ}

\pubyear{2021}

\begin{document}
\label{firstpage}
\pagerange{\pageref{firstpage}--\pageref{lastpage}}
\maketitle

\begin{abstract}
 Single dish observations of AGB star L$_2$ Pup have revealed exceptionally low mass loss rate and expansion velocity, challenging interpretations in terms of standard wind models. Recent VLT and ALMA observations have drawn a detailed picture of the circumstellar envelope within $\sim$20 au from the centre of the star: a nearly edge-on rotating disc of gas and dust, probably hosting a planetary companion near the star. However, these observations provide no direct information on the wind escaping the gravity of the star. The present article uses ALMA observations of the $^{12,13}$CO(3-2), $^{29}$SiO(8-7), $^{12}$CO(2-1) and $^{28}$SiO(5-4) line emissions to shed new light on this issue. It shows the apparent normality of L$_2$ Pup in terms of the formation of the nascent wind, with important line broadening within 4 au from the centre of the star, but no evidence for a wind flowing along the disc axis. At larger distances, up to some 200 au from the centre of the star, the wind morpho-kinematics is dominated by a disc, or equatorial enhancement, expanding isotropically and radially with a velocity not exceeding some 5 \kms, inclined in the north-west/south-east direction with respect to the plane of the sky. In addition, outflows of lower density are observed on both sides of the disc, covering large solid angles about the disc axis, contributing about half the flux of the disc. Such morphology is at strong variance with the expectation of a pair of back-to-back outflows collimated by the central gas-and-dust disc. 
\end{abstract}




\section{Introduction}

Over the past decade, a host of new observations of the circumstellar envelope (CSE) of oxygen-rich AGB stars have considerably improved our understanding of the formation of the nascent wind, yet leaving an important number of questions unanswered \citep{Hofner2018}. From infrared to millimetre wavelengths, the latter probing the proximity of the stellar surface with vibrationally excited molecular lines or continuum emission, they show the importance of pulsation- and shock-induced dynamics in levitating the molecular atmosphere. They reveal a complex dynamics and grain chemistry, influenced by variability, displaying significant inhomogeneity and suggesting the presence of shocks related to star pulsations: evidence is found for short time variability, occasional gas in-fall, hot spots covering a small fraction of the stellar disc, radial velocities at the 10 \kms\ scale.  A consequence on the millimetre observation of the emission of molecular lines is the presence of large Doppler velocity wings near the line of sight crossing the stellar disc in its centre.

Recently, ALMA observations of the $^{12,13}$CO(3-2) and $^{29}$SiO(8-7) emissions of the CSE of such a star, the nearby L$_2$ Pup, have been reported by \citet{Homan2017} and \citet{Kervella2016}, respectively. Both analyses offer a detailed description of the CSE in terms of a nearly edge-on rotating disc of gas and dust, suggesting the presence of a planetary companion at $\sim$2.4 au from the centre of the star. These analyses predict the presence of radial outflows collimated by the disc and do not explicitly account for an important line broadening in the inner region of the CSE, in contrast with what is usually observed in other similar AGB stars. These are the two main issues that the present article is addressing.

L$_2$ Pup is a semi-regular pulsating variable \citep[P$=$141 days][]{Bedding2005} of spectral type M5III corresponding to an effective temperature of $\sim$3500 K. Hipparcos locates it at 64$\pm$4 pc \citep{vanLeeuwen2007}. Its mass was first estimated as 1.7 \msun\ by \citet{Dumm1998}, 2 \msun\ by \citet{Kervella2015} and 0.7 \msun\ by \citet{Lykou2015}; it is now measured as 0.66$\pm$0.05 \msun\ by \citet{Kervella2016} from the Keplerian motion of the gas orbiting it.

The light curve \citep{Bedding2002,Bedding2005, McIntosh2013} has been undergoing a major dimming episode at the end of the past century, in $\sim$1994. This event has triggered detailed studies of the CSE using high angular resolution VLT observations at visible and infrared wavelengths. Using NACO observations, \citet{Kervella2014}, \citet{Lykou2015} and \citet{Ohnaka2015}, the latter authors combining these with AMBER/VLTI observations, have detected and resolved a dust disc, oriented east-west and inclined by 82\dego\ with respect to the plane of the sky, with approximate projected size of 180$\times$50 mas$^2$. While agreeing on the general picture, these authors mention several small differences between their observations. Some of them \citep{Ohnaka2015, Bedding2005} assume that dust clouds were ejected at the end of the past century, causing the appearance of the disc. Others assume that the disc existed before the dimming event and that in 1994 it simply started obscuring significantly the star. \citet{Kervella2015}, using VLT/SPHERE-ZIMPOL observations in the visible, have revealed the possible presence of a companion at $\sim$2.1 au west of the star, but the NACO images and the VINCI interferometer observations do not see trace of it. There had been many earlier speculations about a possible binarity \citep{Goldin2007, Bedding2002, Jorissen2009, McIntosh2013}.

Single dish observations of the CSE use mostly ground based telescopes, with the exception of the most recent, \citet{Danilovich2015}, that uses data from Herschel/HIFI; they quote a mass loss rate of 1.4$\times$10$^{-8}$ \msun\,yr$^{-1}$ obtained from a radiative transfer modelling of the emission of CO lines. A similarly low value of 2$\times$10$^{-8}$ \msun\,yr$^{-1}$ had been obtained by \citet{Olofsson2002}. Typical terminal velocities of $\sim$2 to 3 \kms are quoted \citep{Winters2002, Gonzalez2003}. However these models are crude, in particular they assume a spherically symmetric wind, which is probably far from being the case: they can only provide an order of magnitude estimate of the mass loss rate. Neither the terminal velocity nor the mass loss rate has been measured directly.

The extensions of the CO and SiO line emissions have been evaluated by \citet{Schoier2004} using Australia Telescope Compact Array (ATCA) observations with an angular resolution at the scale of 2 arcsec  and by \citet{Ramstedt2020} using the Atacama Compact Array (ACA). But the morpho-kinematics of the CSE has now been studied in detail using ALMA long baseline millimetre observations of $^{29}$SiO(8-7) and continuum emissions \citep{Kervella2016} and of the $^{12}$CO(3-2) and $^{13}$CO(3-2) line emissions \citep{Homan2017}. These authors draw a detailed picture of the morpho-kinematics of the disc; they find that between 2 au and 6 au from the star, there is only gas and the rotation is Keplerian, while between 6 au and some 25 au from the star, there is both gas and dust and the rotation is sub-Keplerian. Evidence for rotation had already been obtained earlier from the study of SiO masers \citep{McIntosh2013}. \citet{Homan2017} use a radiative transfer model to study the distribution of temperature and density within the disc; they find that the inner rim of the disc has a thickness of $\sim$1.5 au and evaluate a disc mass of $\sim$2$\times$10$^{-4}$ \msun\ and a $^{12}$CO/$^{13}$CO abundance ratio of $\sim$10, however with a large uncertainty. Continuum emission confirms the presence of a likely planetary companion orbiting at only 2.4 au from the centre of the star \citep{Kervella2016}; \citet{Homan2017} estimate its mass at Jupiter mass scale.  \citet{Chen2016} have presented a hydro-dynamical modelling of these observations and \citet{Haworth2018} have studied their implication on the dust properties.

Finally, we note that \citet{Little1987} and \citet{Lebzelter1999} have both reported possible detections of technetium in the atmosphere of L$_2$ Pup, none of which, however, could be firmly established.

\section{Overall picture}

We use archival ALMA observations from project ADS/JAO.ALMA\#2015.1.00141.S (PI P. Kervella) which cover emissions of the $^{12}$CO(3-2), $^{13}$CO(3-2) and $^{29}$SiO(8-7) lines in the close neighbourhood of the star and of the $^{12}$CO(2-1) and $^{28}$SiO(5-4) lines farther away. The first set of observations was previously analysed by \citet{Kervella2016} and \citet{Homan2017}. We refer the reader to these publications for details concerning the observations and their calibration. We used CASA\footnote{https://casa.nrao.edu/} to produce clean maps with natural weighting and no continuum subtraction. The second set was observed on April 29 and August 13, 2016 with 42 and 38 antennas in configurations C36-2 and C36-5, respectively.  The phase centres were shifted to the corresponding peaks of the continuum ($\alpha$$=$07:13:32.4820, $\delta$$=$$-$44:38:17.6336 for April 29 and $\alpha$$=$07:13:32.4835, $\delta$$=$$-$44:38:17.5728 for August 13) before merging; these data have been calibrated using standard scripts of ALMA without continuum subtraction and cleaned using GILDAS\footnote{https://www.iram.fr/IRAMFR/GILDAS/} with natural weighting.

\begin{figure*}
  \centering
  \includegraphics[height=5cm,trim=.5cm 1.5cm 1.5cm 1.5cm,clip]{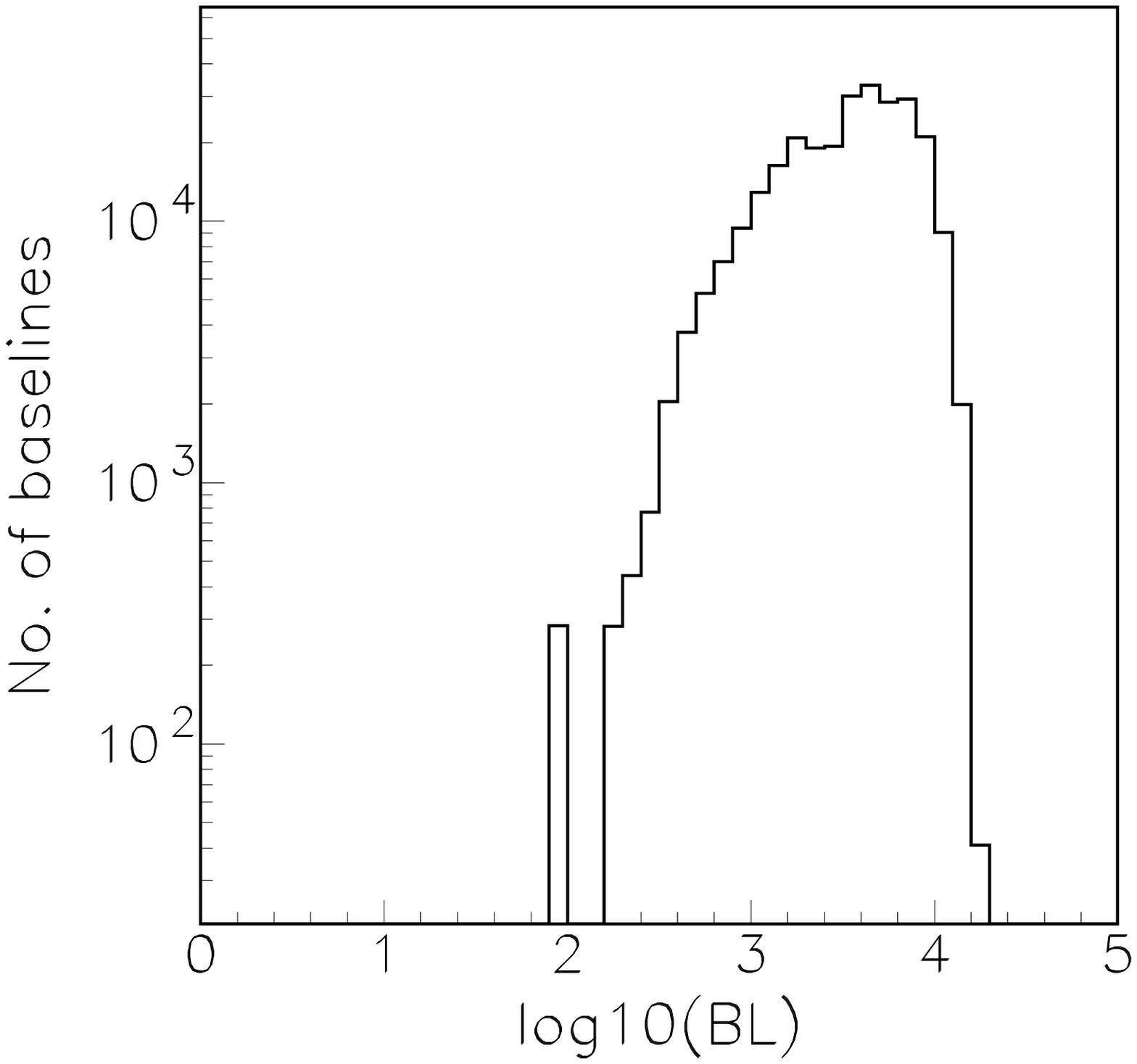}
  \includegraphics[height=5cm,trim=.5cm 1.5cm 1.5cm 1.5cm,clip]{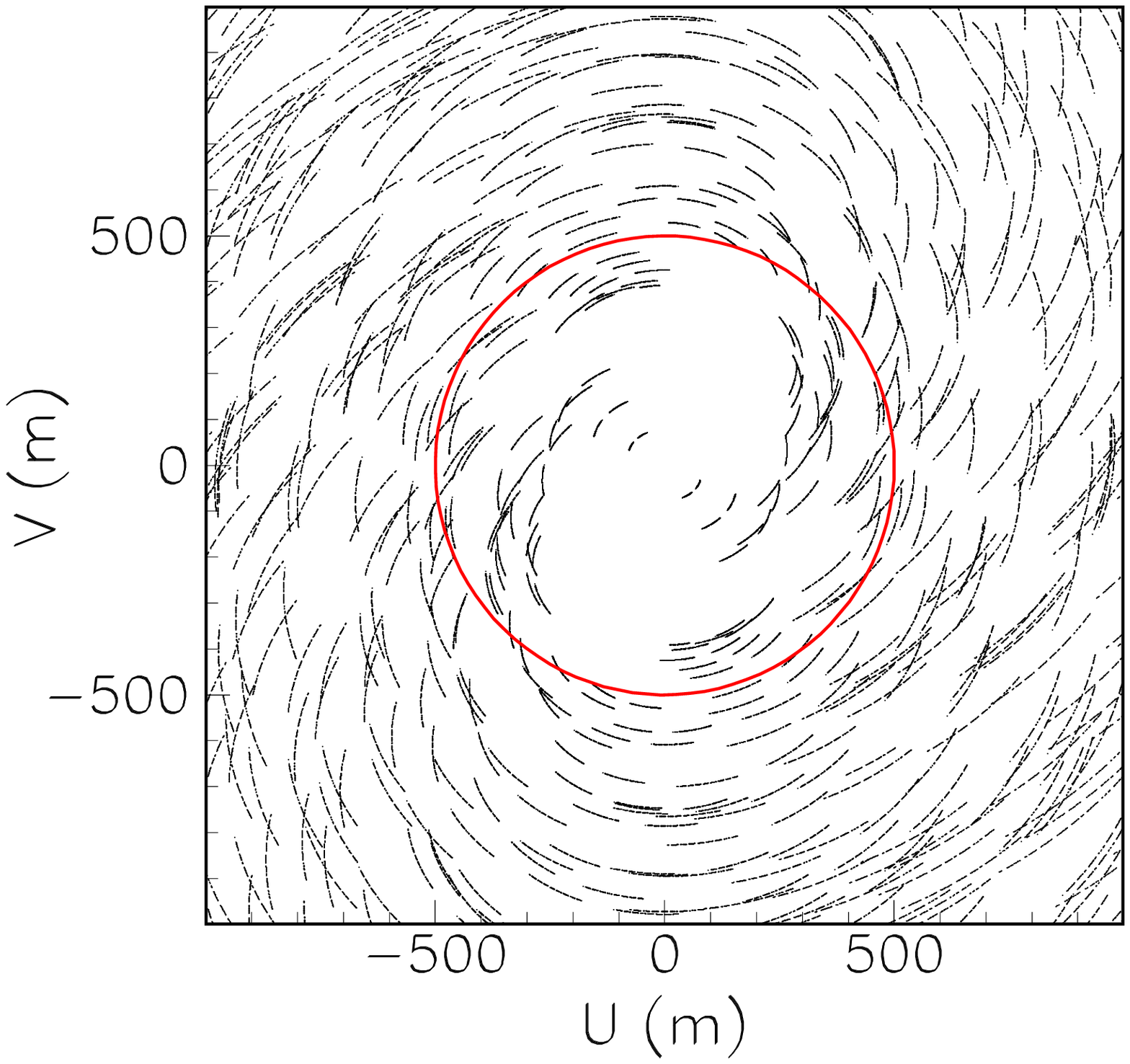}
  \includegraphics[height=5cm,trim=.5cm 1.5cm 1.5cm 1.5cm,clip]{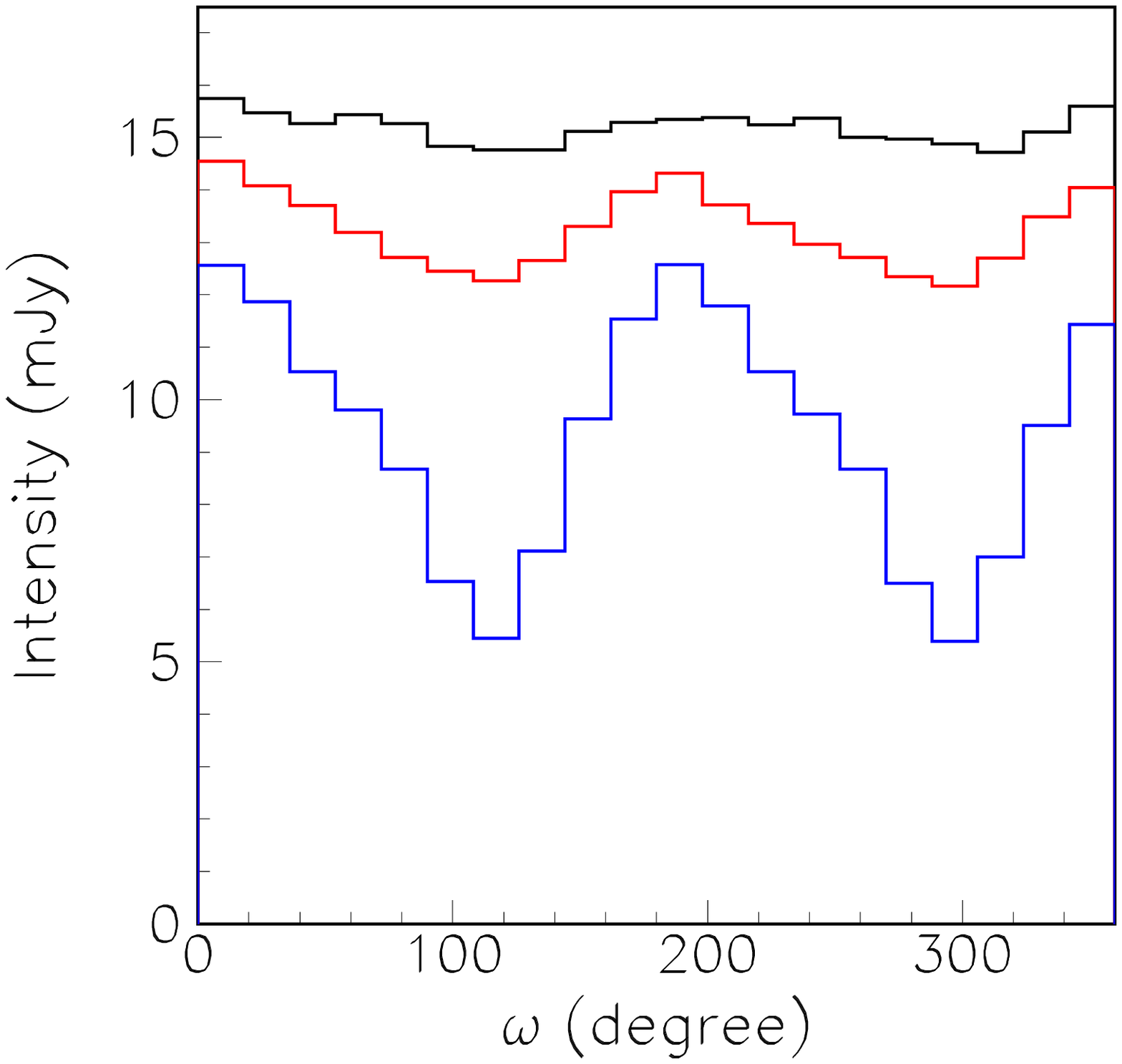}
  \caption{ Left: distribution of the decimal logarithm of the baselines measured in metres. Middle: central $uv$ coverage; the red circle has a radius of 500 m. Right: position angle $\omega$ distributions of the image of an isotropic wind in intervals 0.05$<$$R$$<$0.10 arcsec (black), 0.10$<$$R$$<$0.15 arcsec (red) and 0.15$<$$R$$<$0.20 arcsec (blue).  }
 \label{fig1}
\end{figure*}

Figure \ref{fig1} shows the baseline distribution and the $uv$ coverage for the first set of observations. Baselines smaller than 500 m are sparse and unevenly distributed, limiting the range of reliable imaging to projected distances from the star $R$$<$$\sim$0.2 arcsec. A more detailed evaluation is obtained by producing the $uv$ map of an isotropic wind and inspecting its clean image. The right panel of Figure \ref{fig1} displays the resulting distributions of the position angle $\omega$ in successive intervals of $R$: important distortions are already seen in the 0.15$<$$R$$<$0.20 interval.

\begin{figure*}
  \centering
  \includegraphics[height=3.85cm,trim=0cm .5cm 1.7cm 1.cm,clip]{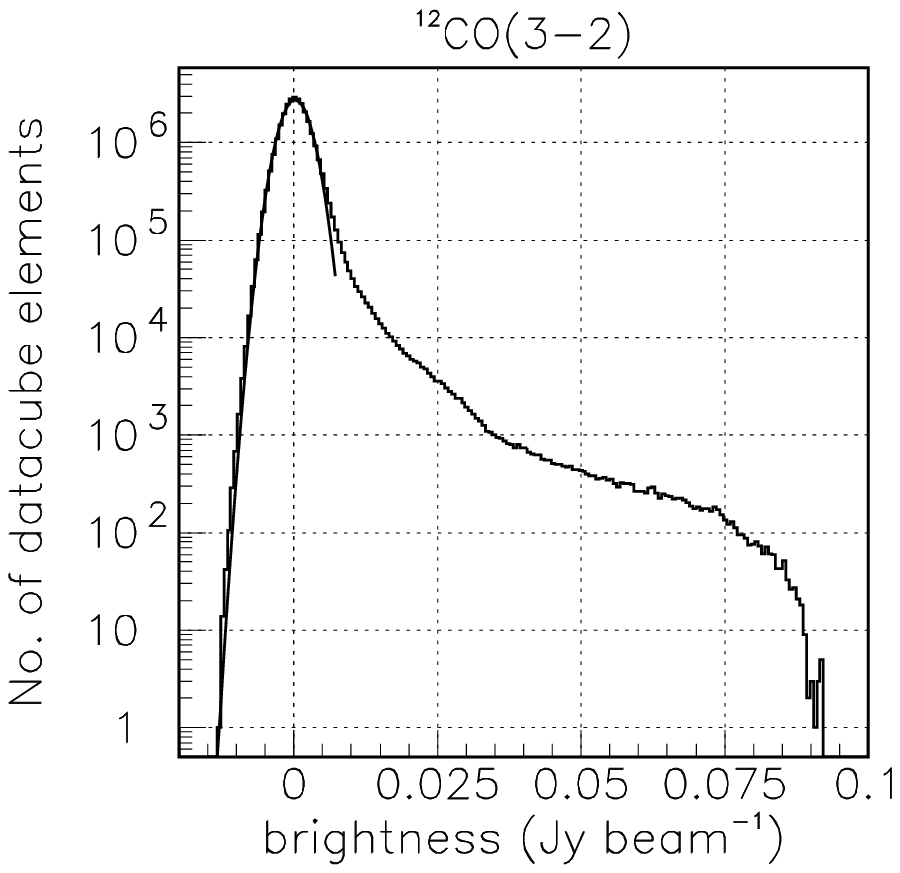}
  \includegraphics[height=3.85cm,trim=1.cm .5cm 1.7cm 1.cm,clip]{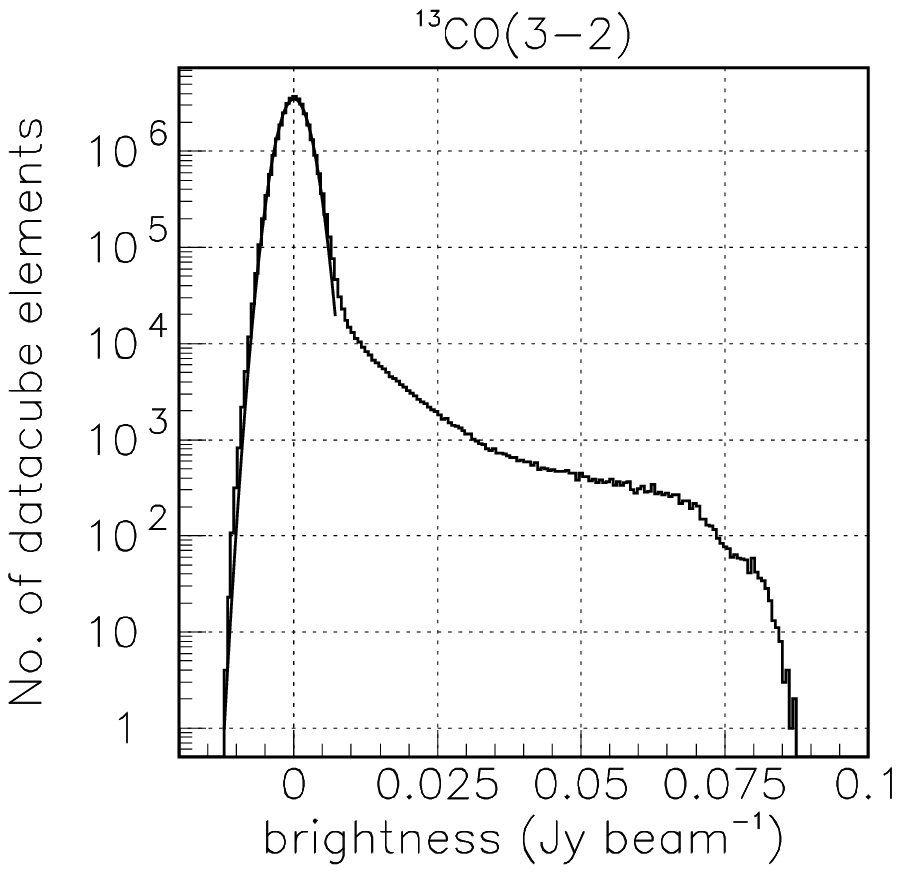}
  \includegraphics[height=3.85cm,trim=1.cm .5cm 1.7cm 1.cm,clip]{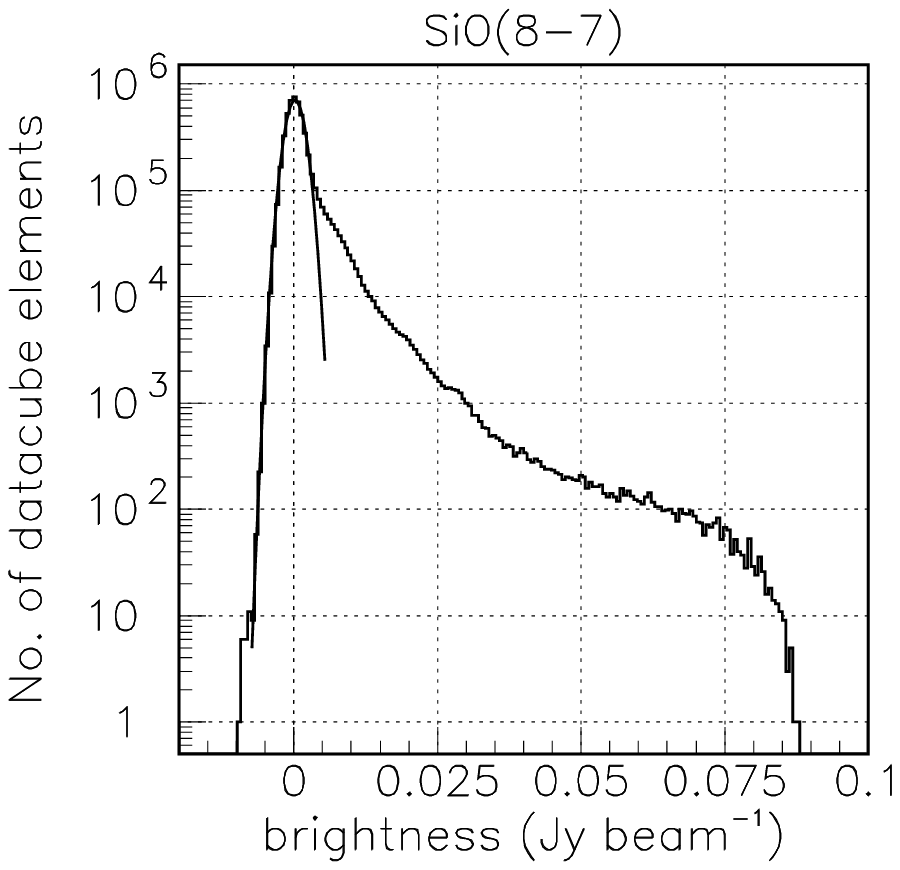}
  \includegraphics[height=3.85cm,trim=1.cm .5cm 1.7cm 1.cm,clip]{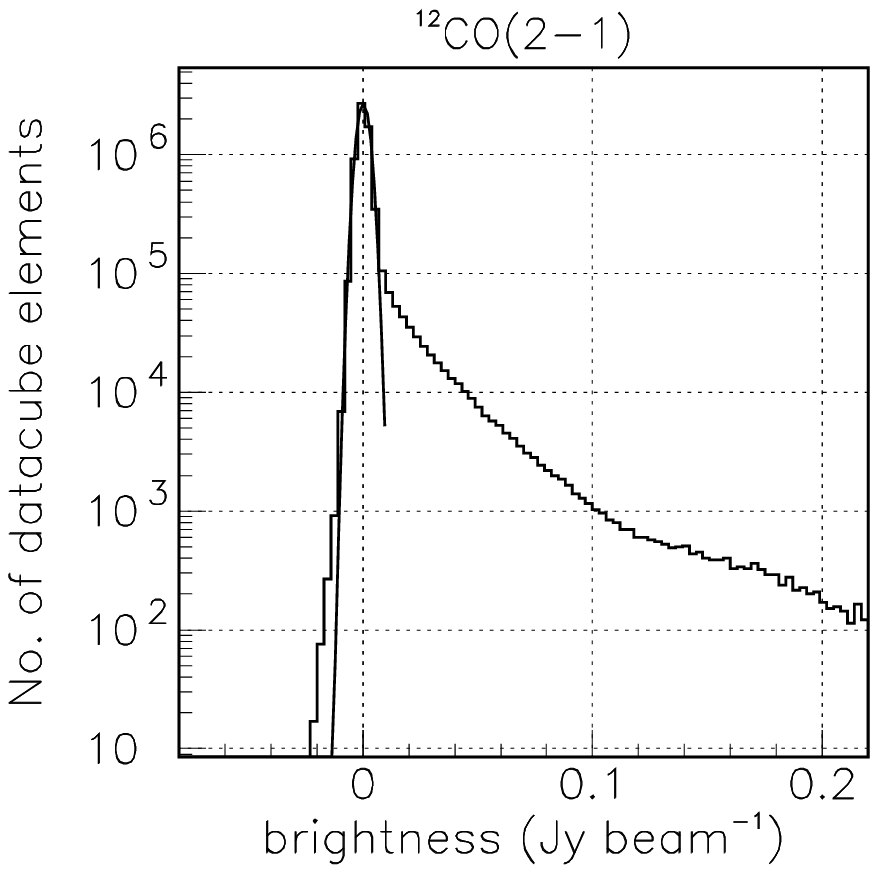}
  \includegraphics[height=3.85cm,trim=1.cm .5cm 1.7cm 1.cm,clip]{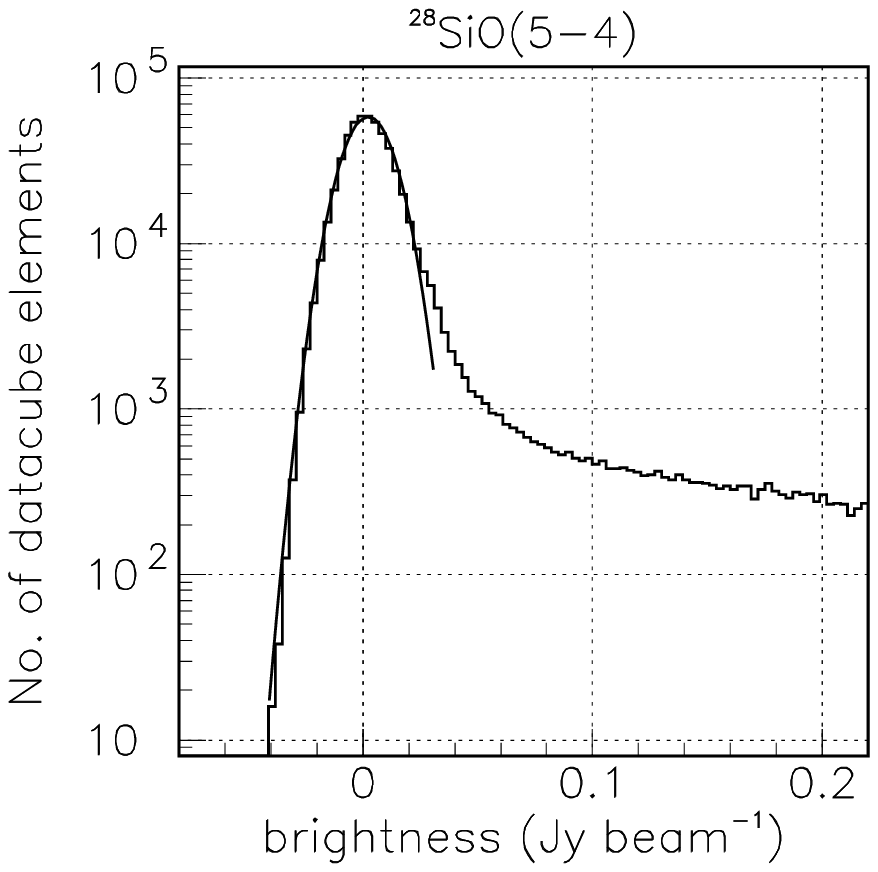}
    \caption{ From left to right: brightness distributions of the $^{12}$CO(3-2), $^{13}$CO(3-2) and $^{29}$SiO(8-7) line emissions in the region of $|x|$, $|y|$$<$0.5 arcsec and $|V_z|$$<$25 \kms; of the $^{12}$CO(2-1) line emission in the region of $R$$<$4 arcsec and $|V_z|$$<$32 \kms; and of the $^{28}$SiO(5-4) line emission in the region of $R$$<$4 arcsec and $|V_z|$$<$5 \kms.}
 \label{fig2}
\end{figure*}

The main relevant parameters are listed in Table \ref{tab1}. They include, for each of the line emissions studied in the article, intensities $f_{1000}$, $f_{100}$ and $f_{10}$ calculated at temperatures of 1000 K, 100 K and 10 K, respectively for an optically thin layer having a column density of 1 molecule cm$^{-3}$ arcsec. They are given in units of Jy arcsec$^{-2}$ \kms\ and are calculated as $f$$=$$hc/(4\upi)NA_{ji} f_{pop}$ with $f_{pop}$$=$$(2J+1)\exp(-E_{up}/T)/Q$ where $Q=T/k$ is the partition function, $h$ the Planck constant, $c$ the light velocity, $N$ the column density and $A_{ji}$ the Einstein coefficient. The beam diameters (FWHM) are typically 20 mas (1.3 au) for the first set of observations and 340 mas (22 au) for the second. The brightness distributions are shown in Figure \ref{fig2}. We use orthonormal coordinates with $x$ pointing east, $y$ pointing north and $z$ pointing away from us. The projected distance from the star is $R=\sqrt{x^2+y^2}$ and the position angle $\omega$ is measured counter-clockwise from north. Doppler velocities $V_z$ are referred to the same radial velocity relative to the Local Standard of Rest as used by \citet{Homan2017}, $V_{lsr}$=33.3 \kms.

\begin{figure*}
  \centering
  \includegraphics[height=5.2cm,trim=.5cm 1.cm 0.cm 1.5cm,clip]{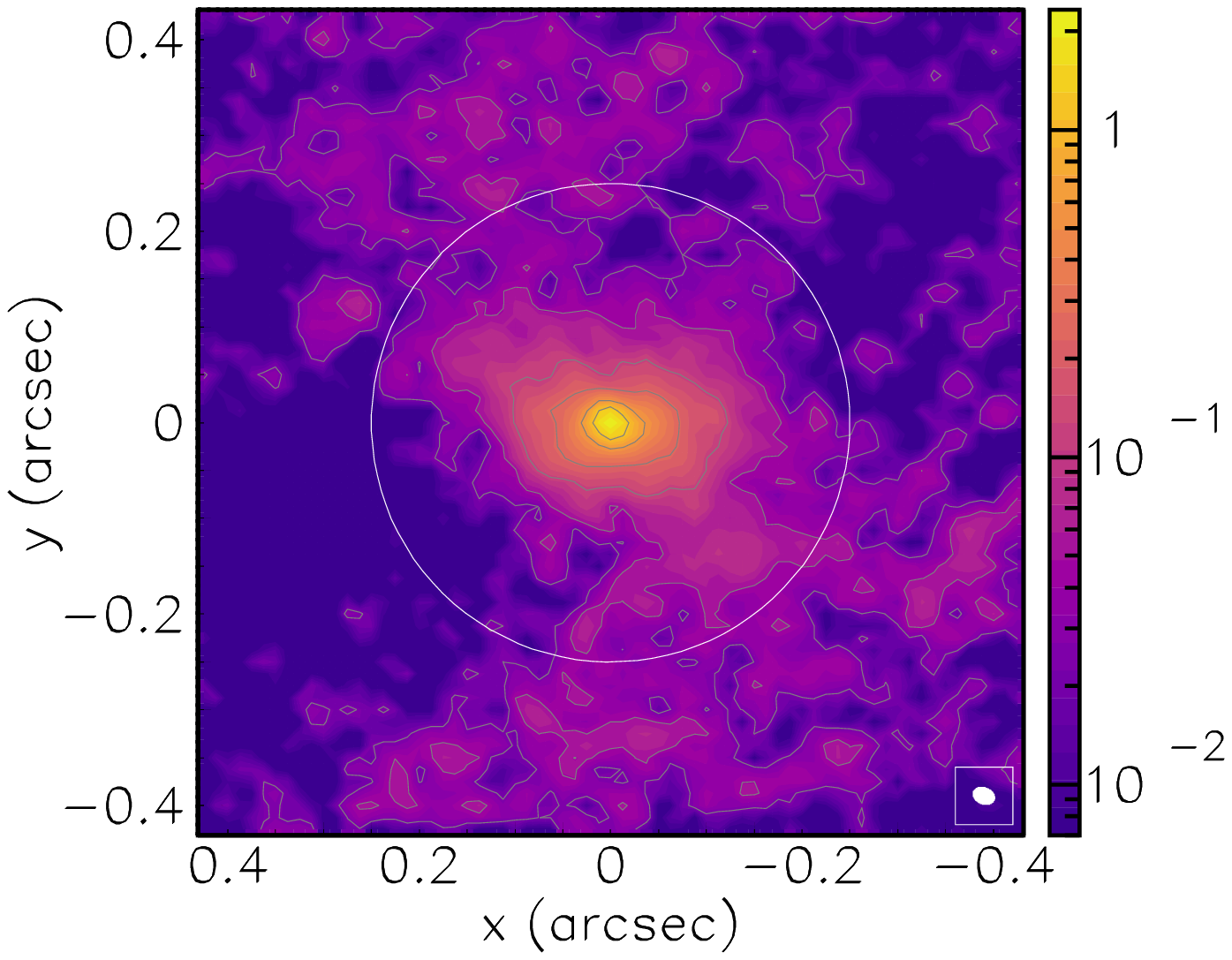}
  \includegraphics[height=5.2cm,trim=.5cm 1.cm .5cm 1.5cm,clip]{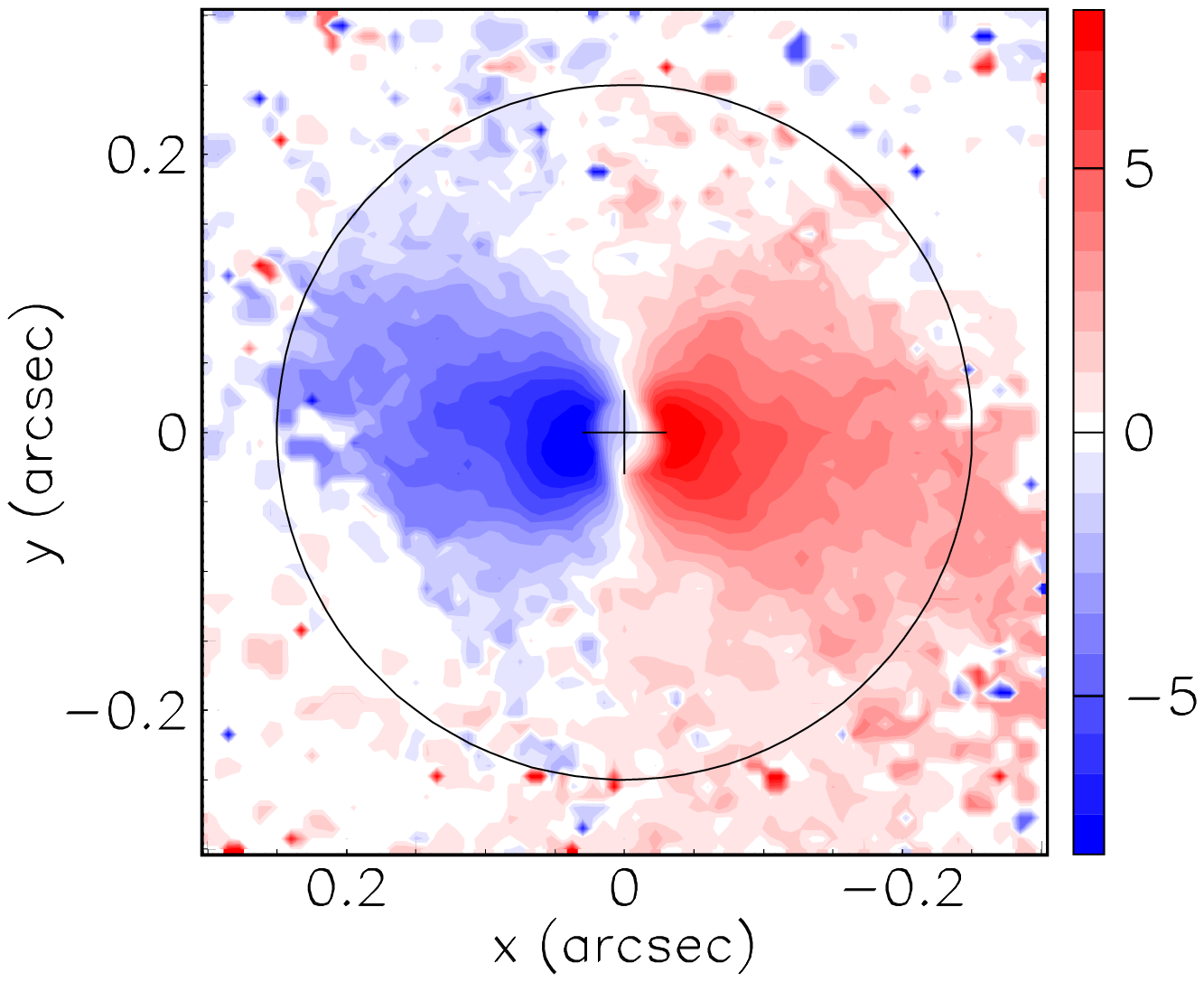}
  \includegraphics[height=5.2cm,trim=.5cm 1.cm 1.5cm 1.5cm,clip]{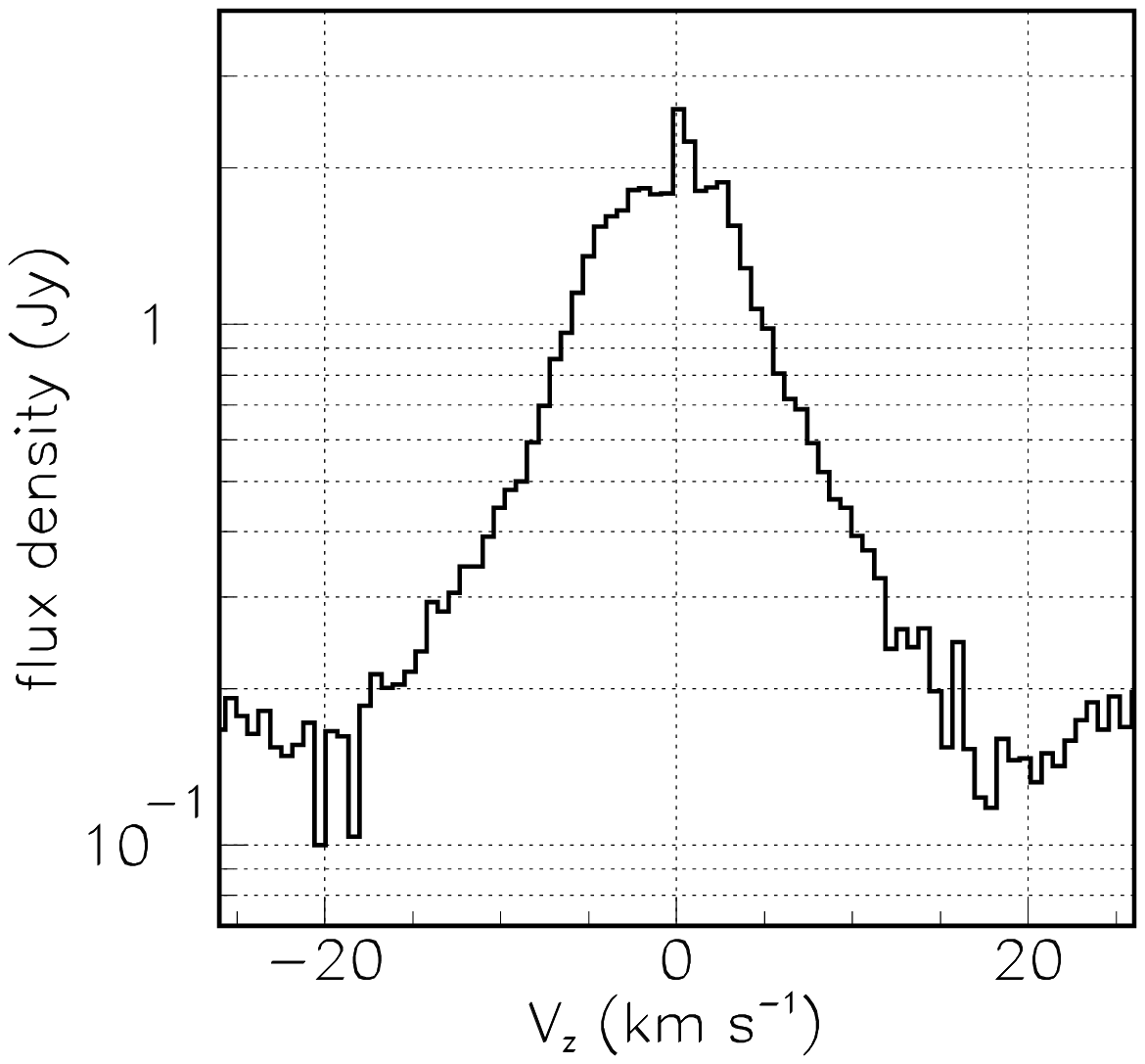}
  \includegraphics[height=5.2cm,trim=.5cm 1.cm 0.cm 1.5cm,clip]{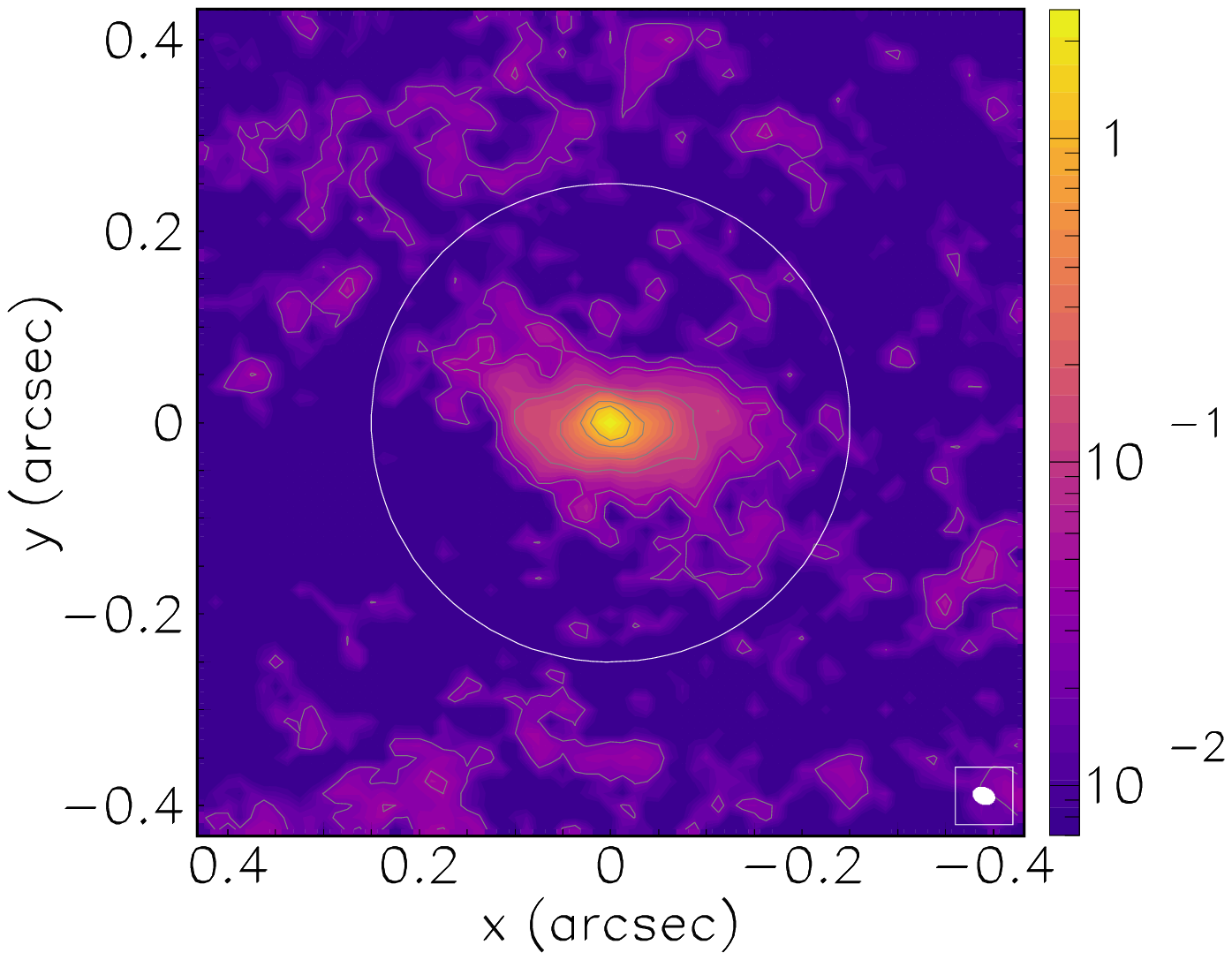}
  \includegraphics[height=5.2cm,trim=.5cm 1.cm .5cm 1.5cm,clip]{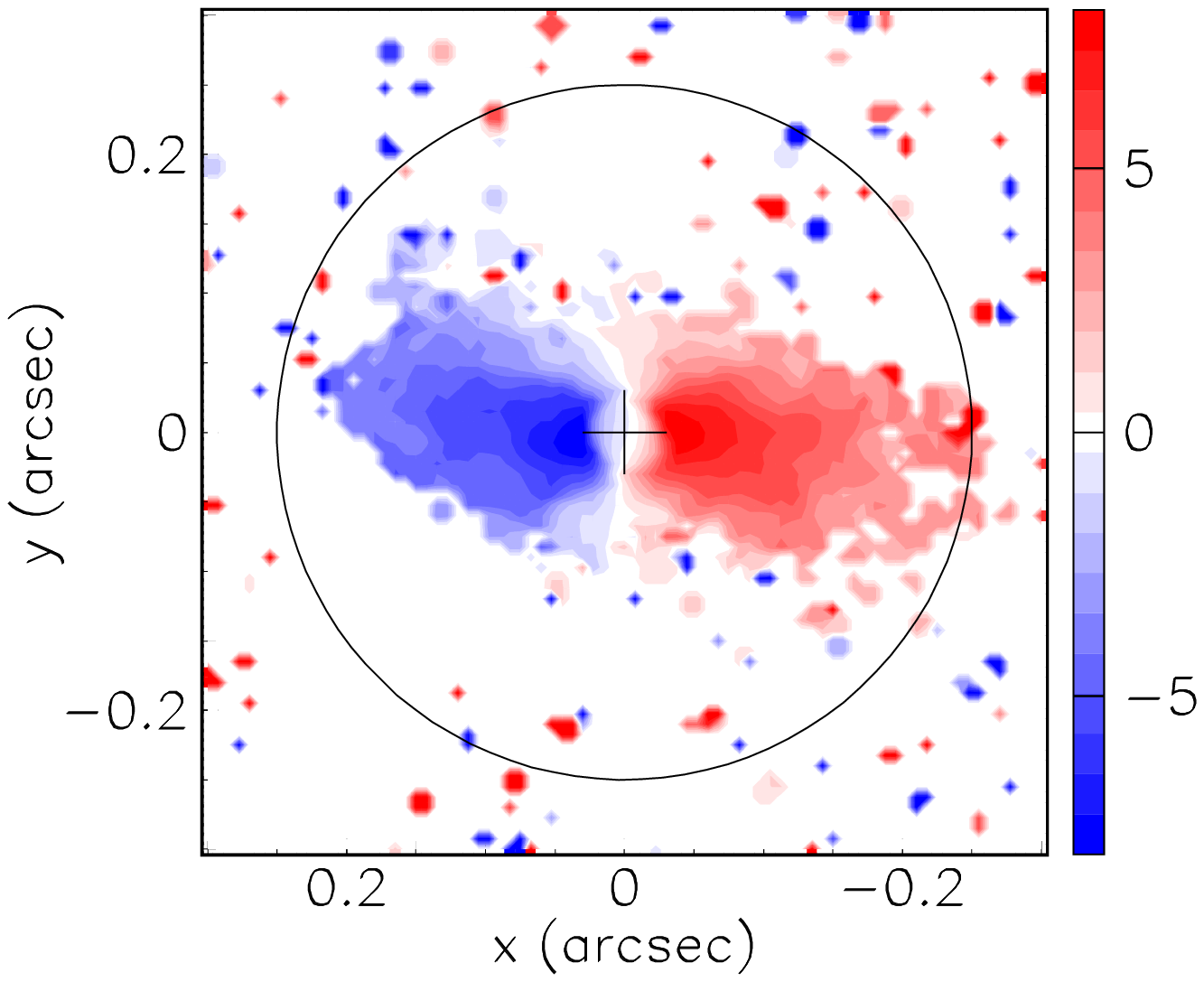}
  \includegraphics[height=5.2cm,trim=.5cm 1.cm 1.5cm 1.5cm,clip]{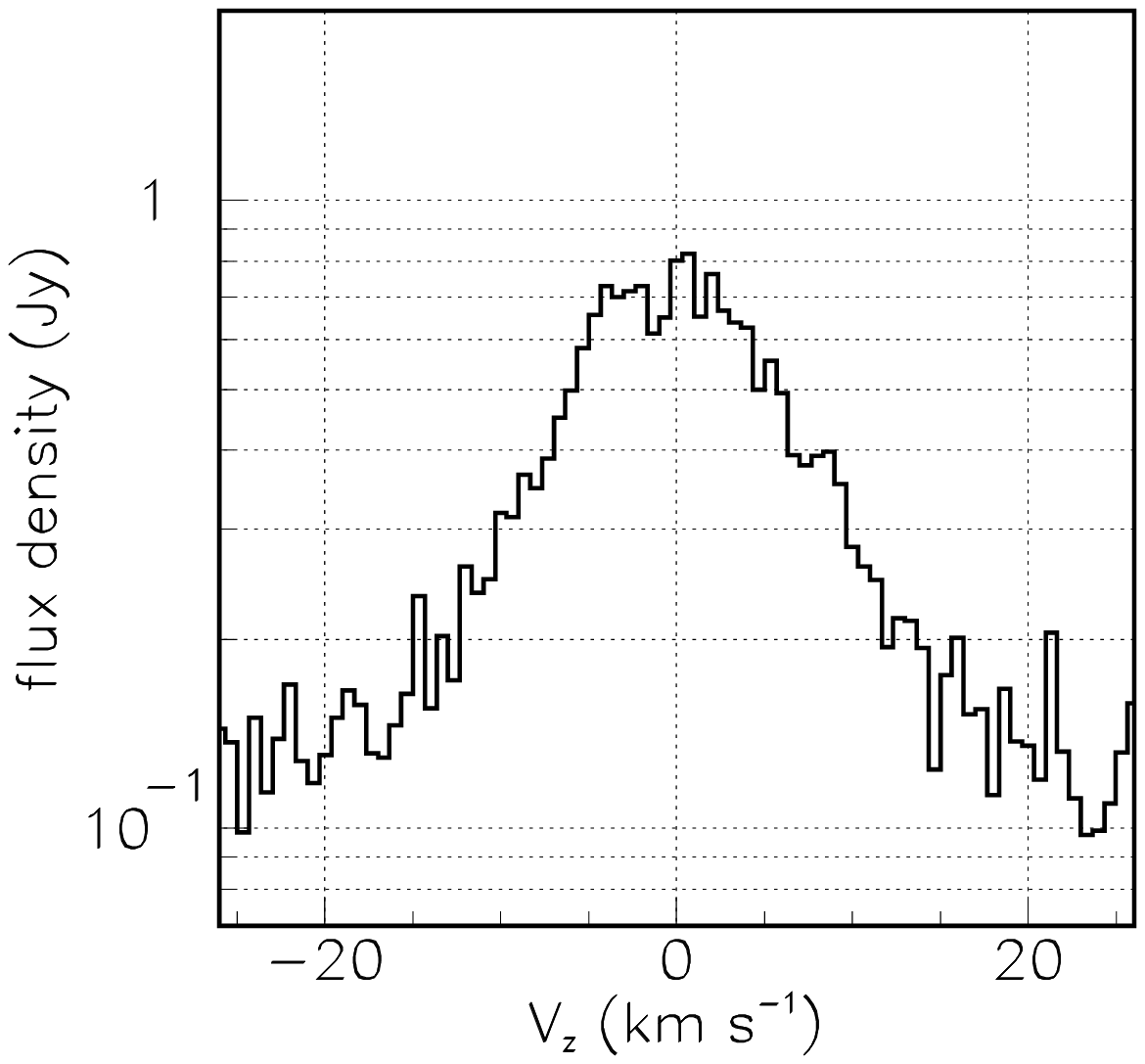}
  \includegraphics[height=5.2cm,trim=.5cm 1.cm 0.cm 1.5cm,clip]{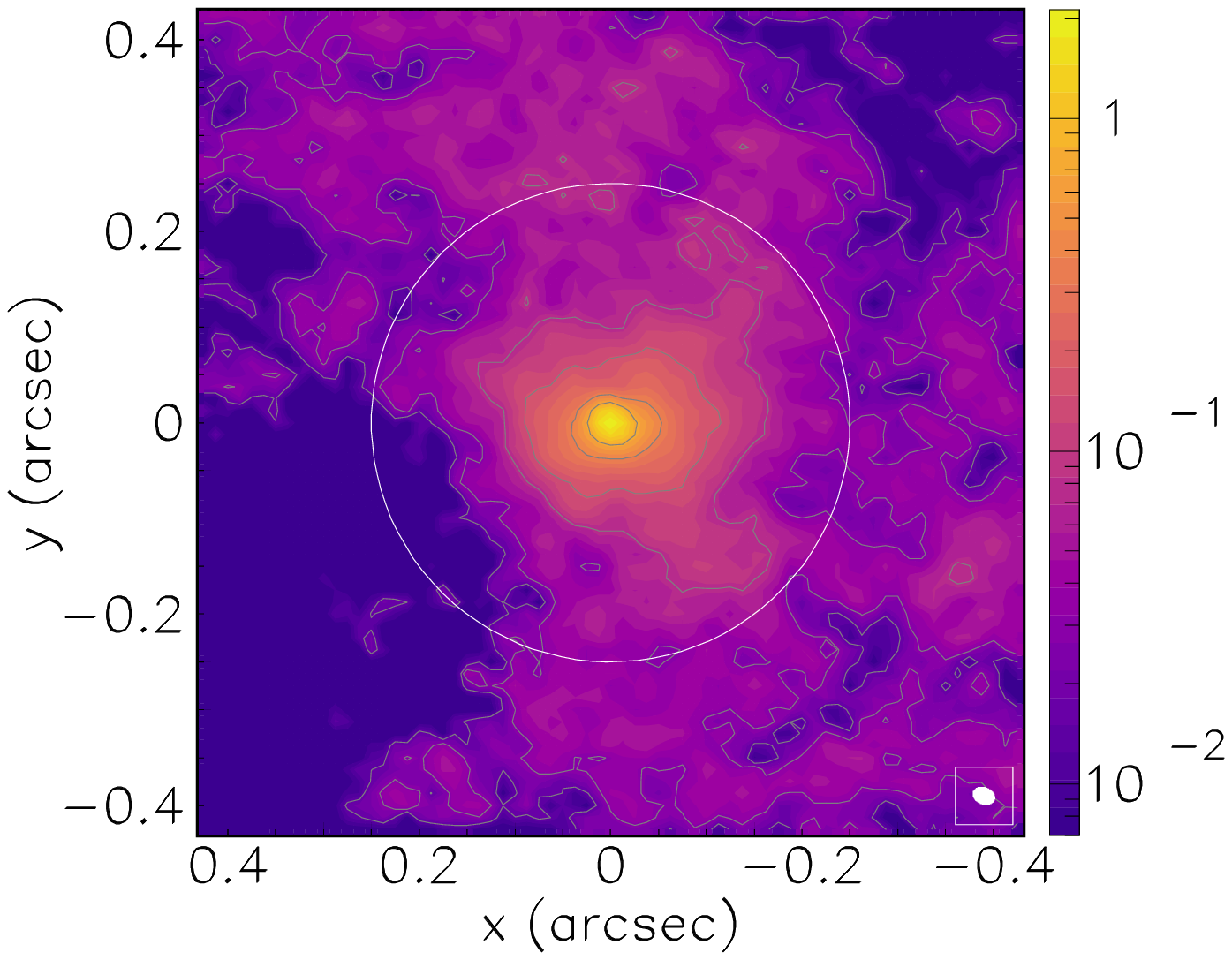}
  \includegraphics[height=5.2cm,trim=.5cm 1.cm .5cm 1.5cm,clip]{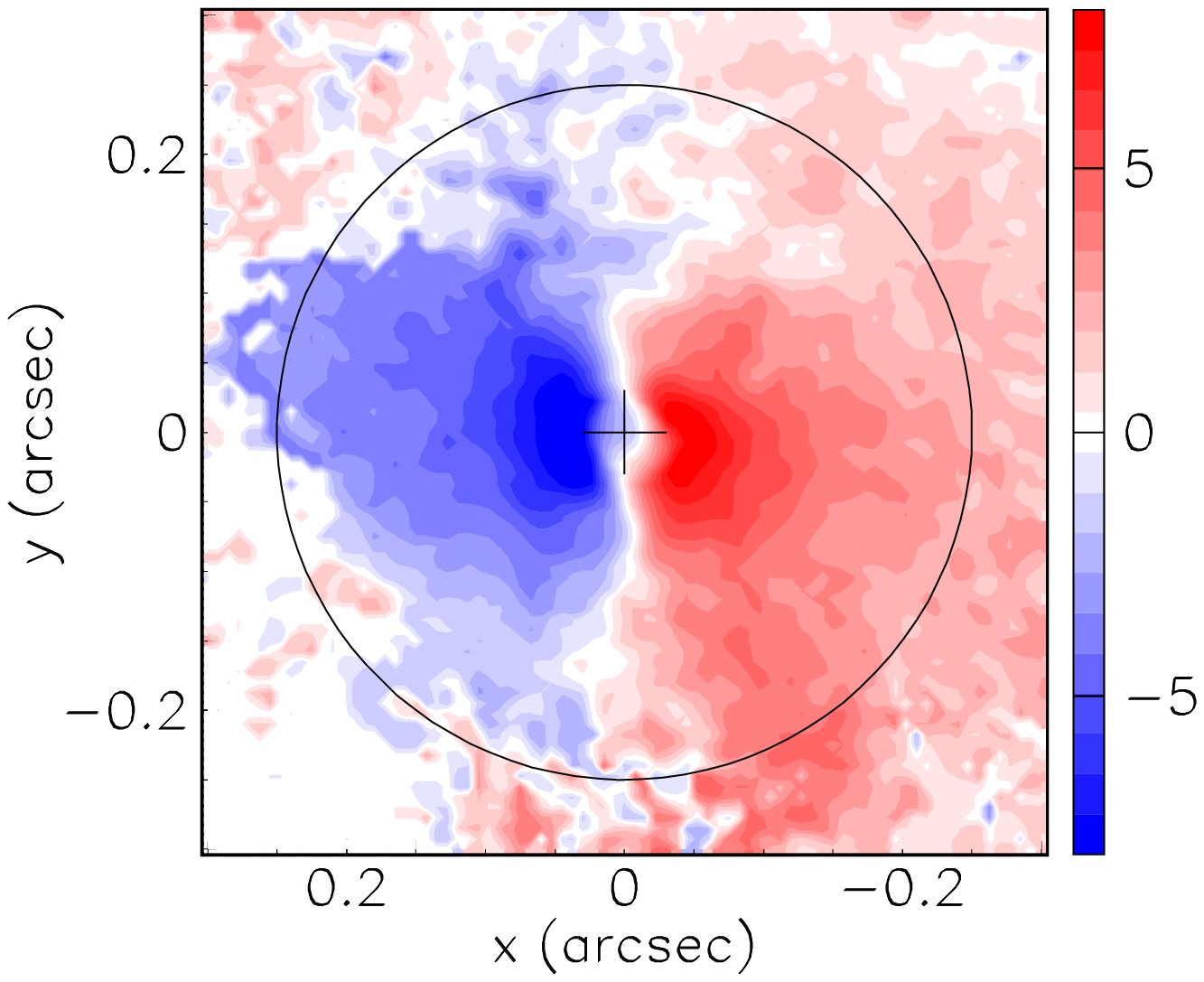}
  \includegraphics[height=5.2cm,trim=.5cm 1.cm 1.5cm 1.5cm,clip]{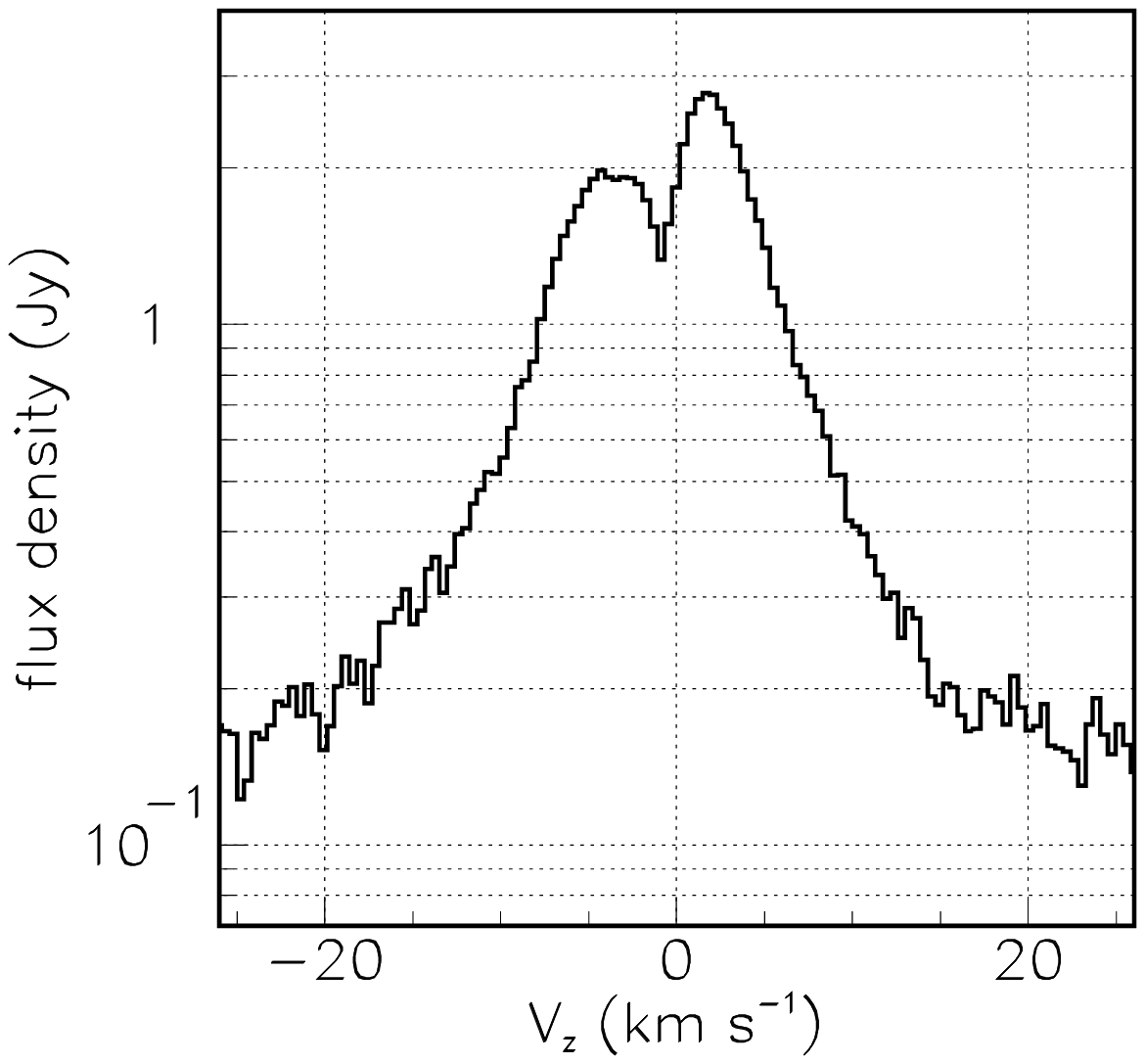}
    \caption{Left: Intensity maps integrated over $|V_z|$$<$20 \kms. Middle: maps of the mean Doppler velocities (moment 1); a 3-$\sigma$ cut has been applied. A circle of radius 0.25 arcsec is shown on each map to indicate the region over which the spectra in the right panels are integrated. Right: Doppler velocity spectra integrated over $R$$<$0.25 arcsec (16 au). The upper row is for $^{12}$CO(3-2) emission, the middle row for $^{13}$CO(3-2) emission and the lower row for $^{29}$SiO(8-7) emission. The units of the colour bars are Jy beam$^{-1}$ \kms\ in the left panels and \kms\ in the middle panels.}
 \label{fig3}
\end{figure*}

In a first step we study the close neighbourhood of the star using the first set of observations. We analyse the second set in Section 5. Figure \ref{fig3} displays intensity maps, mean Doppler velocity <$V_z$> (moment 1) maps and Doppler velocity spectra for each of the CO(3-2) and $^{29}$SiO(8-7) line emissions. They are shown as a reminder: the reader is referred to the detailed analyses of \citet{Kervella2016} and \citet{Homan2017}, which give interpretations in terms of a nearly edge-on rotating disc.

\begin{figure*}
  \centering
  \includegraphics[height=7cm,trim=.5cm 1.cm 1.5cm 1.5cm,clip]{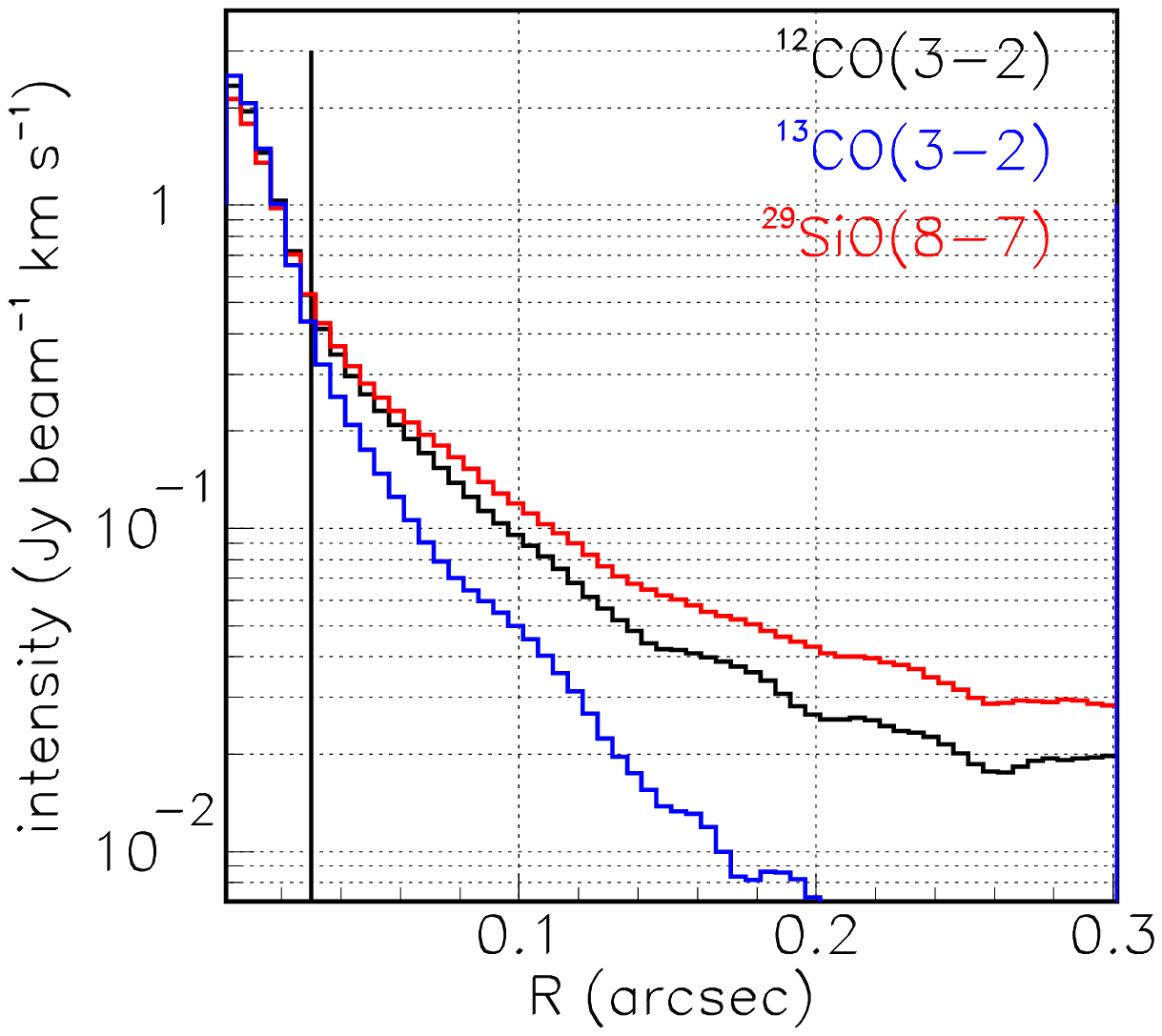}
    \caption{Radial distributions of $^{12}$CO(3-2) (black), $^{13}$CO(3-2) (blue) and $^{29}$SiO(8-7) (red) emissions averaged over position angles and integrated over $|$$V_z$$|$$<$20 \kms. The line at $R$$=$30 mas indicates the region beyond which the contribution of the continuum is negligible \citep{Kervella2016}. Note that the size of the beam is nearly the same for each of the three lines.}
 \label{fig4}
\end{figure*}

Figure \ref{fig4} compares the radial distributions of the $^{12,13}$CO(3-2) and $^{29}$SiO(8-7) line emissions. The contribution of continuum emission is seen to extend to $\sim$0.03 arcsec ($\sim$1.9 au) as expected \citep{Kervella2016}, beyond which $^{12}$CO and SiO emissions have similar intensities. The $^{13}$CO emission is nearly optically thin \citep{Homan2017}. All three line intensities decrease very steeply with $R$, illustrating the confinement of the CO and SiO molecules to the gas and dust disc described by \citet{Homan2017} and \citet{Kervella2016}. The steeper decrease of the $^{13}$CO emission reveals the stronger absorption of the optically thicker $^{12}$CO and SiO line emissions when approaching the star.

\begin{table*}
  \centering
  \caption{Main parameters related to the line emissions studied in the article. The intensities $f$ listed in the lower rows are for an optically thin layer having a column density of 1 molecule cm$^{-3}$ arcsec at temperatures of 1000 K, 100 K and 10 K, respectively. They are given in units of Jy arcsec$^{-2}$ \kms. }
  \label{tab1}
\begin{tabular}{cccccc}
  \hline
Line&
$^{12}$CO(3-2)&
$^{13}$CO(3-2)&
$^{29}$SiO(8-7)&
$^{28}$SiO(5-4)&
$^{12}$CO(2-1)\\
\hline
Beam (mas$^2$)&
24$\times$18&
25$\times$19&
24$\times$18&
370$\times$330&
320$\times$310\\
Frequency (GHz)&
345.796&
330.588&
342.981&
217.105&
230.538\\
Noise ($\sigma$, mJy\,beam$^{-1}$)&
2.4&
2.2&
1.5&
10&
2.5\\

d$V_z$ (\kms)&
0.212&
0.222&
0.427&
0.337&
0.159\\
Pixel size (mas$^2$)&
2.52&
2.52&
2.52&
502&
502\\
$E_{up}$ (K)&
33.19&
31.73&
74.08&
31.26&
16.60\\
$A_{ji}$(10$^{-6}$ s$^{-1}$)&
2.5&
2.2&
2100&
520&
0.69\\
$k$&
2.77&
2.65&
1.03&
1.04&
2.77\\
$f_{1000}$ &
0.016&
0.014&
12.2&
2.0&
0.0034\\
$f_{100}$& 
0.13&
0.11&
62.6&
15.6&
0.029\\
$f_{10}$&
0.062&
0.061&
0.80&
9.3&
0.065\\
\hline
\end{tabular}
\end{table*}

 \section{Absorption over the stellar disc}

\begin{figure*}
  \centering
  \includegraphics[height=6cm,trim=0.cm 0.5cm 1.5cm 1.5cm,clip]{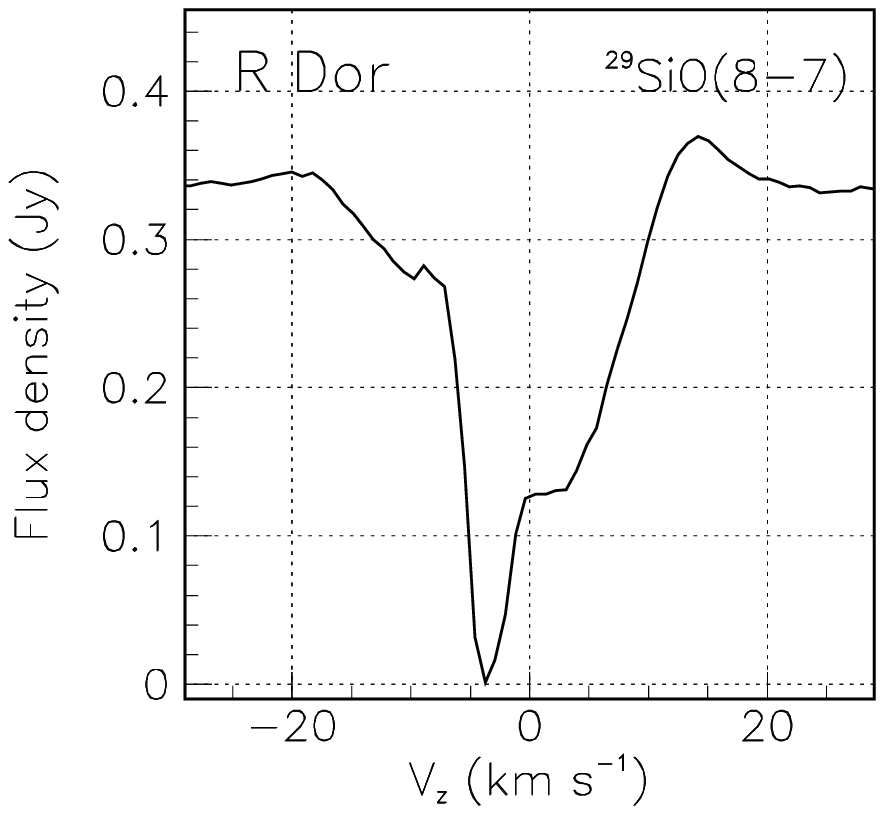}
  \includegraphics[height=6cm,trim=0cm .5cm 1.5cm 1.5cm,clip]{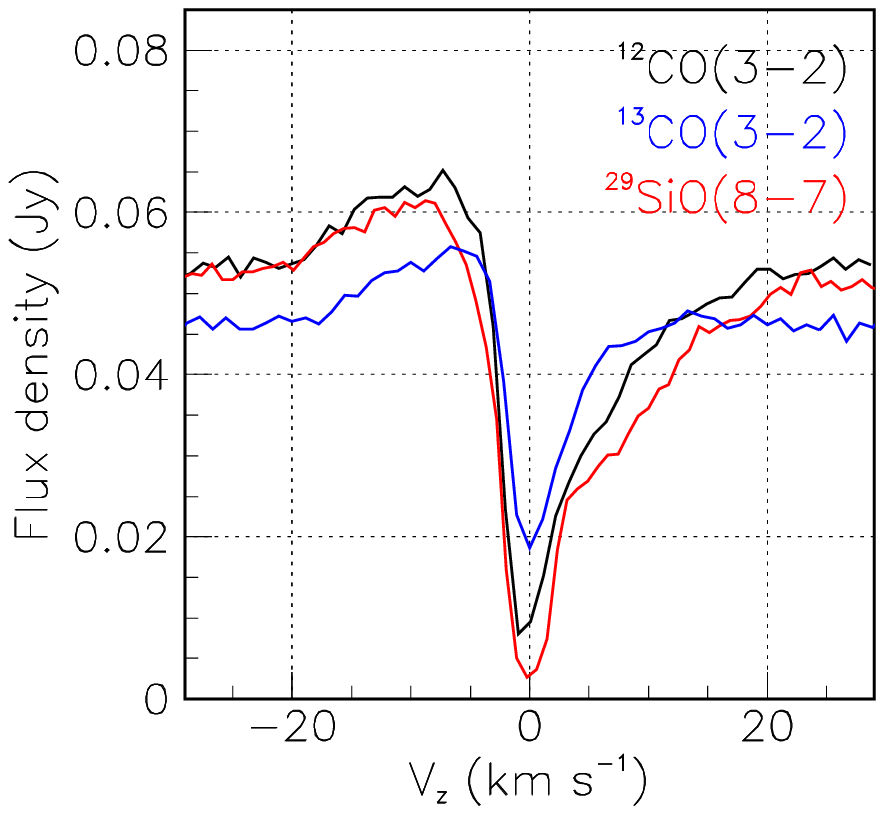}
    \caption{Absorption spectra over the stellar disc, integrated over $R$$<$0.035 arcsec (2.1 au) for R Dor (left) and over $R$$<$0.012 arcsec (0.8 au) for L$_2$ Pup (right). $^{29}$SiO(8-7) data are shown for both stars and CO(3-2) data for L$_2$ Pup only. The slightly different continuum levels of the L$_2$ Pup emission of the two CO isotopologues is the result of the associated frequency difference.}
 \label{fig5}
\end{figure*}

 High angular resolution observations of the emission of molecular lines resolve the stellar disc and reveal the absorption of its continuum emission by the surrounding gas. \citet{Wong2016} were first to present a detailed study of this feature using ALMA observations of Mira Ceti.  It has since been observed on several other oxygen-rich AGB stars: we show in the left panel of Figure \ref{fig5} the case of R Dor, the other nearby such star, measured inside a circle of 35 mas radius \citep{Nhung2021}. The absorption spectrum over the stellar disc is a rich source of information. In particular, it probes the whole CSE, unaffected by maximal resolvable scale constraints: it provides a direct measure of the terminal velocity in the form of a narrow absorption peak, visible at $V_z$$\sim$$-$4 \kms\ in the case of R Dor; moreover, it probes the very close environment of the star, where pulsations and convective cell ejections produce shocks, providing direct evidence for Doppler velocities in excess of 10 \kms.
 
 The absorption spectra measured inside a circle of 12 mas radius centred on L$_2$ Pup are shown in the right panel of Figure \ref{fig5} for both $^{12,13}$CO(3-2) and $^{29}$SiO(8-7) emissions. They are located well within the inner rim of the gas disc, evaluated at $R$$\sim$31 mas by \citet{Homan2017} and should receive no contribution from rotation. Their comparison with the R Dor spectrum invites several comments.
 
The L$_2$ Pup spectra peak at 0, giving evidence for the absence of significant wind along the line of sight crossing the star in its centre. This observation suggests that the wind of L$_2$ Pup, contrary to that of R Dor, lacks spherical symmetry, as will be quantitatively confirmed in Section 5. At first glance, this result is not surprising, the CSE consisting essentially in this direction of a gravitationally bound disc of gas and dust. In the present state of the evolution of L$_2$ Pup, if there is a wind escaping its gravity, it must mostly be in a broad cone around the disc axis, close to the plane of the sky and, therefore, displaying a low mean Doppler velocity. We shall come back to this point in Section 5. However, in the past, before the 1994 dimming episode, there might have been significant mass loss along the line of sight crossing the star in its centre. No trace of it is to be seen on the absorption spectra displayed in Figure \ref{fig5}, however. The $^{12}$CO line emission must be probing the CSE over several arcseconds away from the star and this is therefore evidence for the absence of significant wind, with a velocity in excess of $\sim$1 \kms, along the line of sight crossing the star in its centre.

The L$_2$ Pup spectra reveal a more important component of gas in-fall than for R Dor; they reach Doppler velocities of $\sim$20 \kms in absolute value, suggesting comparable levels of activity in the close neighbourhood of each star. In both R Dor and L$_2$ Pup absorption of the SiO line is nearly complete over the stellar disc, a result of the important optical thickness. The absorptions of the CO lines are slightly less important, larger for $^{12}$CO than for $^{13}$CO as expected. The continuum levels of both stars are similar once accounting for slightly different frequencies and normalisation to a same area, a result to be expected to the extent that both stars have similar temperatures.    

\begin{figure*}
  \centering
  \includegraphics[width=17cm,trim=0cm 1.85cm .5cm 1.cm,clip]{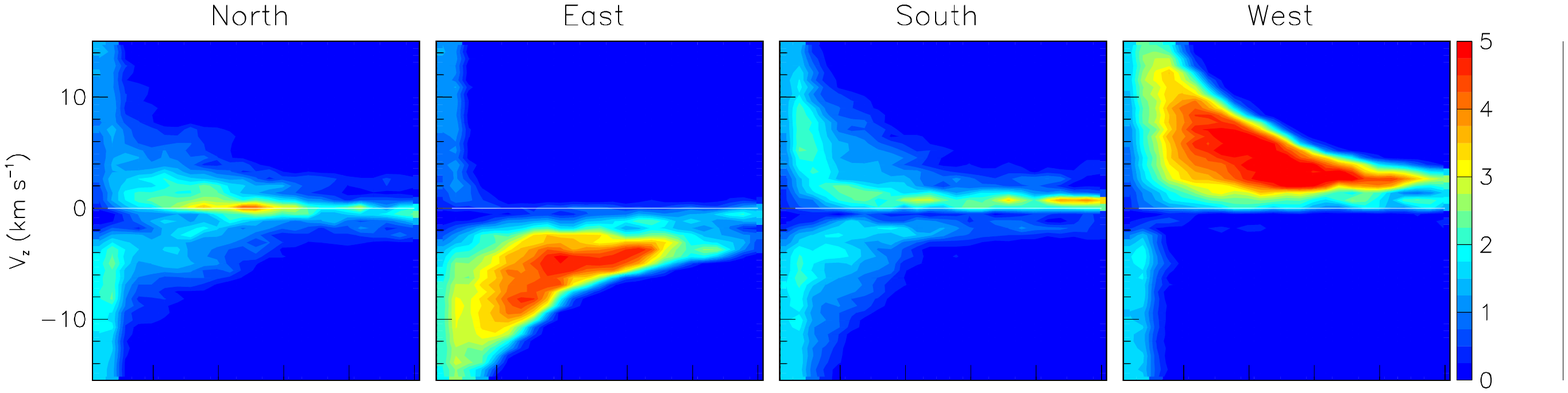}
  \includegraphics[width=17cm,trim=0cm 1.85cm .5cm 1.83cm,clip]{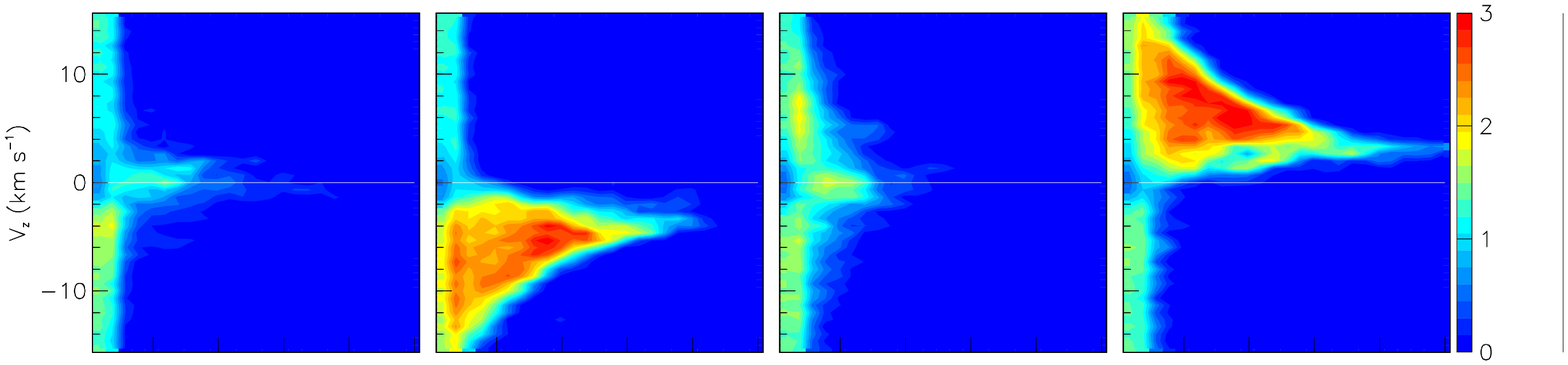}
  \includegraphics[width=17cm,trim=0cm .5cm .5cm 1.83cm,clip]{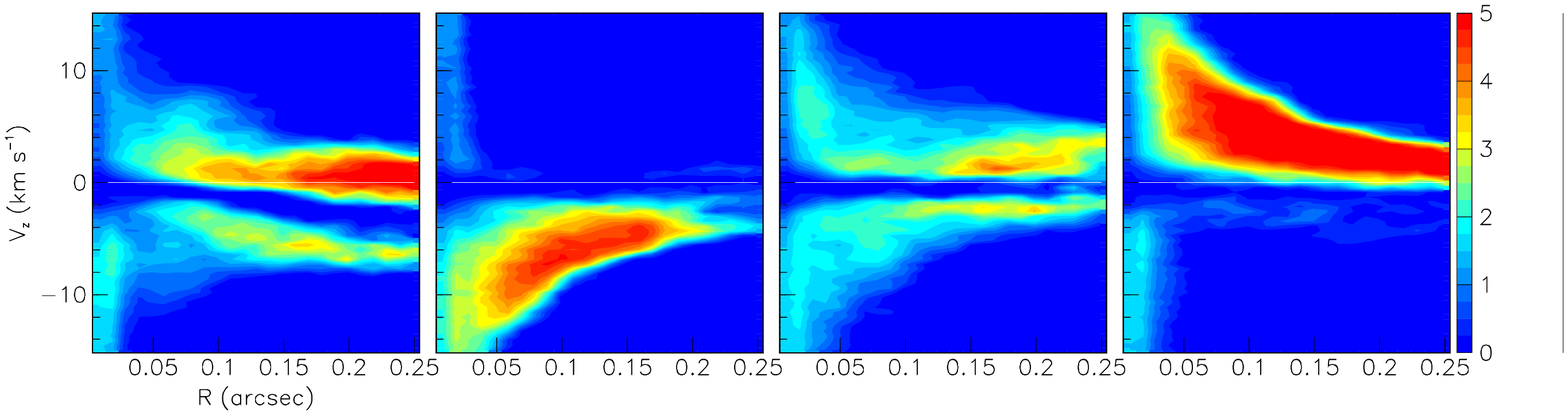}
    \caption{PV maps of $V_z$ vs $R$ for each of the north, east, south and west quadrants (from left to right). The upper row is for $^{12}$CO(3-2) emission, the middle row for $^{13}$CO emission and the lower row for $^{29}$SiO(8-7) emission. The colour scales are in units of Jy arcsec$^{-1}$.}
 \label{fig6}
\end{figure*}

The important optical thickness of the gas and dust disc revealed by the absorption spectra observed over the stellar disc brings up the question of its extension beyond the stellar disc. In other oxygen-rich AGB stars, absorption is usually observed to extend well beyond the stellar disc for $^{28}$SiO emission, in the form of self-absorption, as well as, to a lesser extent, for $^{29}$SiO and $^{12}$CO emissions while being nearly confined to the stellar disc for $^{13}$CO emission. Figure \ref{fig6} displays PV maps of $V_z$ vs $R$ for each of the northern, eastern, southern and western quadrants separately. Significant absorption over $\sim$1 \kms\ below systemic velocity is seen to persist up to $\sim$15 au in most quadrants for both $^{12}$CO and $^{29}$SiO lines. Its slightly negative Doppler shift (also visible on the $^{29}$SiO absorption spectrum of Figure \ref{fig5}) may reveal a possible expansion of the disc, at the level of 0.8$\pm$0.5 \kms. This is smaller than the expansion velocity generally assumed in the earlier literature, of the order of 3 \kms\ \citep{Winters2002}. In contrast, absorption of the $^{13}$CO line is limited to the stellar disc.

\section{High Doppler velocity wings}

The absorption spectra observed over the stellar disc and discussed in the preceding section give evidence for large velocities in the neighbourhood of the star, in line with the standard picture of a region dominated by shocks from pulsations and convection cell ejections, boosting the gas to distances where acceleration from dust grains becomes effective. These high velocities are known to be seen as high Doppler velocity wings on the profiles of molecular lines observed on lines of sight crossing the star near its centre. Their evolution when moving away from the star centre provides an estimate of the thickness of the layer where shocks are important. Such an evaluation was done in the case of Mira Ceti \citep{Hoai2020}, locating the layer between 5 and 15 au from the centre of the star.  In the case of R Dor, a confinement of the layer within $\sim$12 au from the centre of the star has been obtained \citep{Nhung2021}.

\begin{figure*}
  \centering
  \includegraphics[height=4.5cm,trim=.5cm 1cm 2cm 1.5cm,clip]{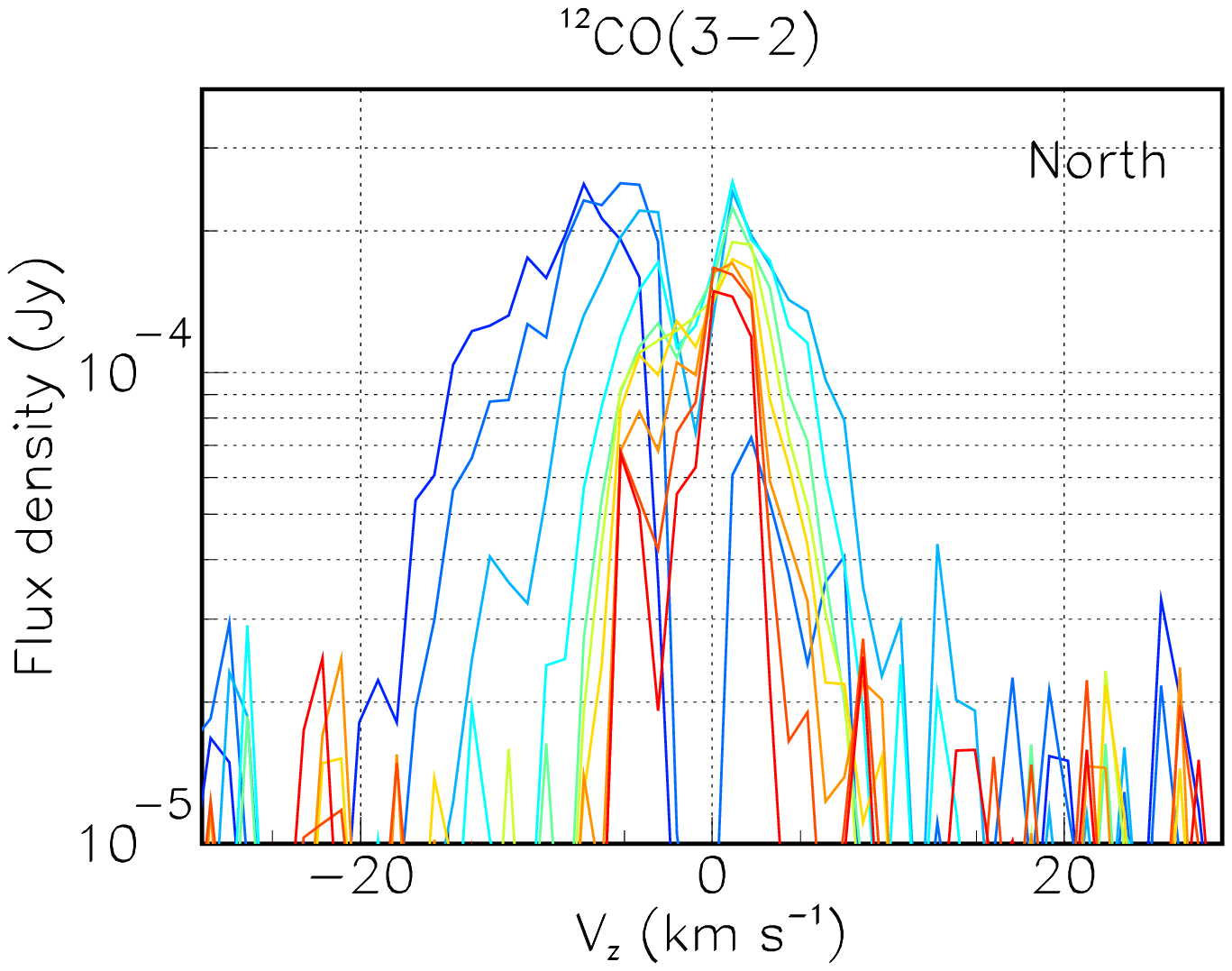}
  \includegraphics[height=4.5cm,trim=.5cm 1cm 2cm 1.5cm,clip]{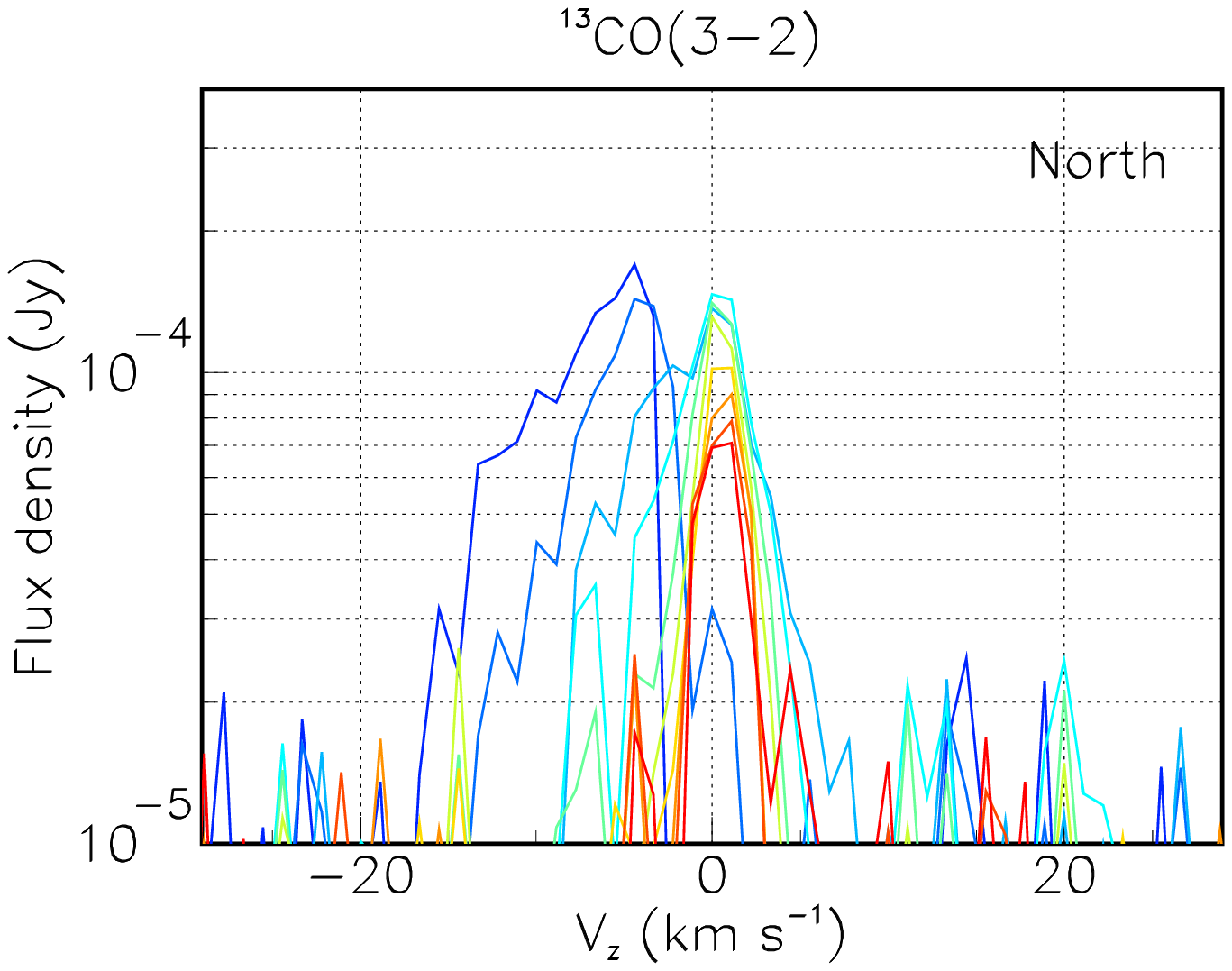}
  \includegraphics[height=4.5cm,trim=.5cm 1cm 2cm 1.5cm,clip]{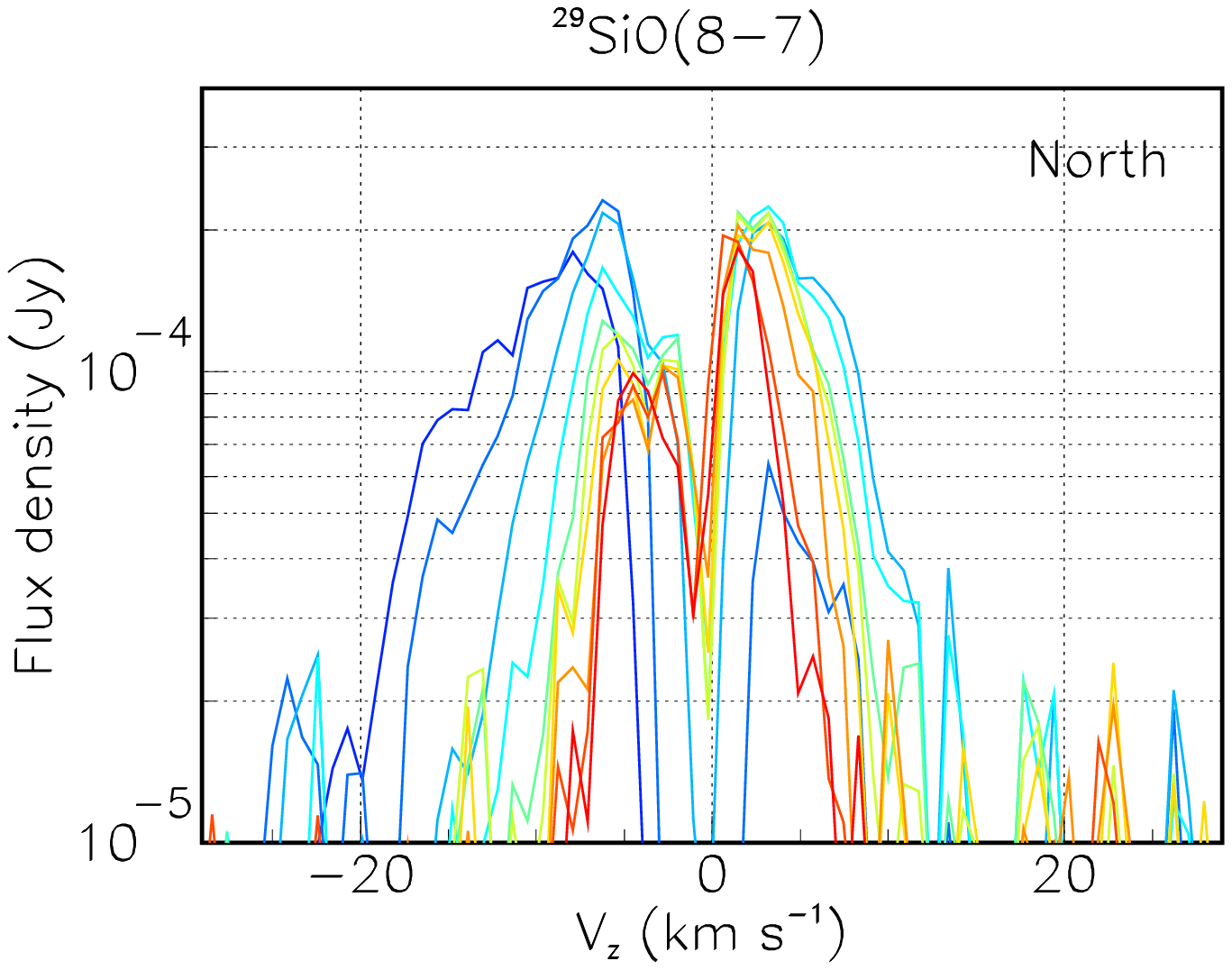}
  \includegraphics[height=4.5cm,trim=.5cm 1cm 2cm 1.5cm,clip]{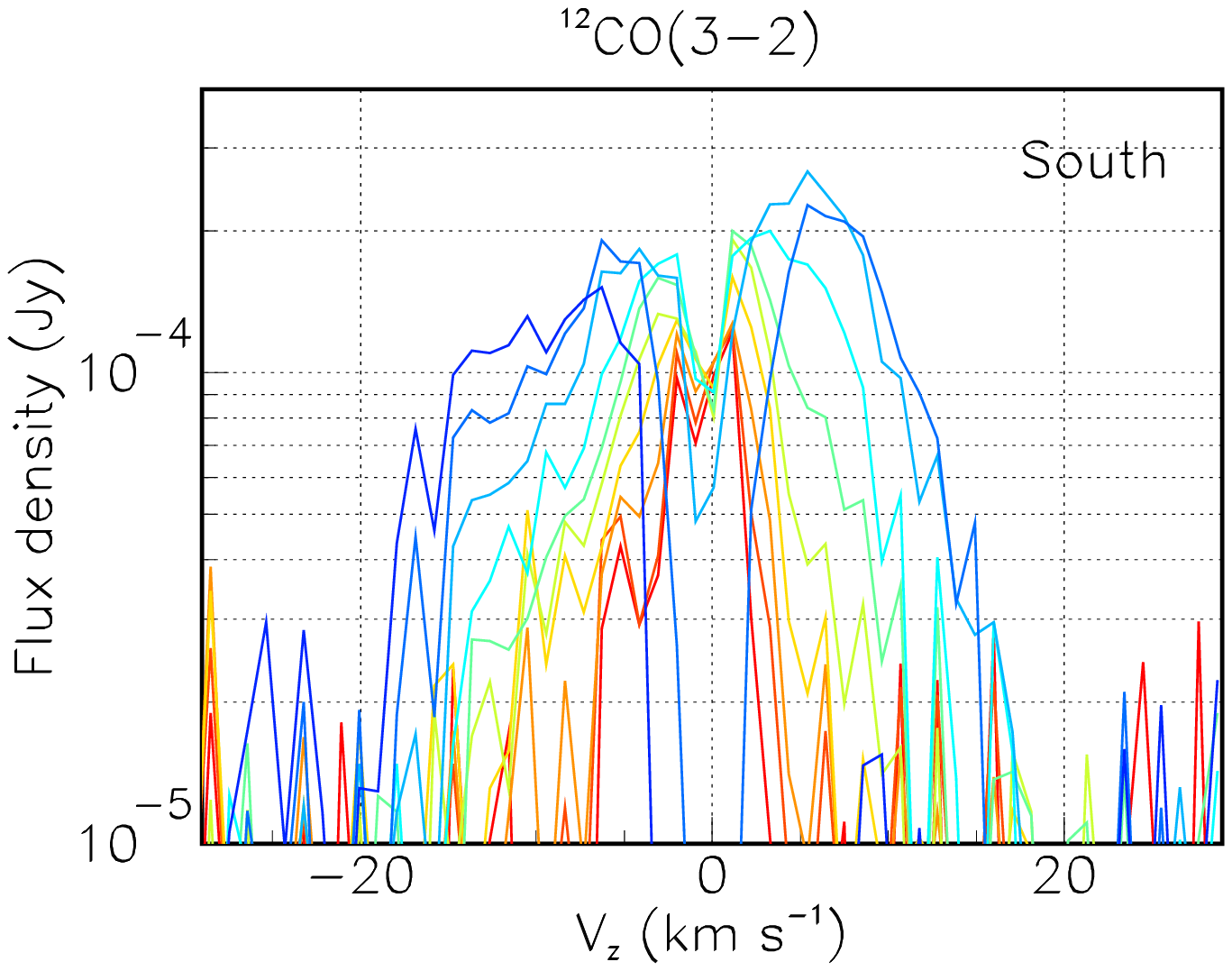}
  \includegraphics[height=4.5cm,trim=.5cm 1cm 2cm 1.5cm,clip]{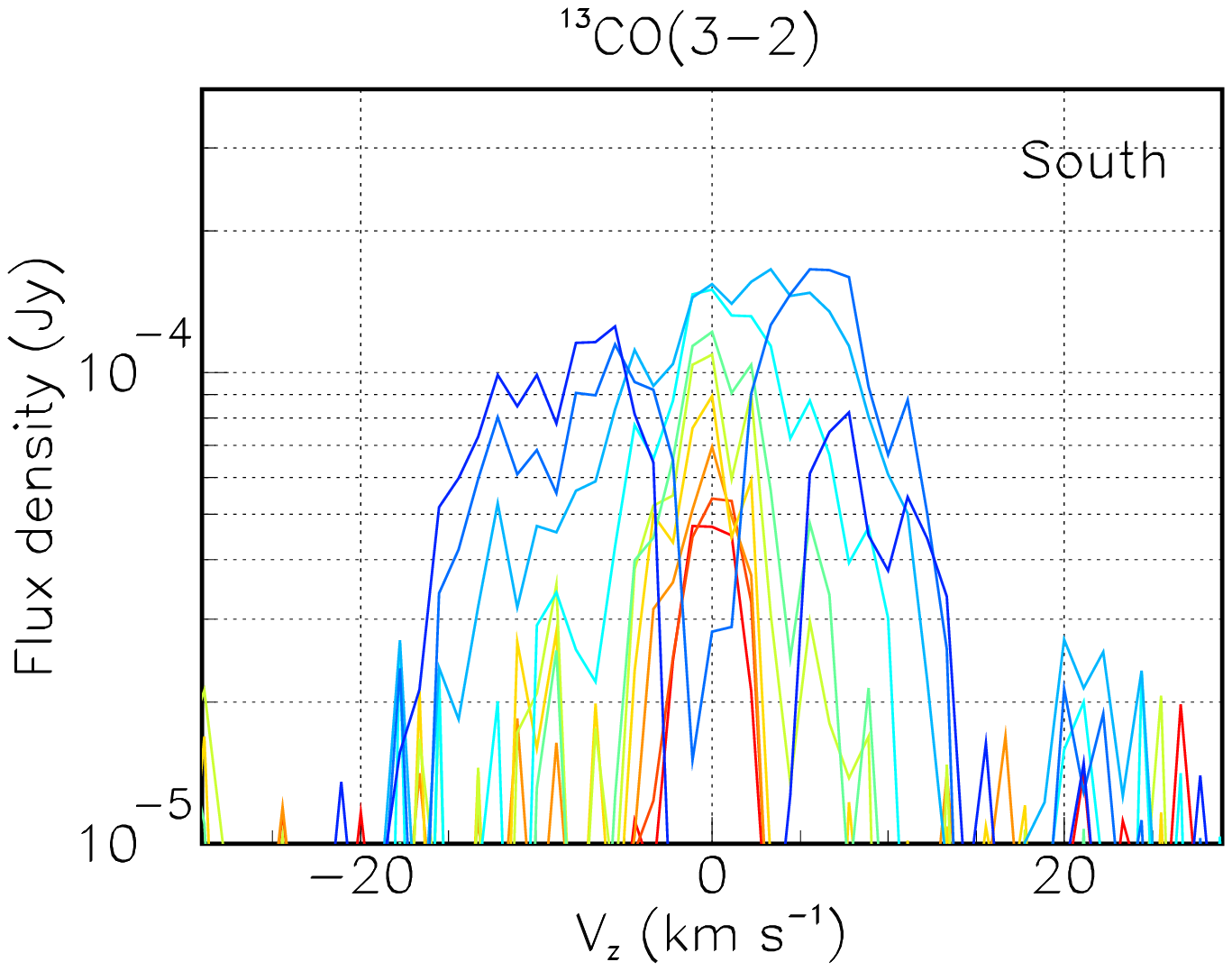}
  \includegraphics[height=4.5cm,trim=.5cm 1cm 2cm 1.5cm,clip]{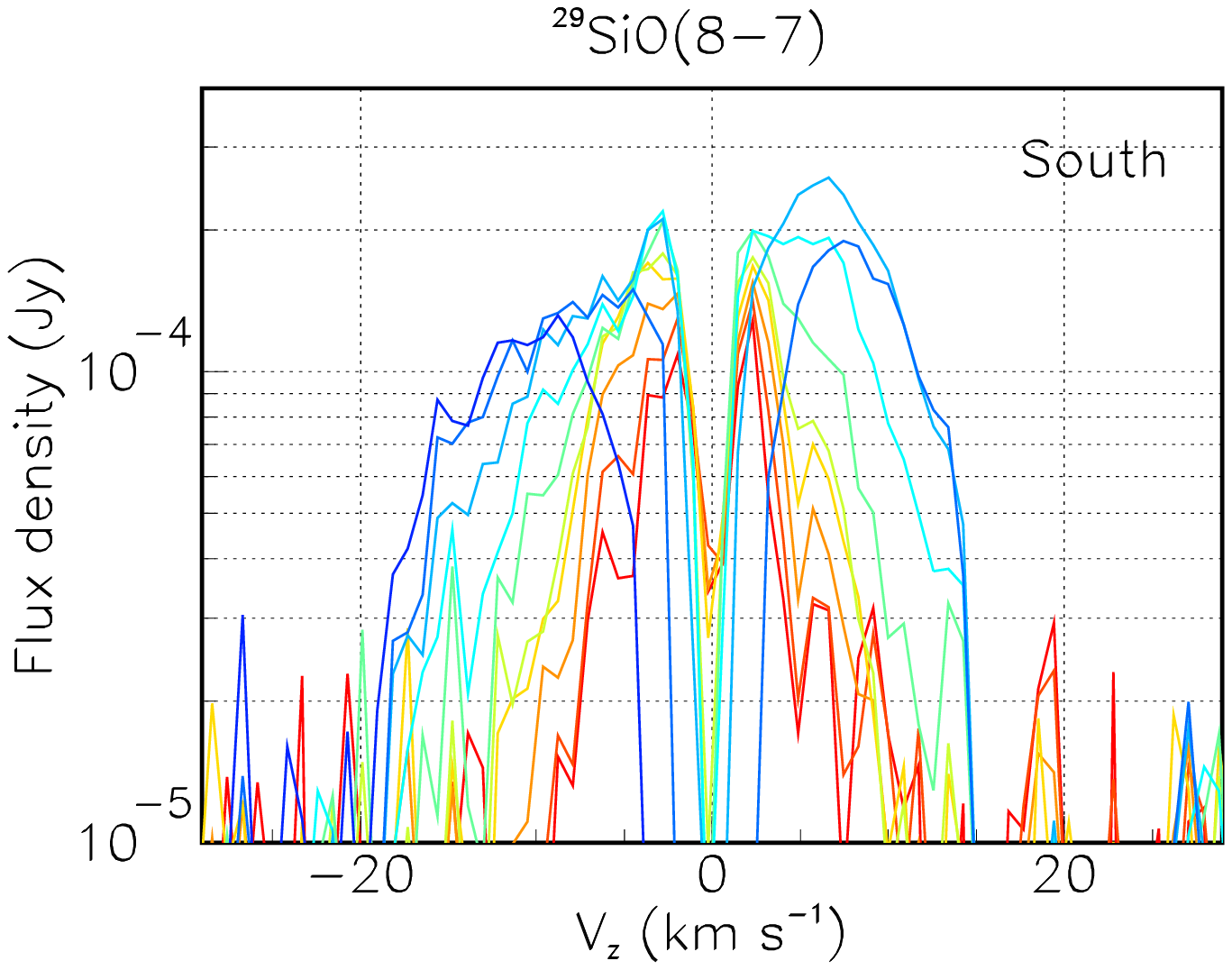}
    \caption{Evolution of the Doppler velocity spectrum when scanning at $x$$=$0 in 10 mas steps of $y$ (see text). The northern and southern scans are illustrated separately in the upper and lower rows, respectively. From left to right: $^{12}$CO(3-2), $^{13}$CO(3-2) and $^{29}$SiO(8-7). Colours evolve from blue to red when $|y|$ increases from 10 to 90 mas.}
 \label{fig7}
\end{figure*}

\begin{figure*}
  \centering
  \includegraphics[height=4.1cm,trim=.5cm 1.5cm 0.5cm 1.5cm,clip]{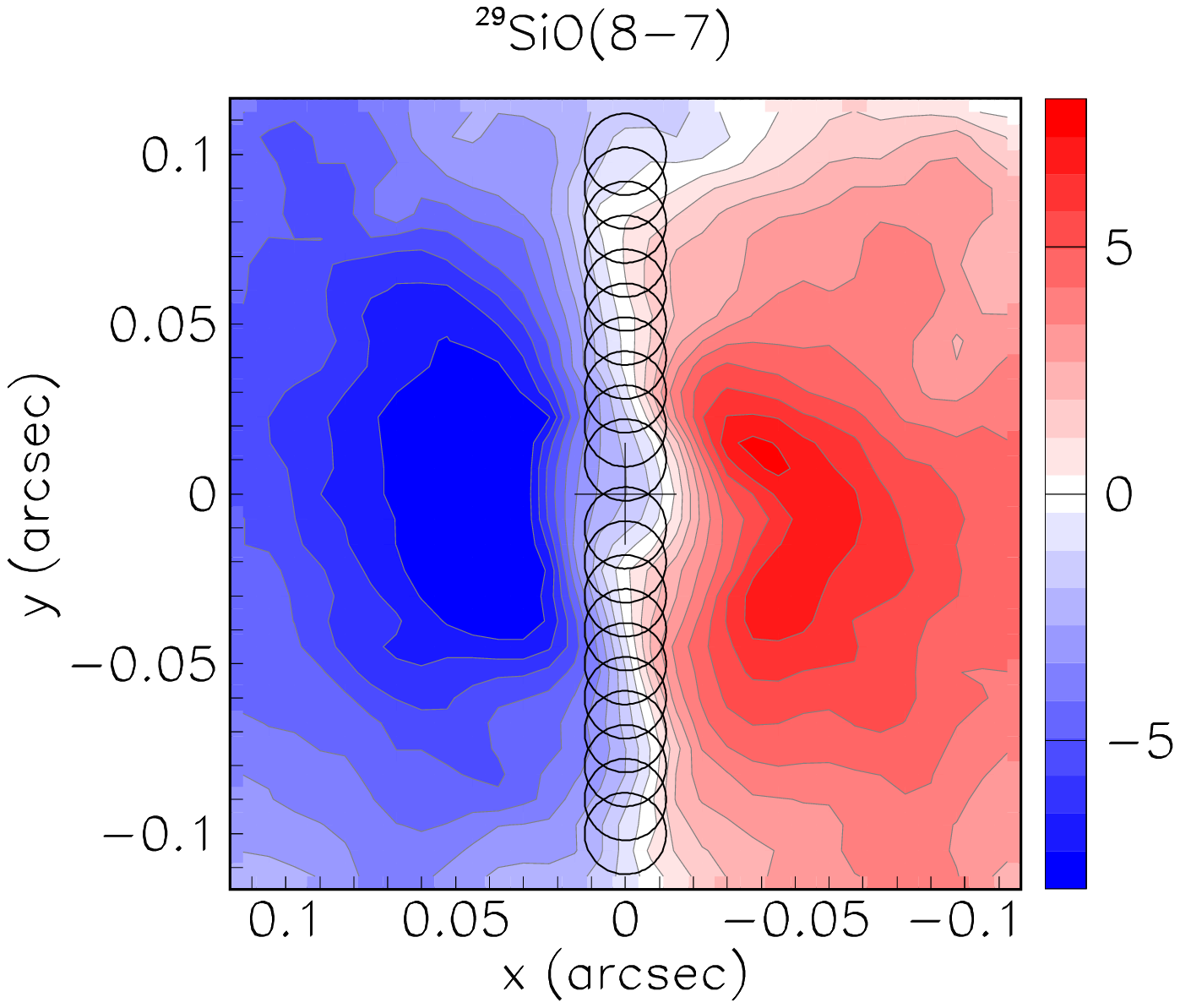}
  \includegraphics[height=4.1cm,trim=.5cm 1.5cm 2cm 1.5cm,clip]{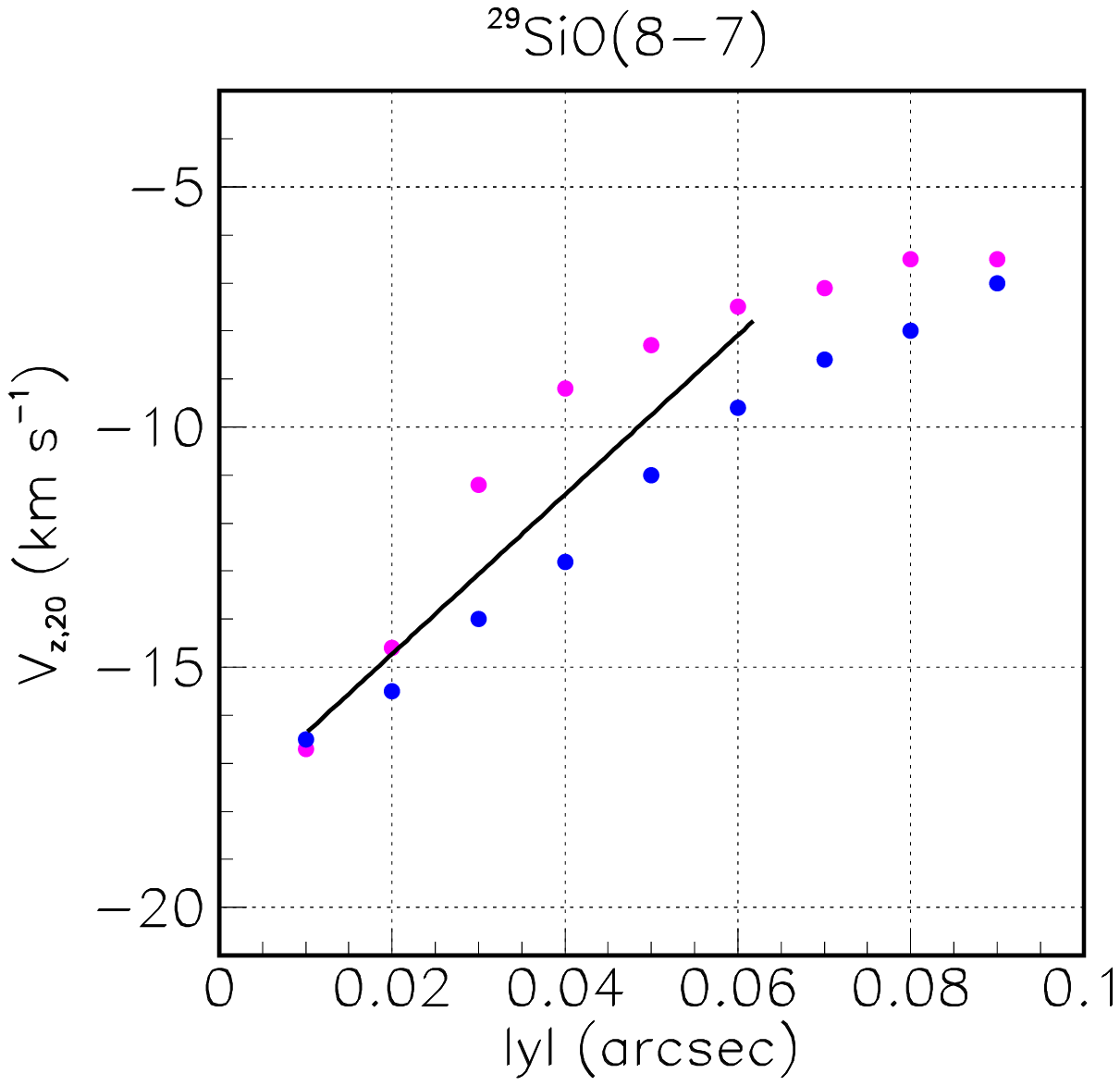}
  \includegraphics[height=4.1cm,trim=.5cm 1.5cm 2cm 1.5cm,clip]{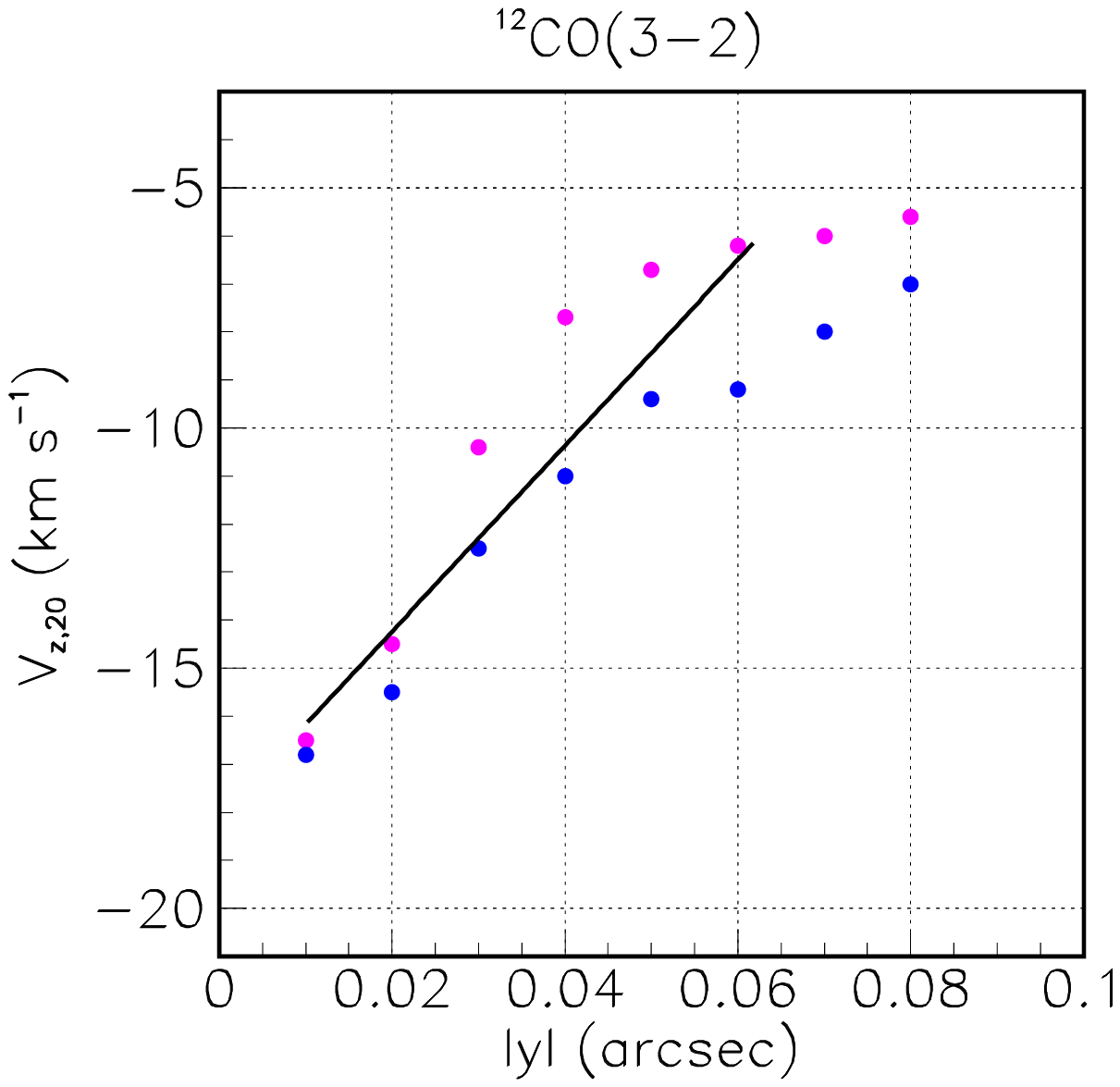}
  \includegraphics[height=4.1cm,trim=.5cm 1.5cm 2cm 1.5cm,clip]{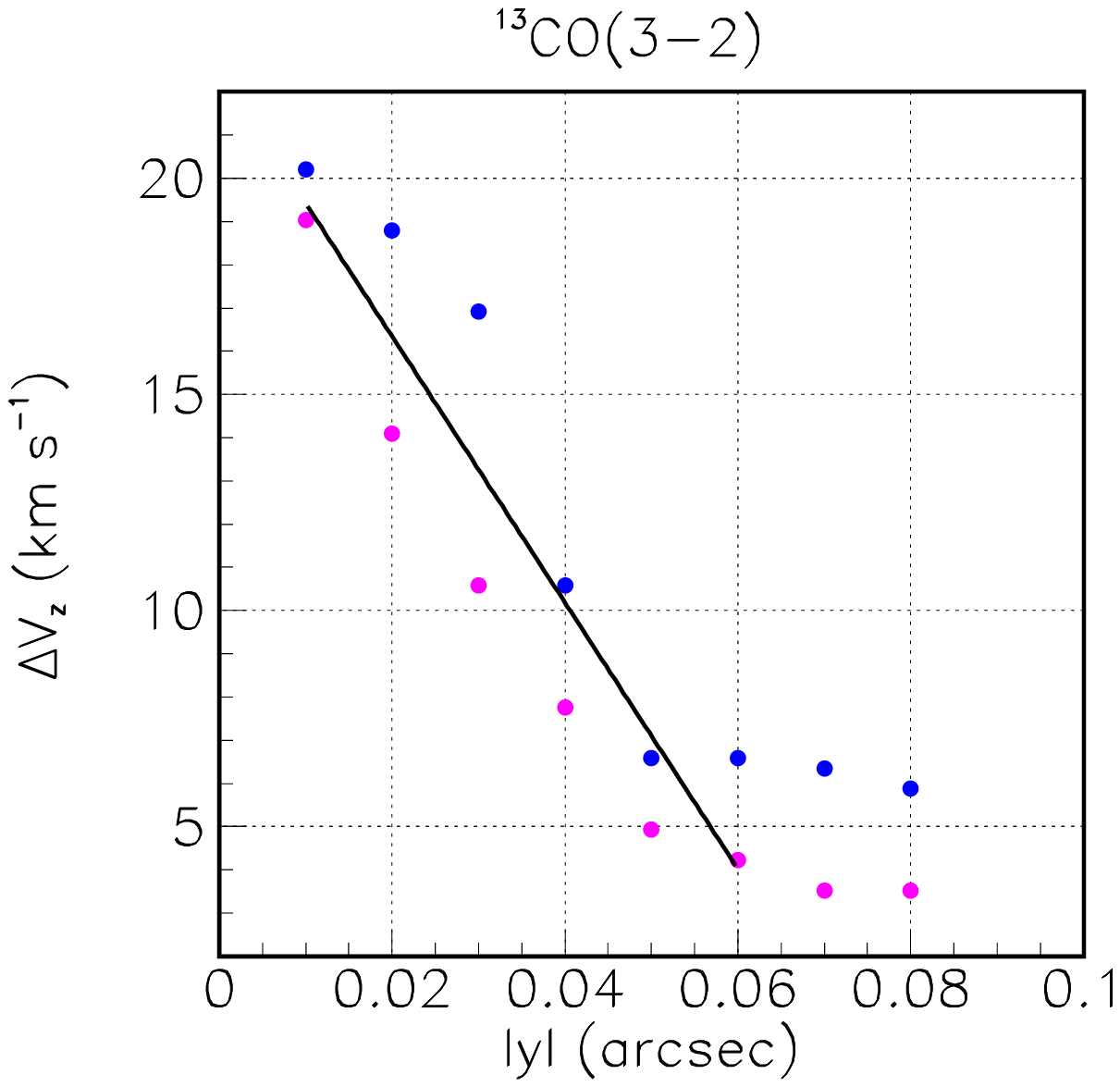}
    \caption{ Left: scanning steps superimposed on the $^{29}$SiO moment 1 map. Middle: dependence of $V_{z,20}$ on $|y|$ for the north (magenta) and south (blue) scans of $^{29}$SiO(8-7) (middle-left) and $^{12}$CO(3-2) (middle-right) emissions. Right: evolution of the $^{13}$CO line width (FWHM). In the three rightmost panels the lines show linear fits.}
 \label{fig8}
\end{figure*}

\begin{figure*}
  \centering
  \includegraphics[height=5cm,trim=.5cm 1.5cm 2cm 1.5cm,clip]{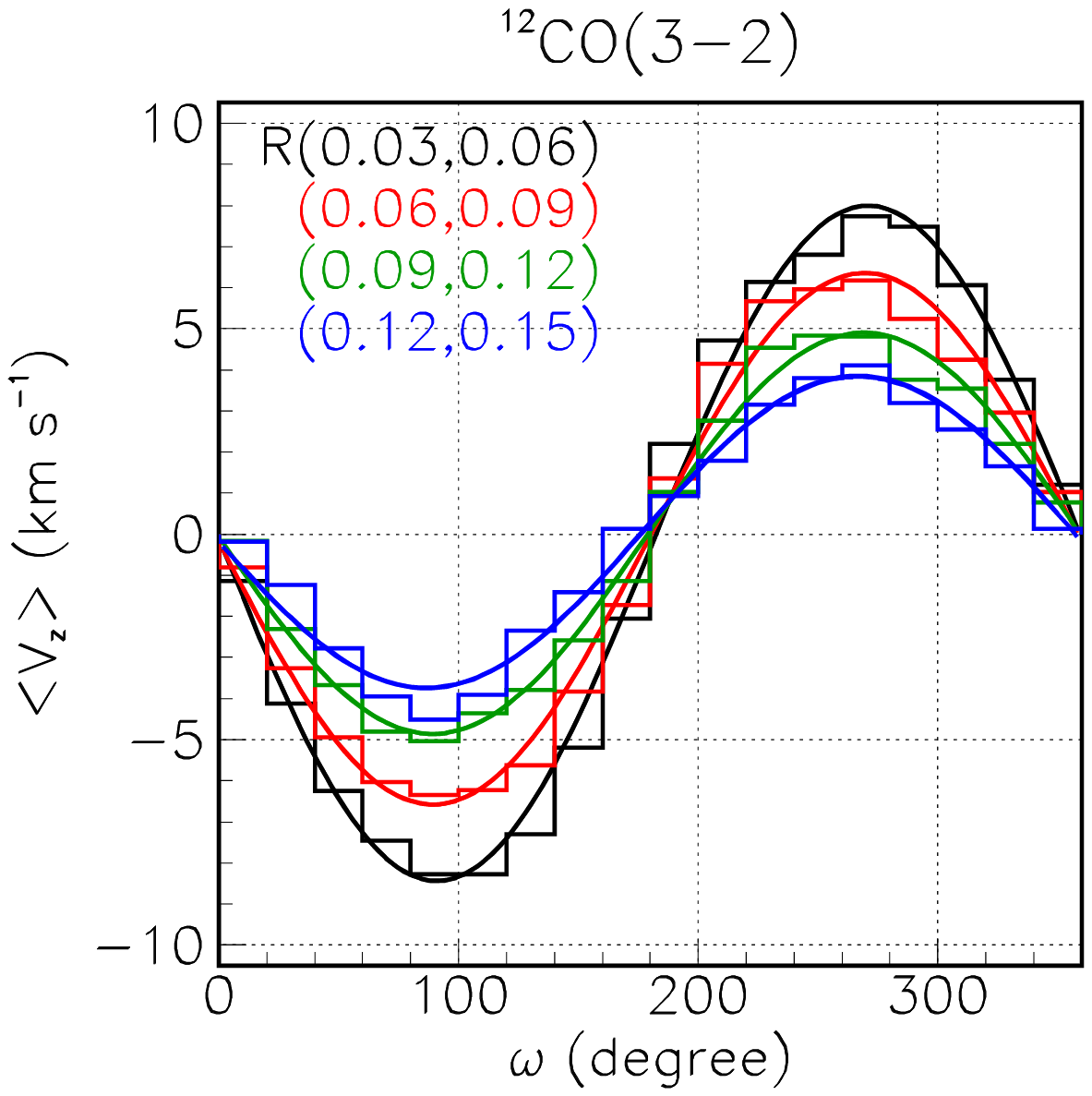}
  \includegraphics[height=5cm,trim=.5cm 1.5cm 2cm 1.5cm,clip]{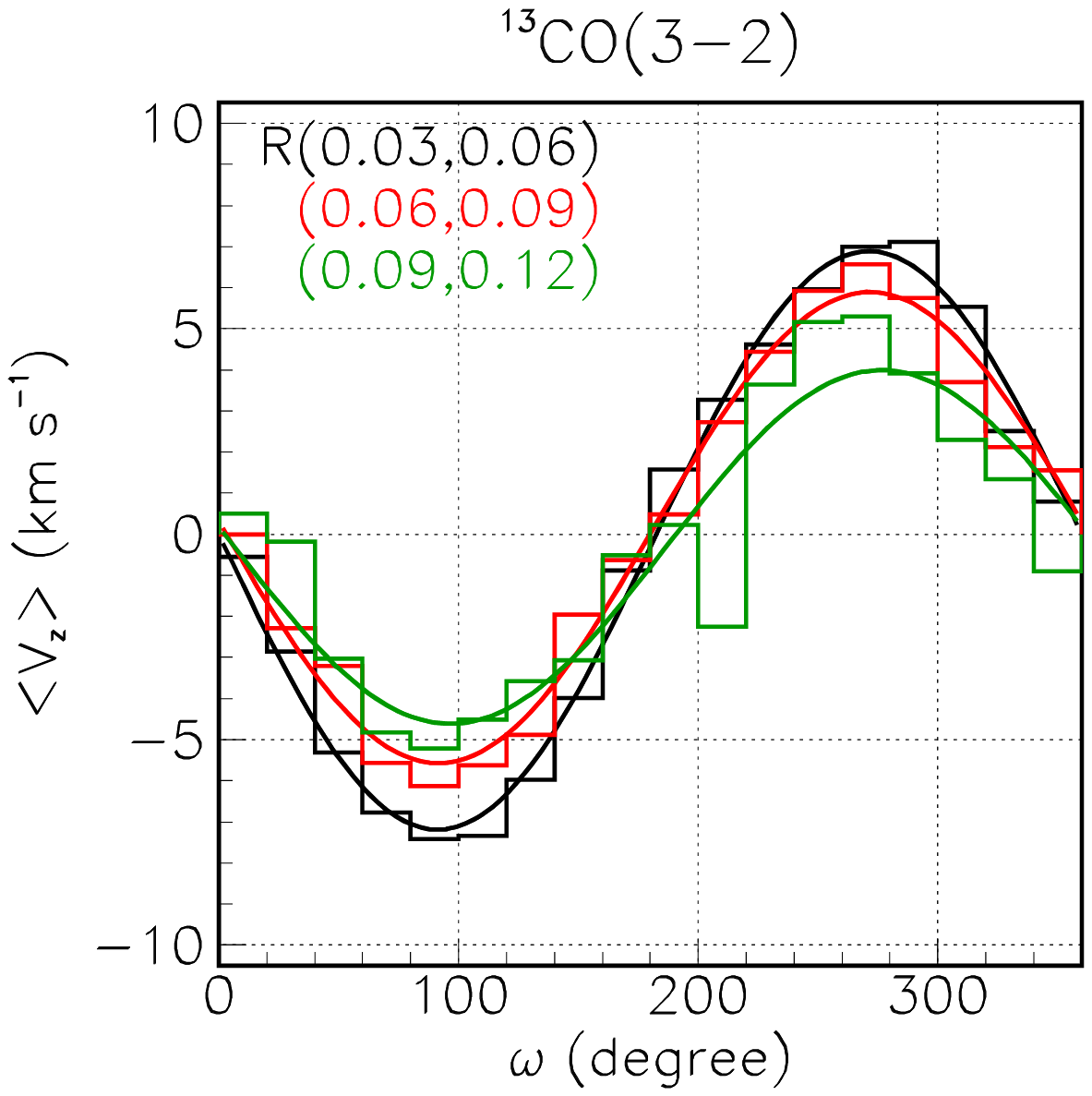}
  \includegraphics[height=5cm,trim=.5cm 1.5cm 2cm 1.5cm,clip]{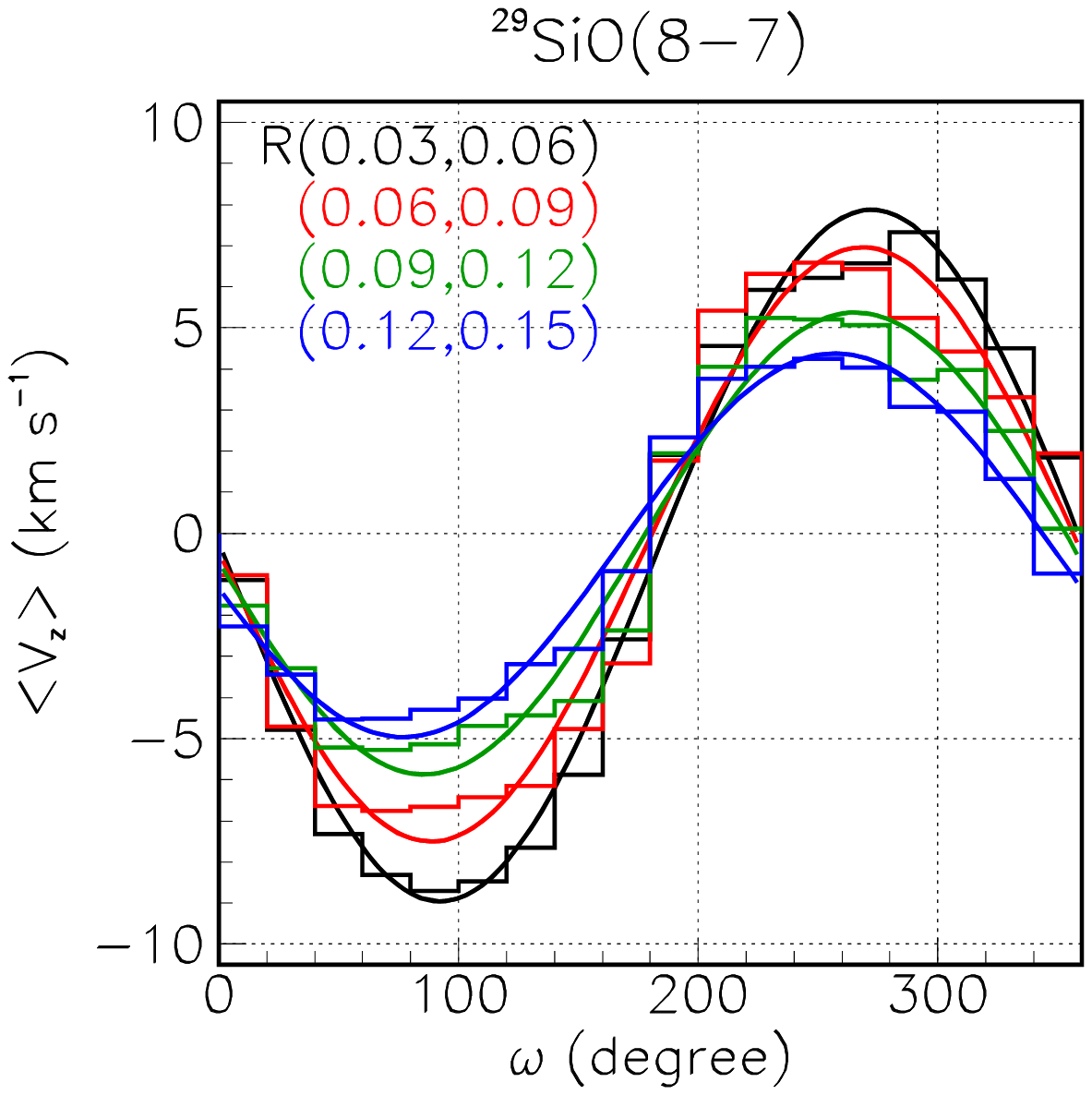}
    \caption{Dependence on position angle of the mean Doppler velocity, averaged over $R$ in successive rings as indicated in the inserts (in arcsec) and integrated over $|V_z|$$<$20 \kms. The lines show sine wave fits described in Table \ref{tab2}.}
 \label{fig9}
\end{figure*}

The present section aims at obtaining such an evaluation in the case of L$_2$ Pup, but the presence of high rotation velocities reaching close to the star makes the exercise more difficult. The inner rim of the gas disc is located at $\sim$2 au from the centre of the star \citep{Homan2017} and the diameter of the beam (FWHM) is $\sim$1.2 au: when staying on the $x$$=$0 line, one may then expect to be free of significant contribution from rotation. Figure \ref{fig7} illustrates the evolution of the line profile when scanning along this line in steps of $y$$=$10 mas (0.64 au). Very strong distortions caused by absorption illustrate the difficulty of making a faithful radiative transfer modelling of a very complex reality: speaking of a line width is but a crude approximation. Yet, to be free of the effect of the strong red-side absorption, we evaluate its evolution from the value $V_{z,20}$ of $V_z$ at which 20\% of the maximal flux density is reached on the blue side.  
In both south and north directions, Figure \ref{fig8} shows that, for both $^{29}$SiO(8-7) and $^{12}$CO(3-2) emissions, this value increases by $\sim$8 \kms when $|y|$ increases from 0.8 au to 4 au, confining broad effective line widths within $\sim$4 au from the centre of the star, slightly beyond the inner cavity of the gas disc as measured by \citet{Homan2017}. Uncertainties at the scale of $\pm$1 \kms are estimated from the results obtained using different methods to evaluate the values of $V_{z,20}$. We have checked that this evaluation is consistent with the results obtained on the red side and with a scan made at $y$$=$0 in small steps of $x$ (Figure \ref{fig8}). The evolution of $V_{z,20}$ depicted in Figure \ref{fig8} is reminiscent of similar results presented by \citet{Homan2017} in their Figures 4 and 12. These authors have studied critically the detailed configuration of the velocity field in the inner disc; however their interpretation differs from ours. To quantify the line broadening is made difficult by the importance of absorption. In the approximation of Gaussian line profiles, we estimate that the effective line width (FWHM) remains below 10 \kms down to projected distances of $\sim$4 au and then increases linearly to reach $\sim$20 \kms\ near the surface of the star. The values obtained for $^{13}$CO (right panel of Figure \ref{fig8}) are less affected by absorption and are $\sim$5 \kms\ and $\sim$20 \kms, respectively . Such line broadening is not explicitly accounted for in the analysis of \citet{Homan2017}; they use an rms turbulent velocity of only 0.5 \kms\ and a modest thermal broadening (their Figure 12); including an explicit line broadening in the model would probably produce lower rotation velocities. To illustrate this point we show in Figure \ref{fig9} the dependence on position angle of the mean Doppler velocity, averaged over $R$ in successive rings and integrated over $|V_z|$$<$20 \kms. The results are summarized in Table \ref{tab2} and show a rotation velocity reaching up to some 8 \kms\ in the inner ring, 30$<$$R$$<$60 mas, smaller than obtained by \citet{Homan2017} and \citet{Kervella2016}.

In summary, in spite of being qualitative, this analysis has shown reliably that significant line broadening is observed at projected distances from the star below $\sim$4 au and is absent above. Over this radial range, the effective line width increases (FWHM) from $\sim$5 \kms\ to $\sim$20 \kms. Compared with R Dor and Mira Ceti, the extension of the line broadening region is three times smaller for L$_2$ Pup. We did not find in the analyses of the infrared and visible observations presented by \citet{Kervella2014, Kervella2015, Lykou2015} and \citet{Ohnaka2015} any argument that might either comfort or contradict this result.

\begin{table*}
  \centering
  \caption{Sine wave fits of the form $<$$V_z$$>$$=$$A+B\sin(\omega-\omega_0)$. Typical uncertainties are 0.2 \kms\ on A, 0.5 \kms\ on B and 4\dego\ on $\omega_0$.}
    \label{tab2}
    \begin{tabular}{cccccccccc}
      \hline
      &
      \multicolumn{3}{c}{$A$ (\kms)}&
      \multicolumn{3}{c}{$B$ (\kms)}&
      \multicolumn{3}{c}{$\omega_0$ (degrees)}\\
      $R$ range (mas)&  
      $^{12}$CO&
      $^{13}$CO&
      $^{29}$SiO&
      $^{12}$CO&
      $^{13}$CO&
      $^{29}$SiO&
      $^{12}$CO&
      $^{13}$CO&
      $^{29}$SiO\\
      \hline
      30-60&
      $-$0.2&
      $-$0.2&
      $-$0.5&
      $-$8.2&
      $-$7.0&
      $-$8.4&
      1&
      1&
      2\\
      60-90&
      $-$0.1&
      0.2&
      $-$0.3&
      $-$6.5&
      $-$5.7&
      $-$7.2&
      $-$1&
      2&
      $-$1\\
      90-120&
      0.0&
      $-$0.3&
      $-$0.3&
      $-$4.9&
      $-$4.3&
      $-$5.6&
      $-$1&
      7&
      $-$4\\
      120-150&
      0.0&
      0.4&
      $-$0.3&
      $-$3.8&
      $-$4.3&
      $-$4.7&
      $-$3&
      $-$4&
      $-$13\\
      \hline
    \end{tabular}
\end{table*}

\section{The wind of L$_2$ Pup} 

\subsection{Searching for north/south bipolar outflows at short distance from the star}
All recent studies of the L$_2$ Pup CSE \citep{Kervella2014, Kervella2015, Kervella2016, Lykou2015, Ohnaka2015, Homan2017} focus on the morpho-kinematics of the gas and dust disc and the possible presence of a companion but none of these addresses directly the question of the probable existence of a wind escaping the gravity of the star. \citet{Kervella2015} note the presence of features such as plumes and spirals at large angles from the disc median plane and speculate \citep{Kervella2015,Kervella2016} about the existence of broad bipolar outflows along the disc axis, but no direct observation of such outflows has ever been made. Yet, from what we have seen in Sections 3 and 4 the presence of large velocities near the star suggests that L$_2$ Pup, like other oxygen-rich AGB stars, should give a primordial boost to the gas near the surface of the star that would bring it to a region where dust grains get accelerated and share with the gas some of their radial momentum. Part of this wind would be absorbed by the disc, which covers a significant fraction of the solid angle, but the other part should produce broad outflows of gas and dust having their axes near the sky plane and pointing in the north and south directions.

\citet{Chen2016} have performed a three-dimensional simulation of the CSE that is meant to account for the most recent VLT and ALMA observations. They do not model the launching of the wind but assume instead the ejection of a pulse of dense wind (9.3$\times$10$^{-6}$ \msun yr$^{-1}$) lasting 11.6 years and superimposed on a steady wind of 39.3 \kms\ producing a mass loss rate of 1.67$\times$10$^{-6}$ \msun yr$^{-1}$. These choices are meant to cope with the lack of simulation of the launching mechanism while maintaining an expansion velocity of 15 to 20 \kms\ in the region where the wind interacts with the companion. The pulse of wind is meant to result from the ingestion of a planet by the AGB star. The model predicts a gravitationally bound disc and wide polar outflows along the disc axis with velocities reaching 20 \kms.

\begin{figure*}
  \centering
  \includegraphics[height=4.5cm,trim=.5cm 1.cm 1.5cm 1.5cm,clip]{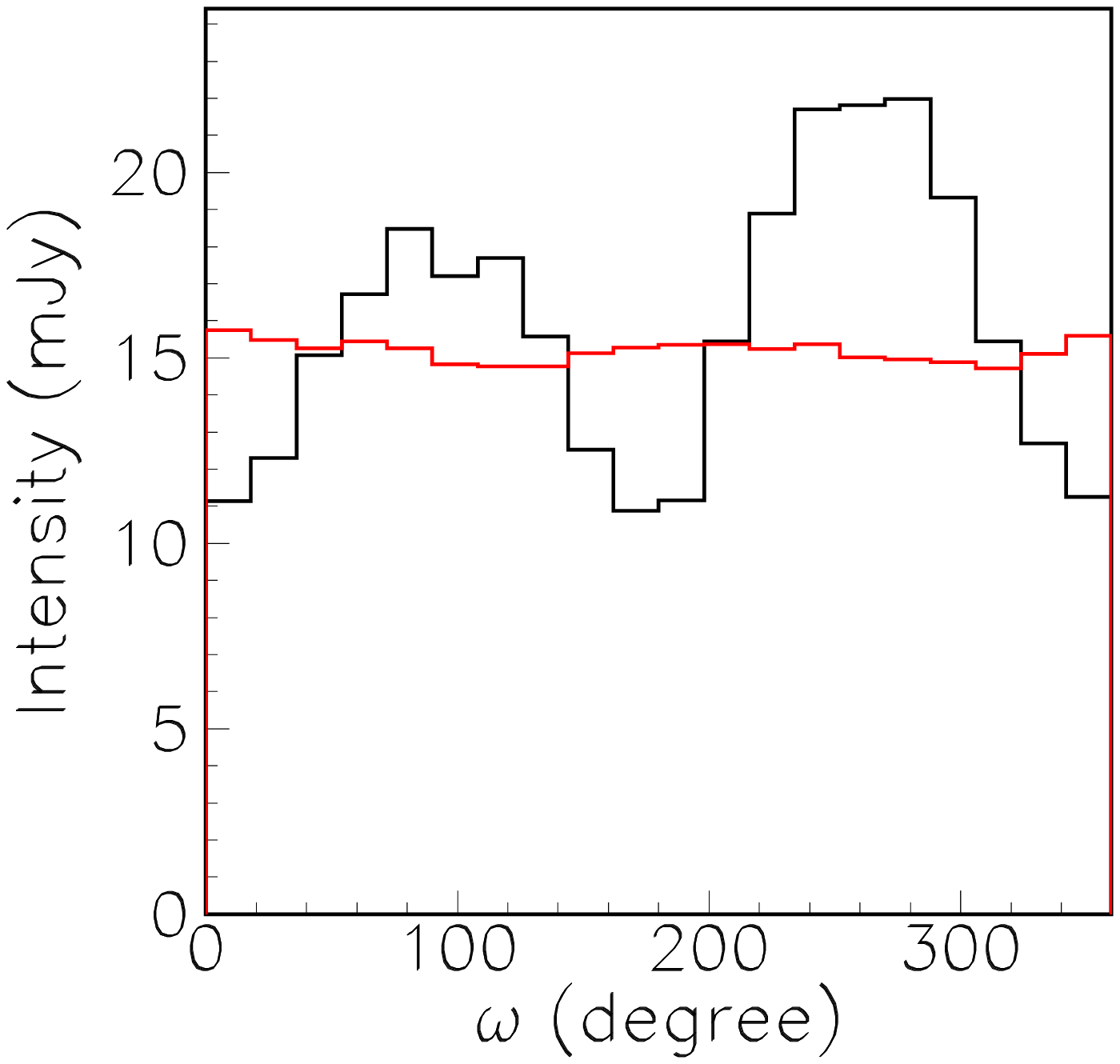}
  \includegraphics[height=4.5cm,trim=1.8cm 1.cm 1.5cm 1.5cm,clip]{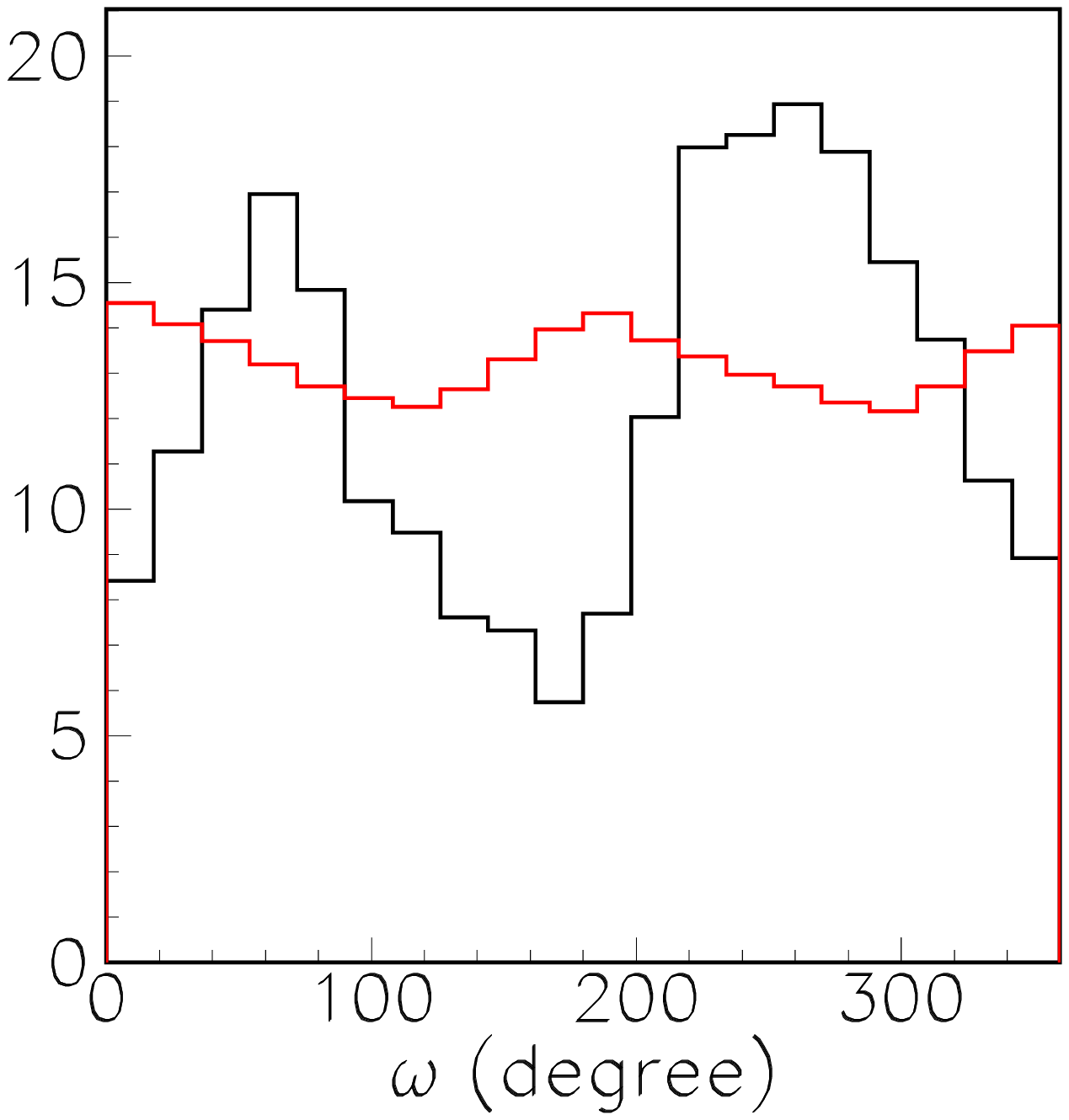}
  \includegraphics[height=4.5cm,trim=1.8cm 1.cm 1.5cm 1.5cm,clip]{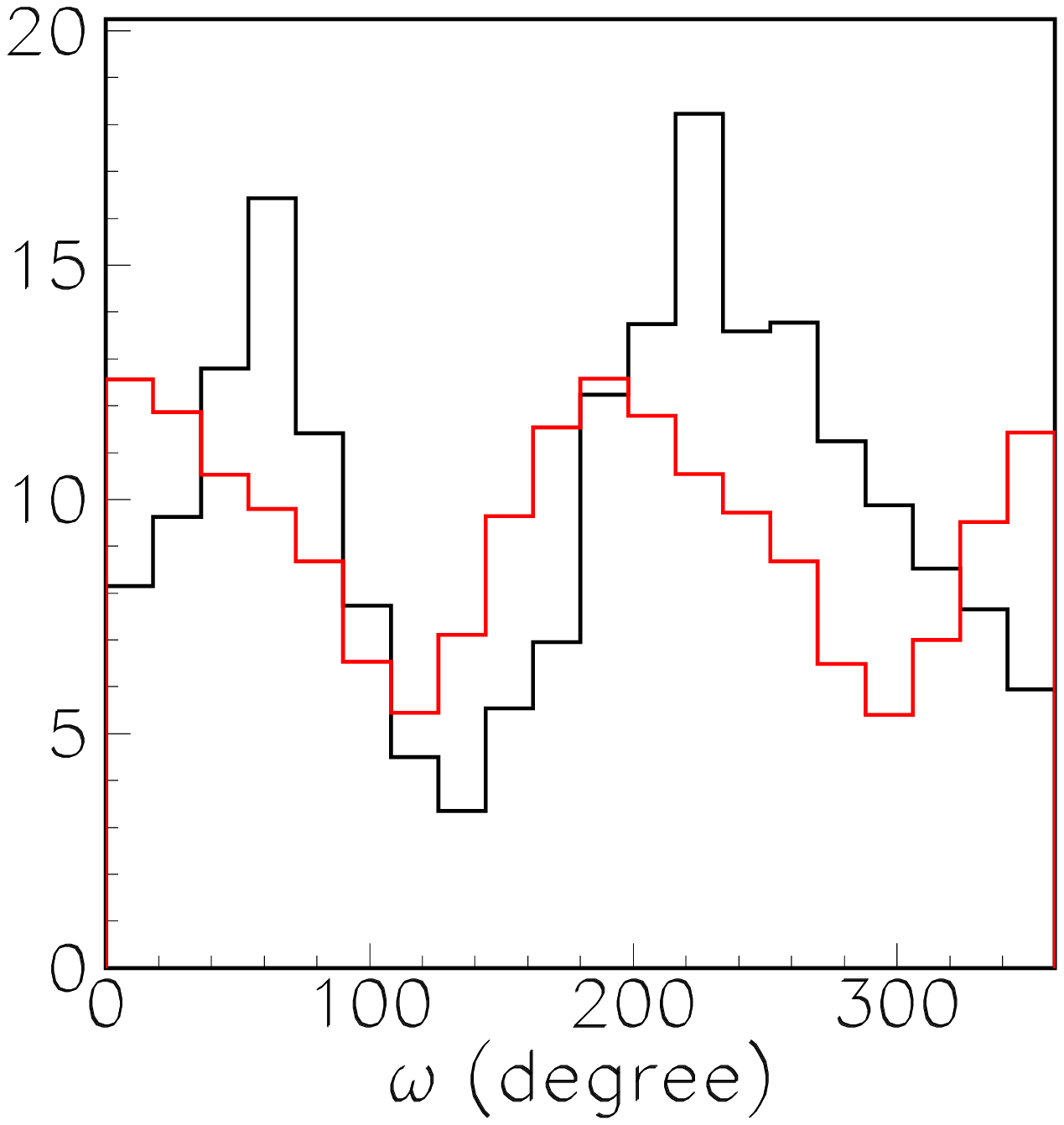}
  \includegraphics[height=4.5cm,trim=1.8cm 1.cm 1.5cm 1.5cm,clip]{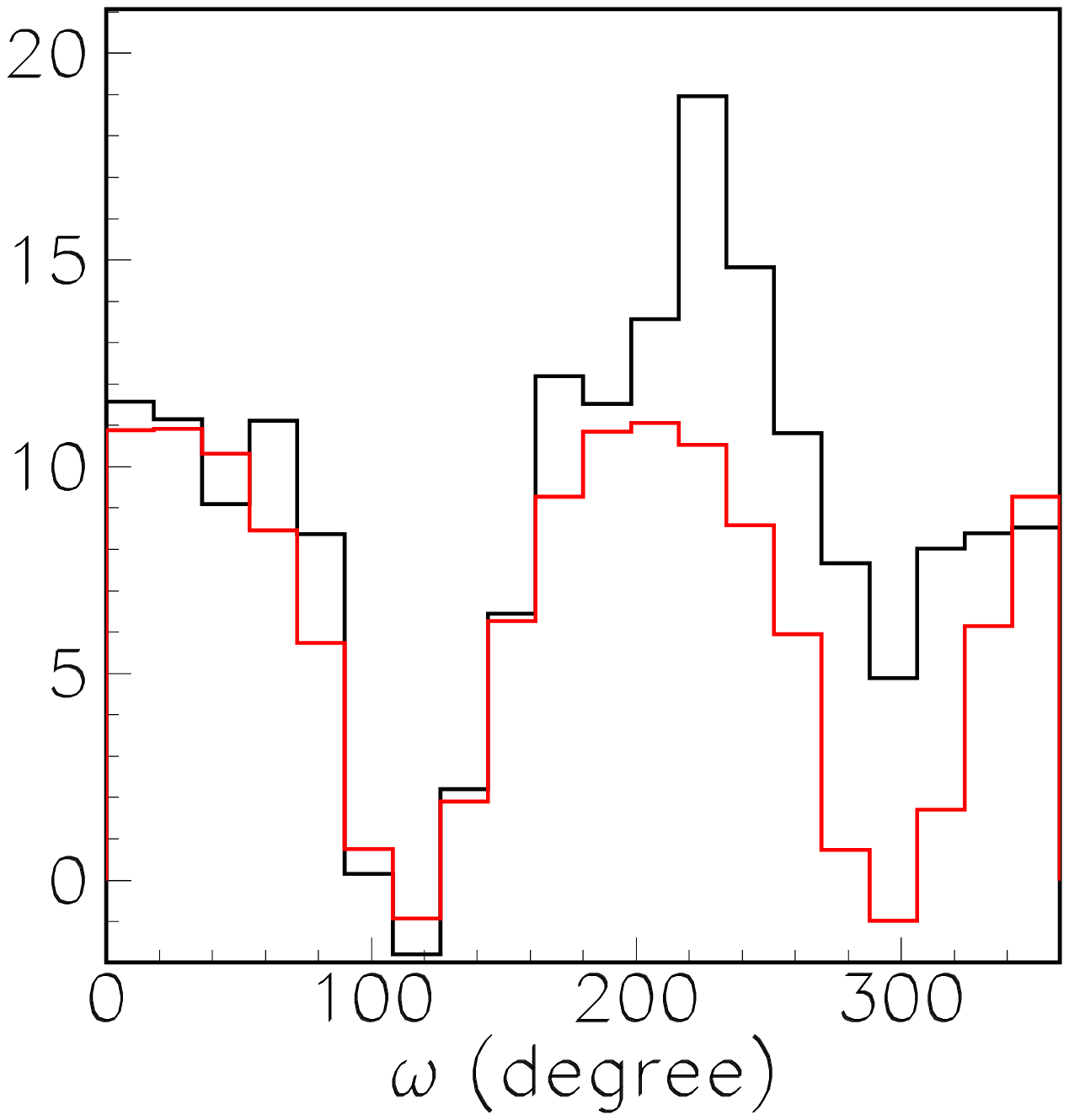}
    \caption{Comparing the position angle distributions of the $^{12}$CO(3-2) emission (black) and that of an isotropic wind (red) in successive intervals of $R$. From left to right: 0.05$<$$R$$<$0.10 arcsec, 0.10$<$$R$$<$0.15 arcsec, 0.15$<$$R$$<$0.20 arcsec and 0.20$<$$R$$<$0.25 arcsec.}
 \label{fig10}
\end{figure*}

However, beyond $R$$\sim$0.15 arcsec, such outflows cannot be detected using the data analysed in the preceding sections, the uv coverage being inadequate. This is illustrated in Figure \ref{fig10}, which compares the position angle distributions of the $^{12}$CO(3-2) emission in different intervals of $R$ with that produced by an isotropic wind. Beyond $R$$\sim$0.15 arcsec, the poor $uv$-coverage at short baseline lengths causes an isotropic wind to appear as a north-south emission as illustrated in Figure \ref{fig10}. Therefore, we cannot distinguish the $^{12}$CO and $^{29}$SiO emission of Figure \ref{fig3} in the interval [0.15, 0.4] arcsec from such artefacts. 

\begin{figure*}
  \centering
  \includegraphics[height=5.cm,trim=.5cm 1.5cm .5cm 1.5cm,clip]{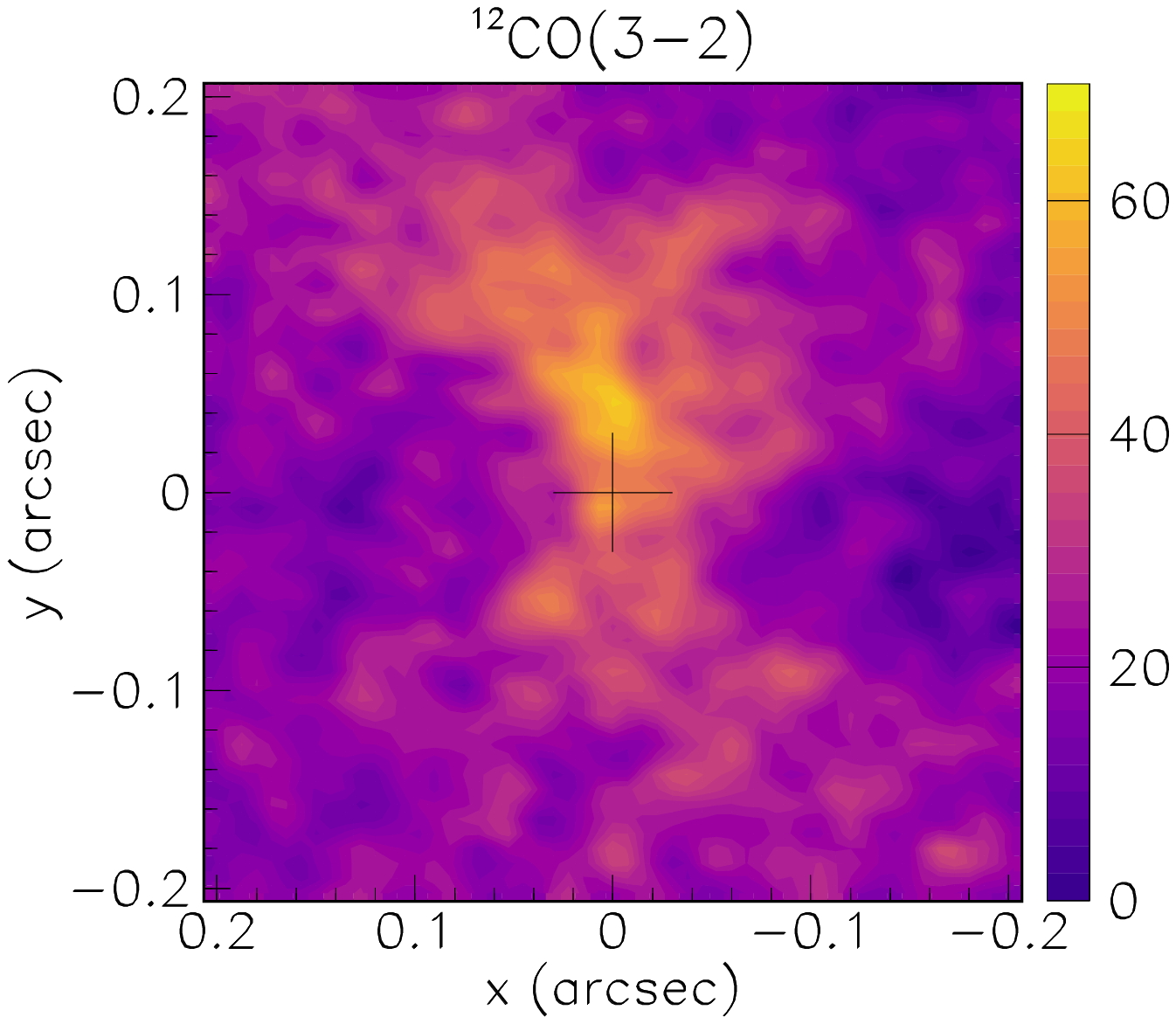}
  \includegraphics[height=5.cm,trim=.5cm 1.5cm .5cm 1.5cm,clip]{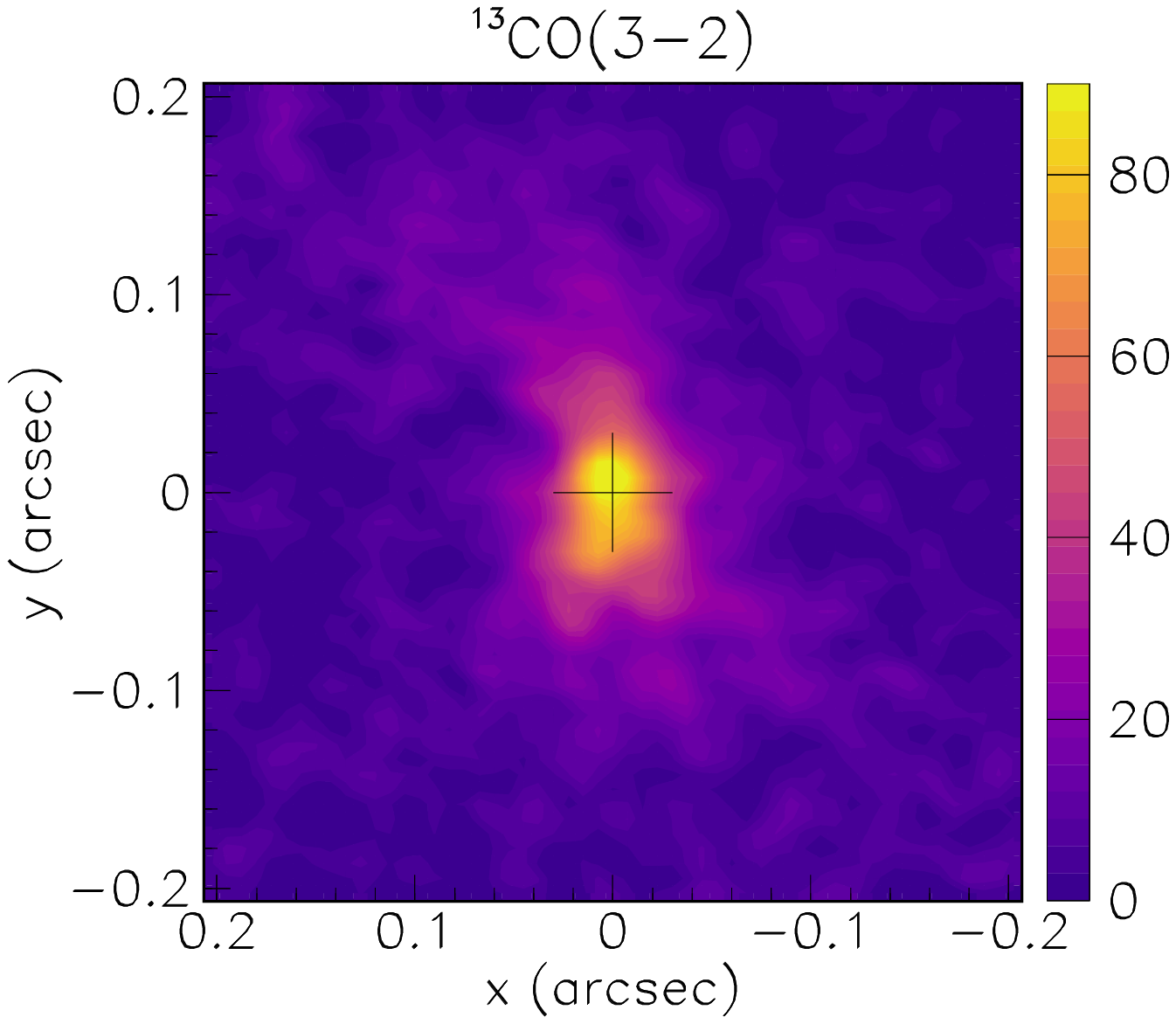}
  \includegraphics[height=5.cm,trim=.5cm 1.5cm .5cm 1.5cm,clip]{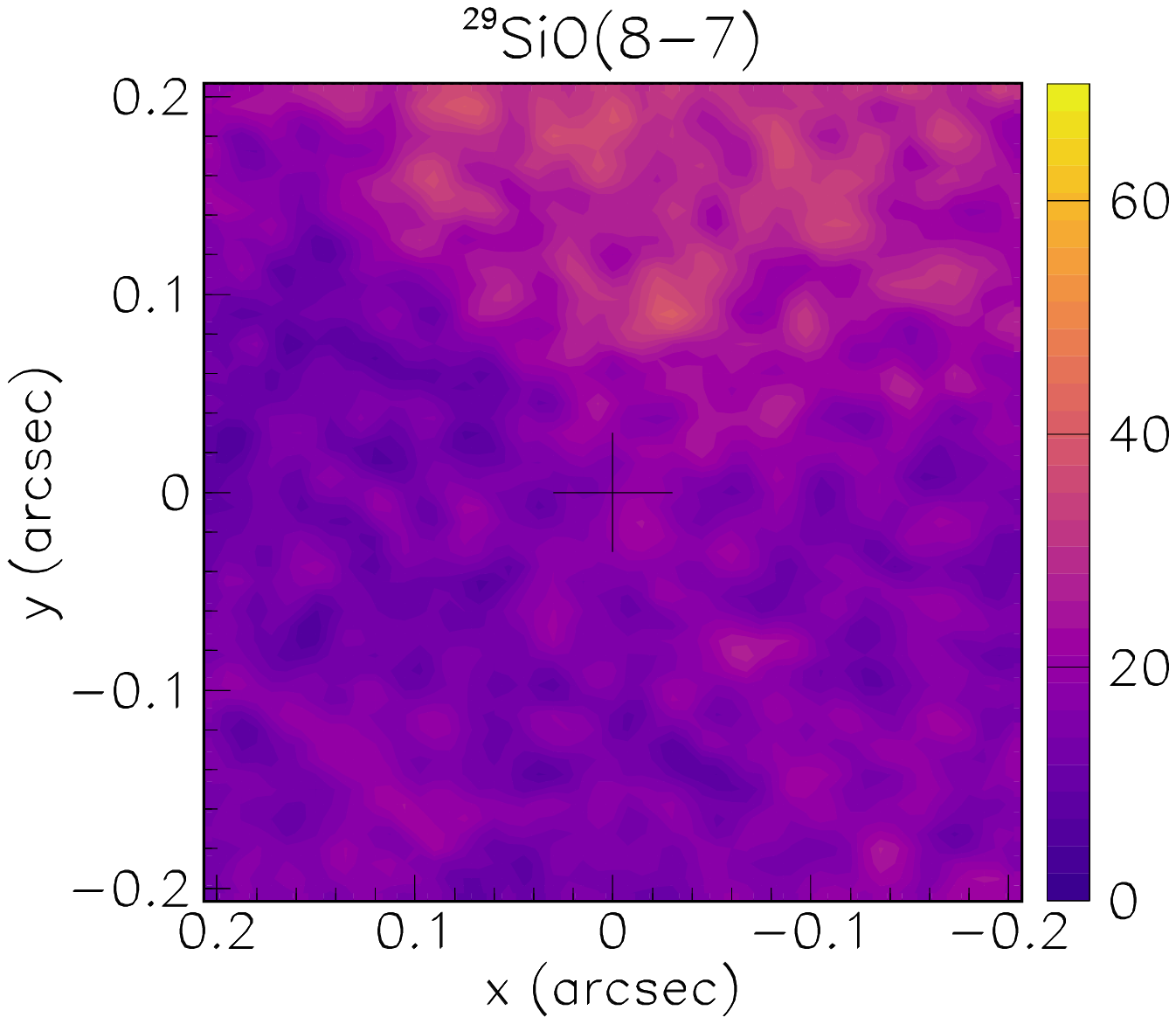}
    \caption{Intensity maps integrated over |$V_z|$$<$1 \kms. From left to right: $^{12}$CO(3-2), $^{13}$CO(3-2) and $^{29}$SiO(8-7). The colour scales are in units of Jy arcsec$^{-2}$ \kms.}
 \label{fig11}
\end{figure*}

Staying within $R$$<$0.2 arcsec, we might hope to see evidence for a north/south outflow close to the plane of the sky by confining the absolute value of the Doppler velocity to very small values. Figure \ref{fig11} displays intensity maps integrated over |$V_z|$$<$1 \kms\ for each of the three lines. The $^{12}$CO and $^{29}$SiO maps show emission that can be interpreted as a northern outflow, the southern part being obscured by the tilt of the disc, the cold outer surface of which is absorbing, particularly strongly in the case of the SiO line. The $^{13}$CO map displays a north-south elongation that is also compatible with an interpretation in terms of an optically thin north-south outflow. However, these comments can only be taken as qualitative.  \citet{Homan2017} have given a detailed discussion of the mechanisms at stake in this region in their Section 6.2.2. They underline their complexity, with steep temperature gradients and hydrodynamical boundary instabilities at the disc/outflow interface. In such a context it is difficult to ascertain that the maps displayed in Figure \ref{fig11}, which also include emission from the upper part of the disc, are indeed evidence for a north/south outflow and we refrain from attempting from these an evaluation of the mass loss rate. Moreover, the present observations of the CO(3-2) and $^{29}$SiO(8-7) emissions show that the FWHM of the line profiles is $\sim$5 \kms\ beyond $\sim$4 au from the centre of the star (Section 4), contrary to what could be expected from broad outflows with velocities reaching 20 \kms\ as predicted by \citet{Chen2016}. 

\subsection{Overall wind kinematics within 200 au from the centre of the star}
\begin{figure*}
  \centering
  \includegraphics[height=4.5cm,trim=.5cm 1.cm 1.5cm .5cm,clip]{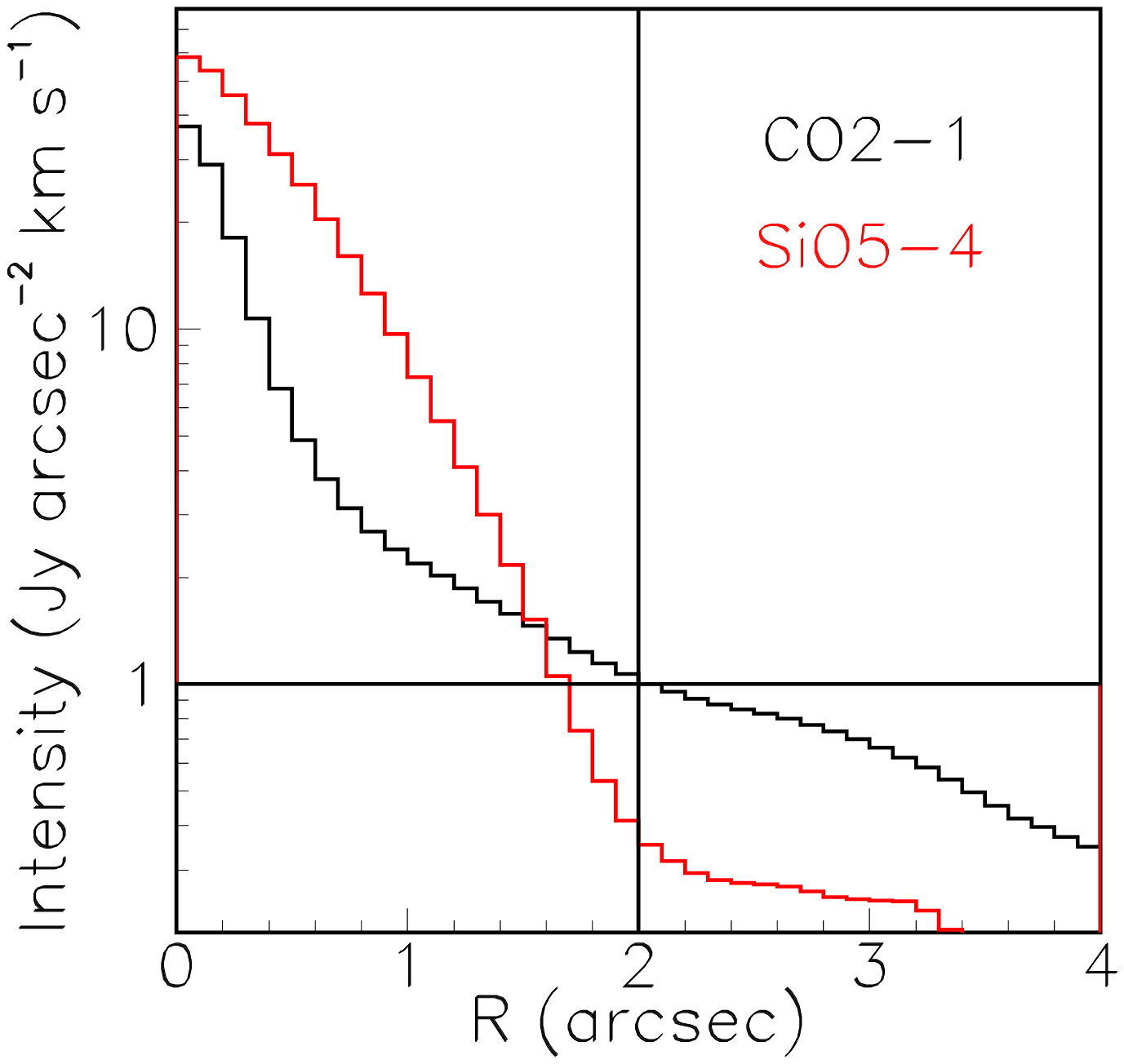}
  \includegraphics[height=4.5cm,trim=.5cm 1.cm 1.5cm .5cm,clip]{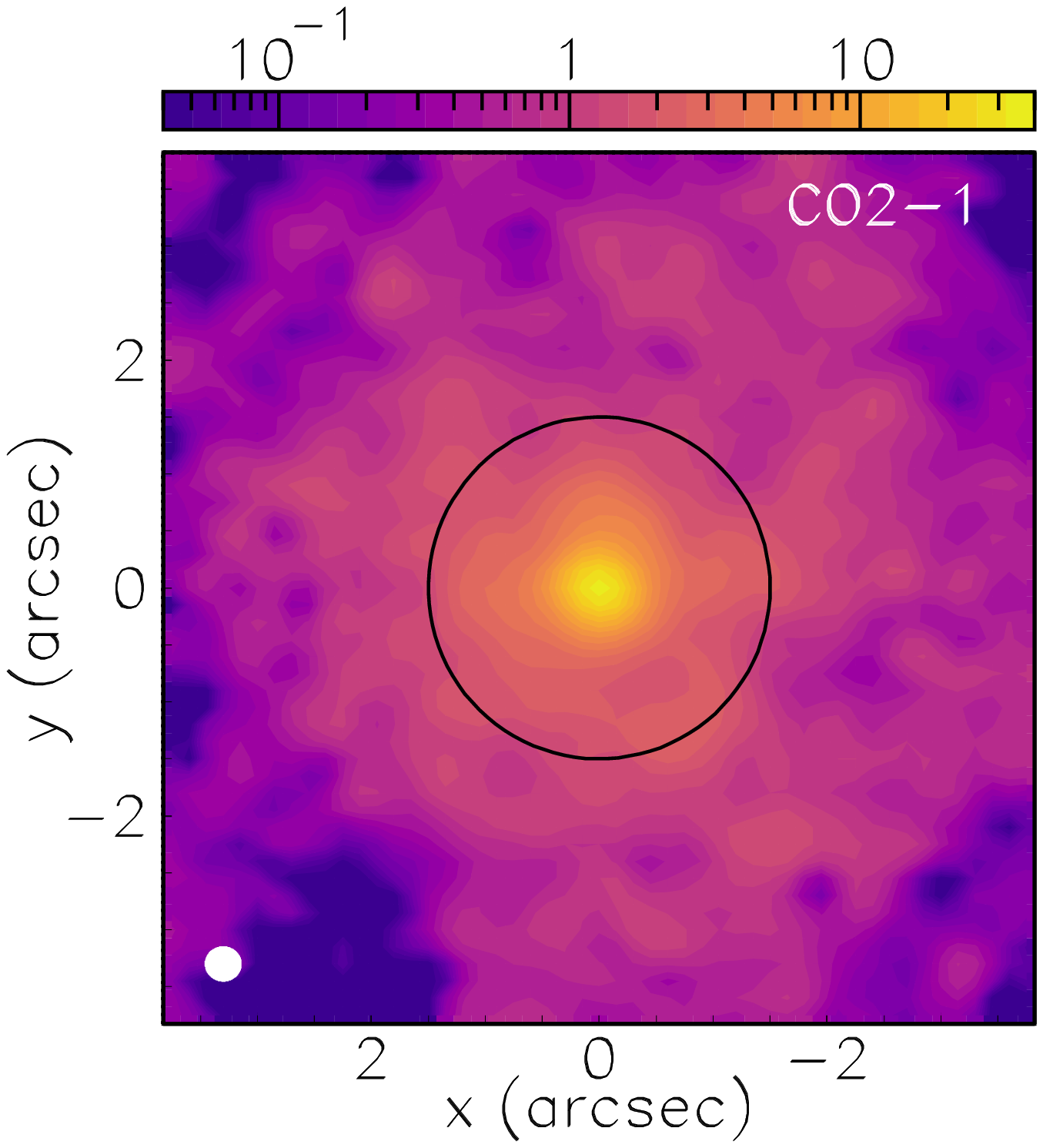}
  \includegraphics[height=4.5cm,trim=.5cm 1.cm 1.5cm .5cm,clip]{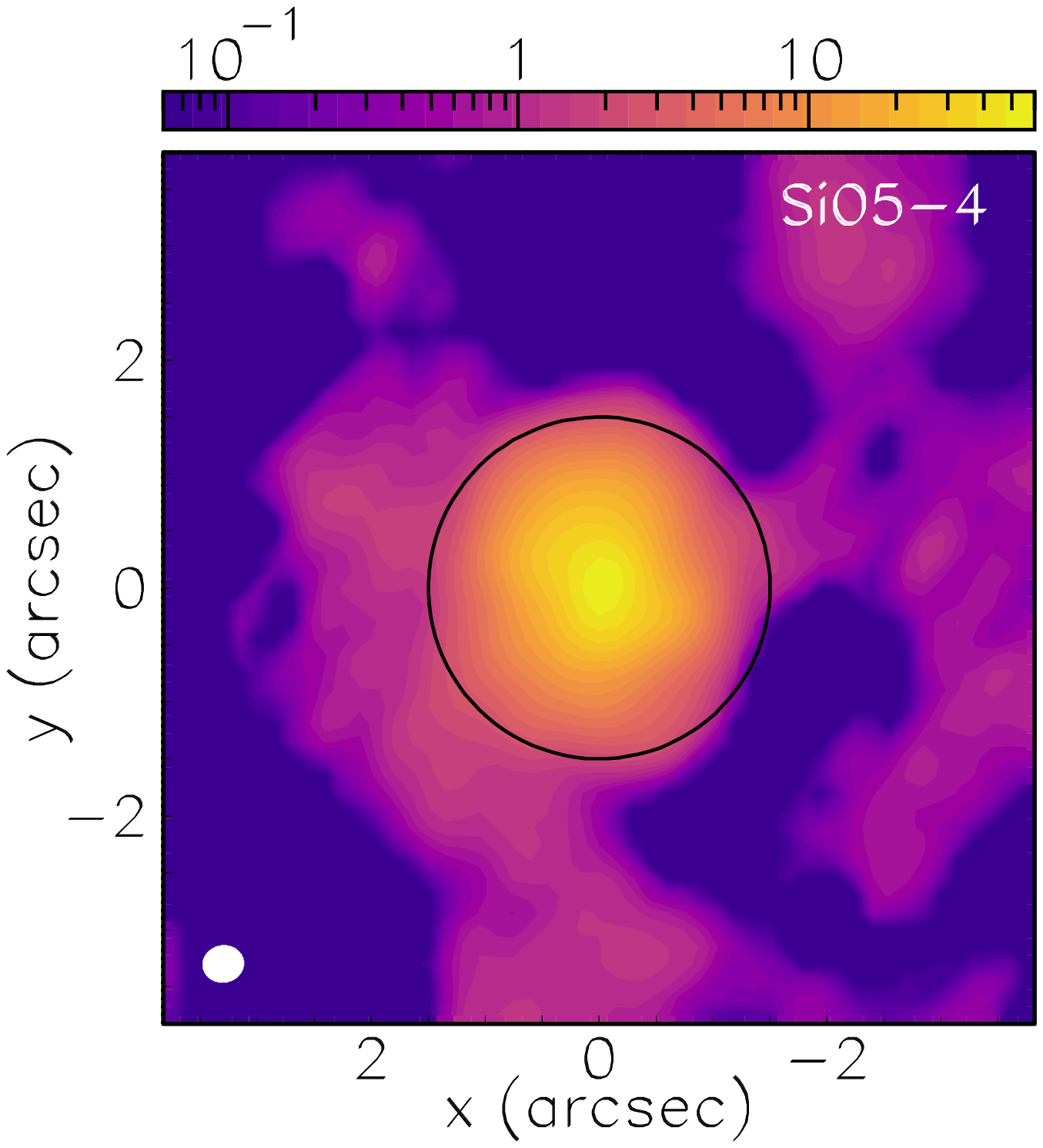}
  \includegraphics[height=4.5cm,trim=.5cm 1.cm 1.5cm .5cm,clip]{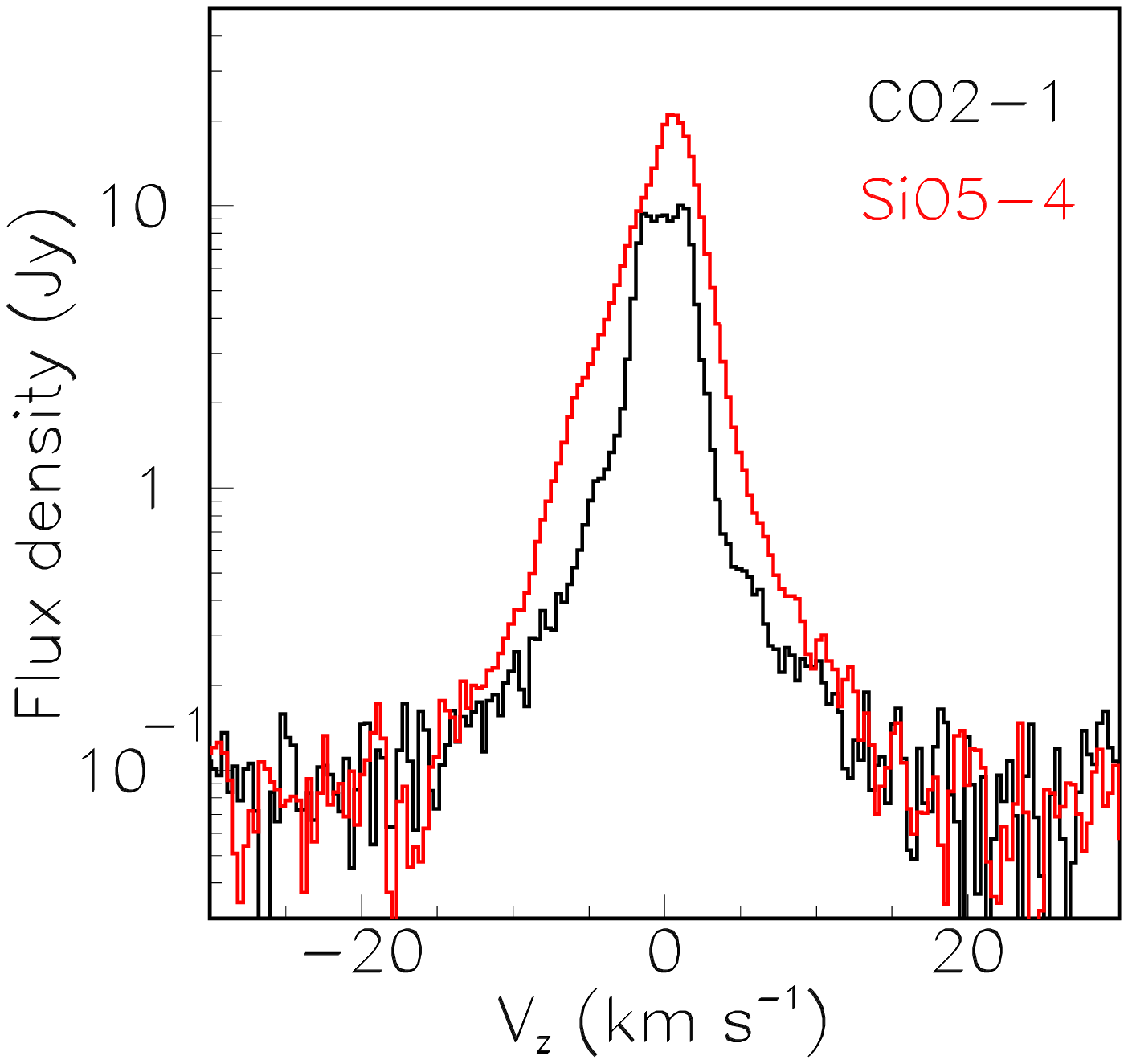}
    \caption{$^{12}$CO(2-1) and $^{28}$SiO(5-4) emissions. Left: radial distributions averaged over position angle and integrated over $|V_z|$$<$5 \kms. Middle: intensity maps integrated over $|V_z|$$<$5 \kms; the circle has a radius of 1.5 arcsec. The colour scales are in units of Jy arcsec$^{-2}$ \kms. Right: Doppler velocity spectra integrated over $R$$<$3 arcsec. }
 \label{fig12}
\end{figure*}

Having shown the difficulty to detect reliably north/south bipolar outflows from the $^{12,13}$CO(3-2) and $^{29}$SiO(8-7) observations studied in the preceding sections, we now analyse the second set of observations presented in Section 2. They are of the emissions of the $^{12}$CO(2-1) and $^{28}$SiO(5-4) lines, the antenna configuration providing reliable imaging up to distances of $\sim$3 arcsec from the star. To our knowledge, there exists no published analysis of these data. The channel maps are shown in the Appendix. 
At a distance of 64 pc from the Earth, with a projected velocity of 10 \kms\ it takes $\sim$30 yr to cover 1 arcsec; we can therefore assume that most of what is seen beyond $R$$\sim$1 arcsec happened before the dimming event at the end of the past century.
Figure \ref{fig12} compares the intensity maps, radial distributions and Doppler velocity spectra of both line emissions. As commonly observed in oxygen-rich AGB stars, the SiO emission is confined near the star, here within $\sim$80 au, and displays therefore a slightly broader Doppler velocity spectrum; this confinement is usually interpreted \citep{Schoier2004} as the result of the condensation of the gas molecules on dust grains; moreover, the SiO emission is cut-off by photo-dissociation earlier than the CO emission. These results suggest an overall standard behaviour of the wind.

\begin{figure*}
  \centering
  \includegraphics[height=4.5cm,trim=.5cm 1.cm 1.5cm .5cm,clip]{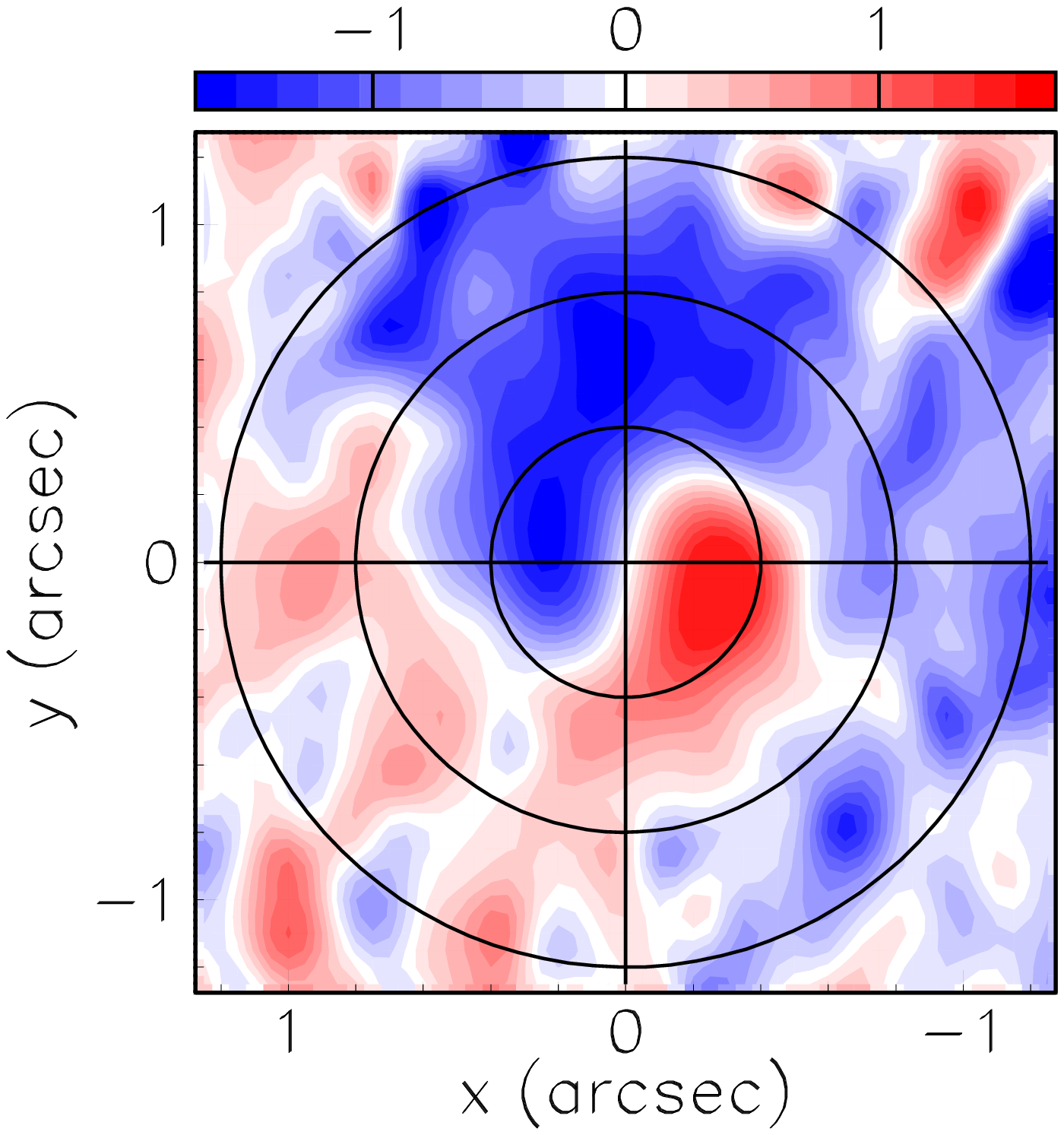}
  \includegraphics[height=4.5cm,trim=.5cm 1.cm 1.5cm .5cm,clip]{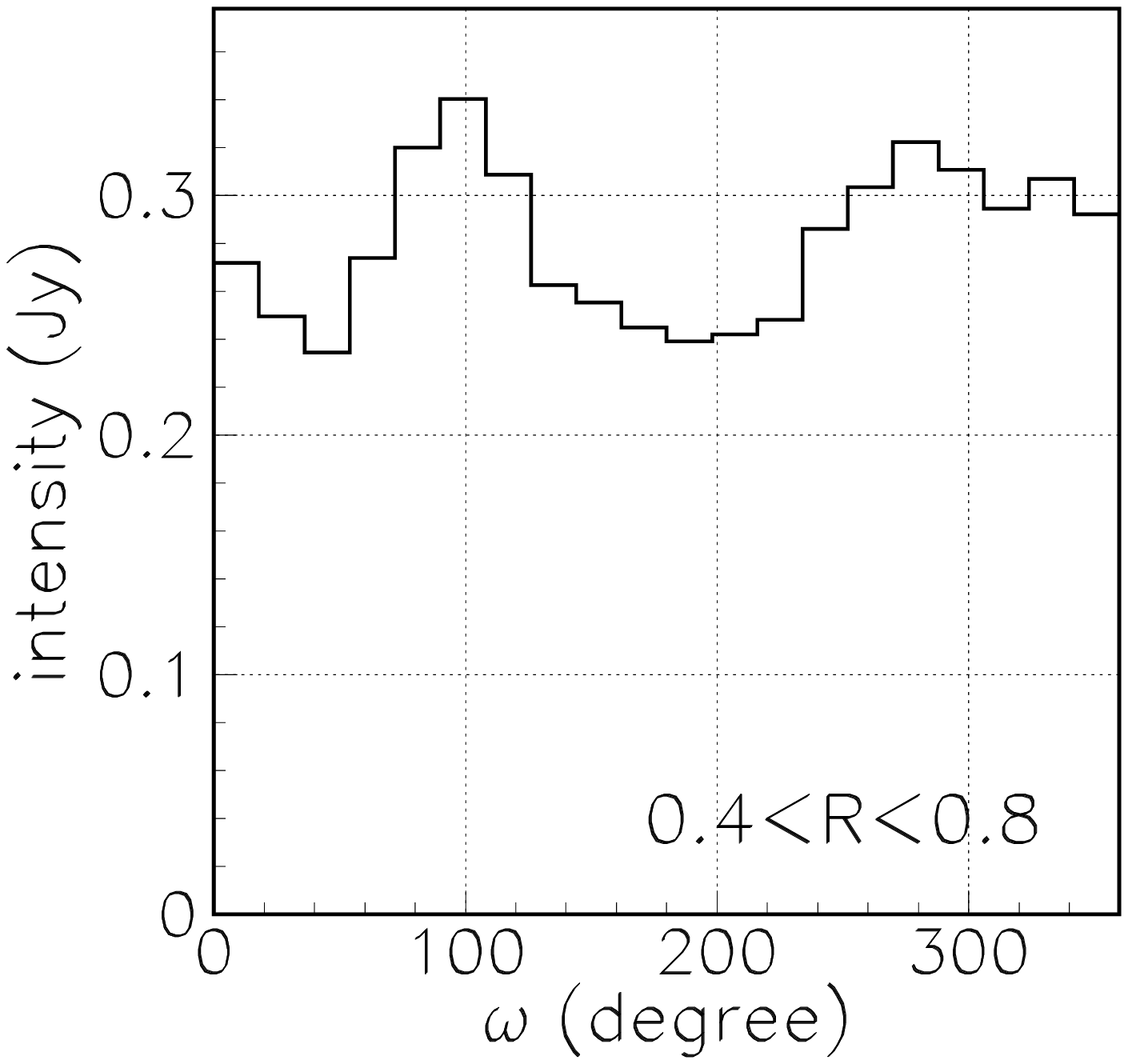}
  \includegraphics[height=4.5cm,trim=1.5cm 1.cm 1.5cm .5cm,clip]{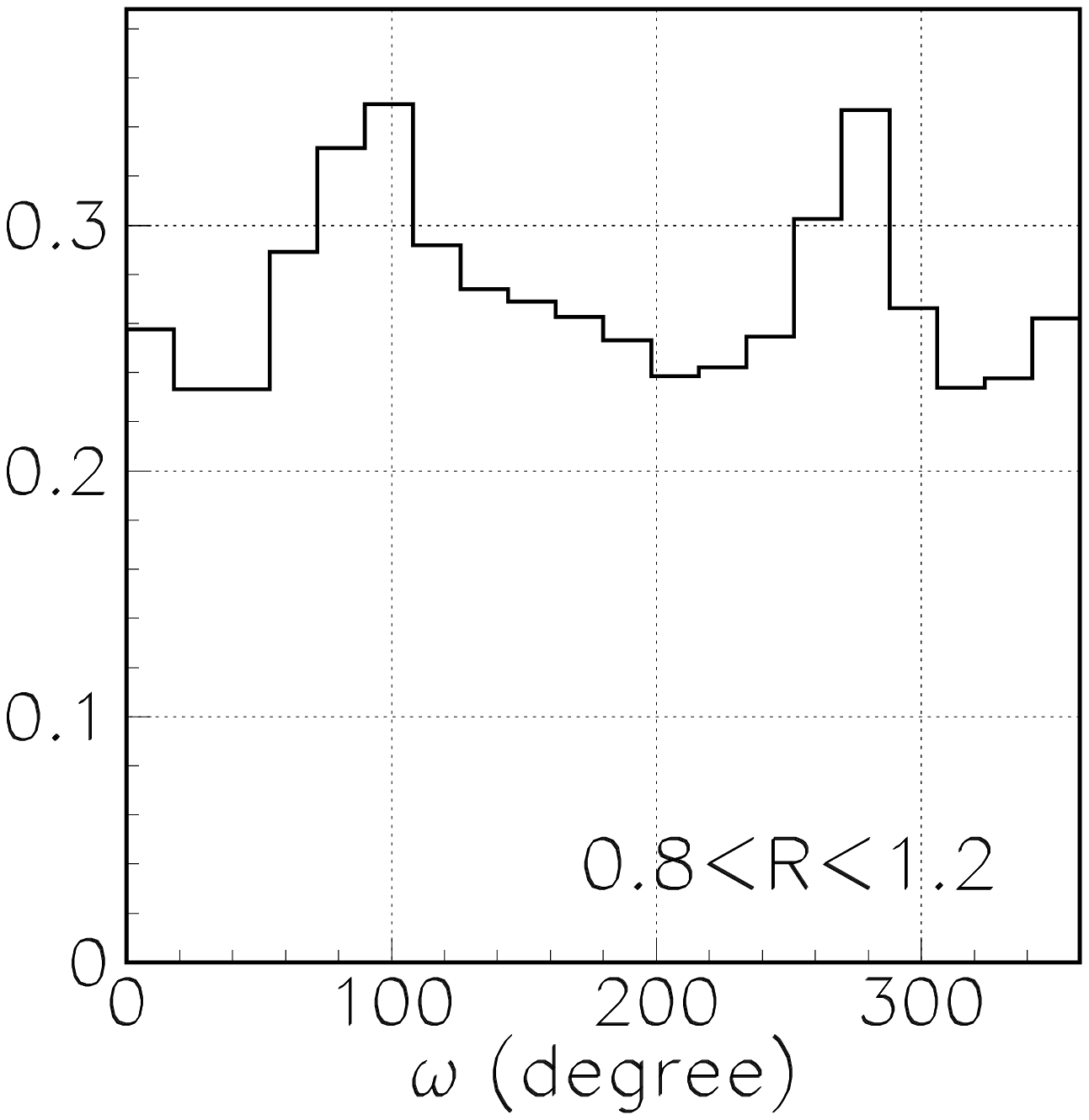}
  \includegraphics[height=4.5cm,trim=1.5cm 1.cm 1.5cm .5cm,clip]{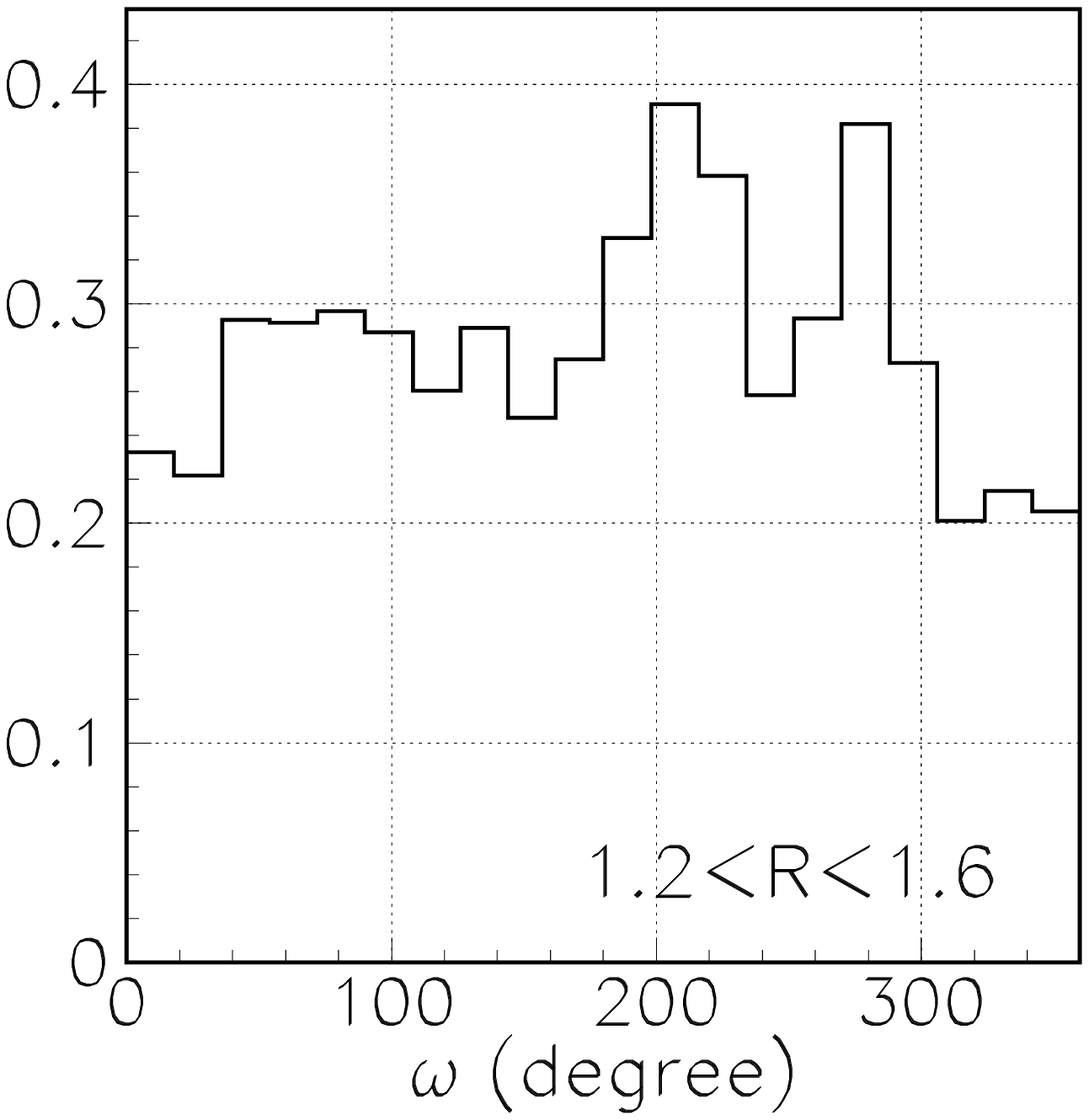}
    \caption{$^{12}$CO(2-1) line emission close to the star. Left: map of the mean Doppler velocity (moment 1) (\kms) in the interval $|V_z|$$<$10 \kms; circles of 0.4, 0.8 and 1.2 arcsec radius are shown. Right: Dependence on position angle $\omega$ of the intensity integrated over $|V_z|$$<$2.5 \kms\ and, from left to right, 0.4$<$$R$$<$0.8 arcsec, 0.8$<$$R$$<$1.2 arcsec and 1.2$<$$R$$<$1.6 arcsec. }
 \label{fig13}
\end{figure*}

\begin{figure*}
  \centering
  \includegraphics[height=5cm,trim=0.5cm 1.cm 1.5cm .5cm,clip]{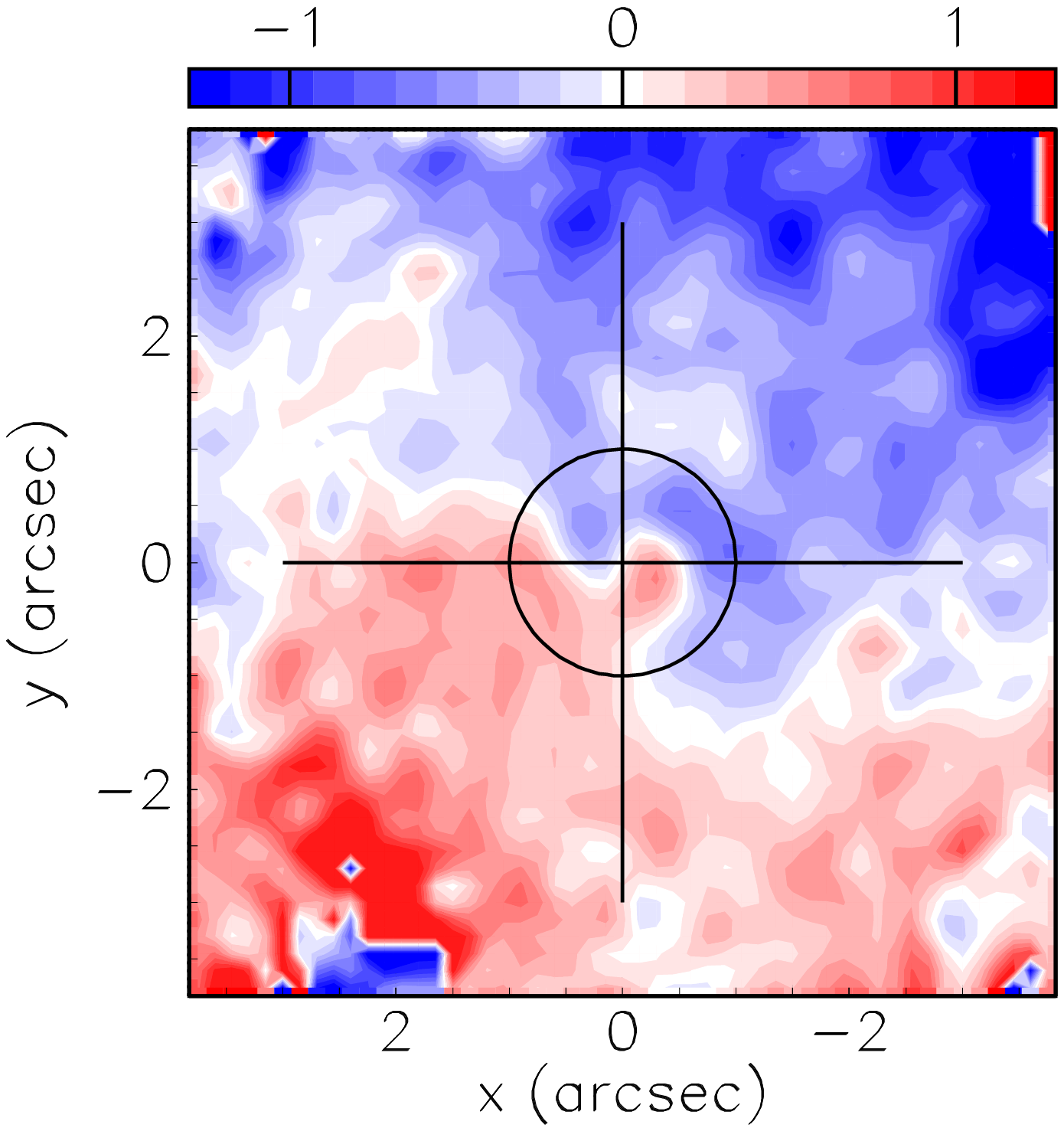}
  \includegraphics[height=5cm,trim=0.5cm 1.cm 1.5cm .5cm,clip]{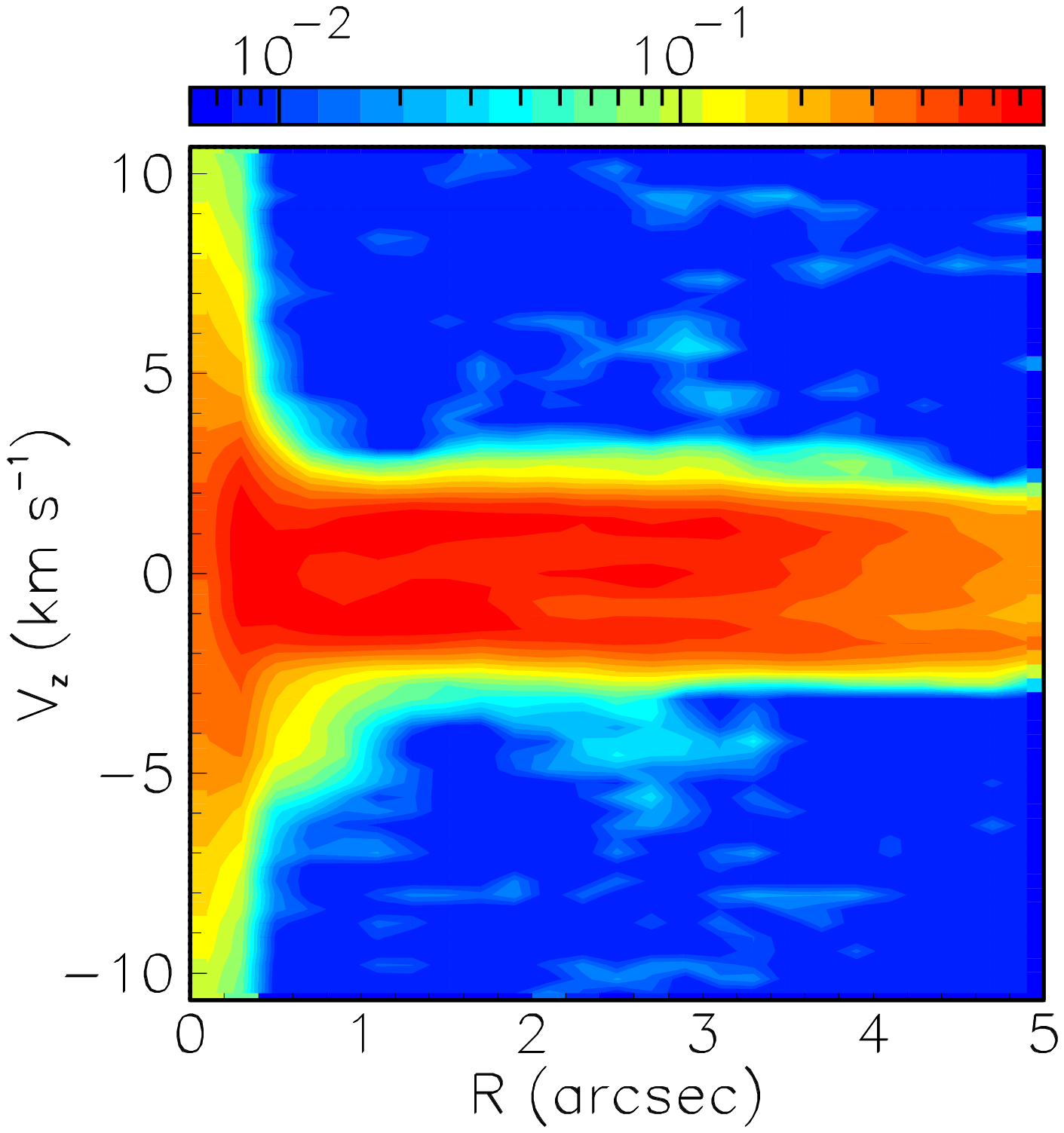}
  \includegraphics[height=5cm,trim=0.5cm 1.cm 1.5cm .5cm,clip]{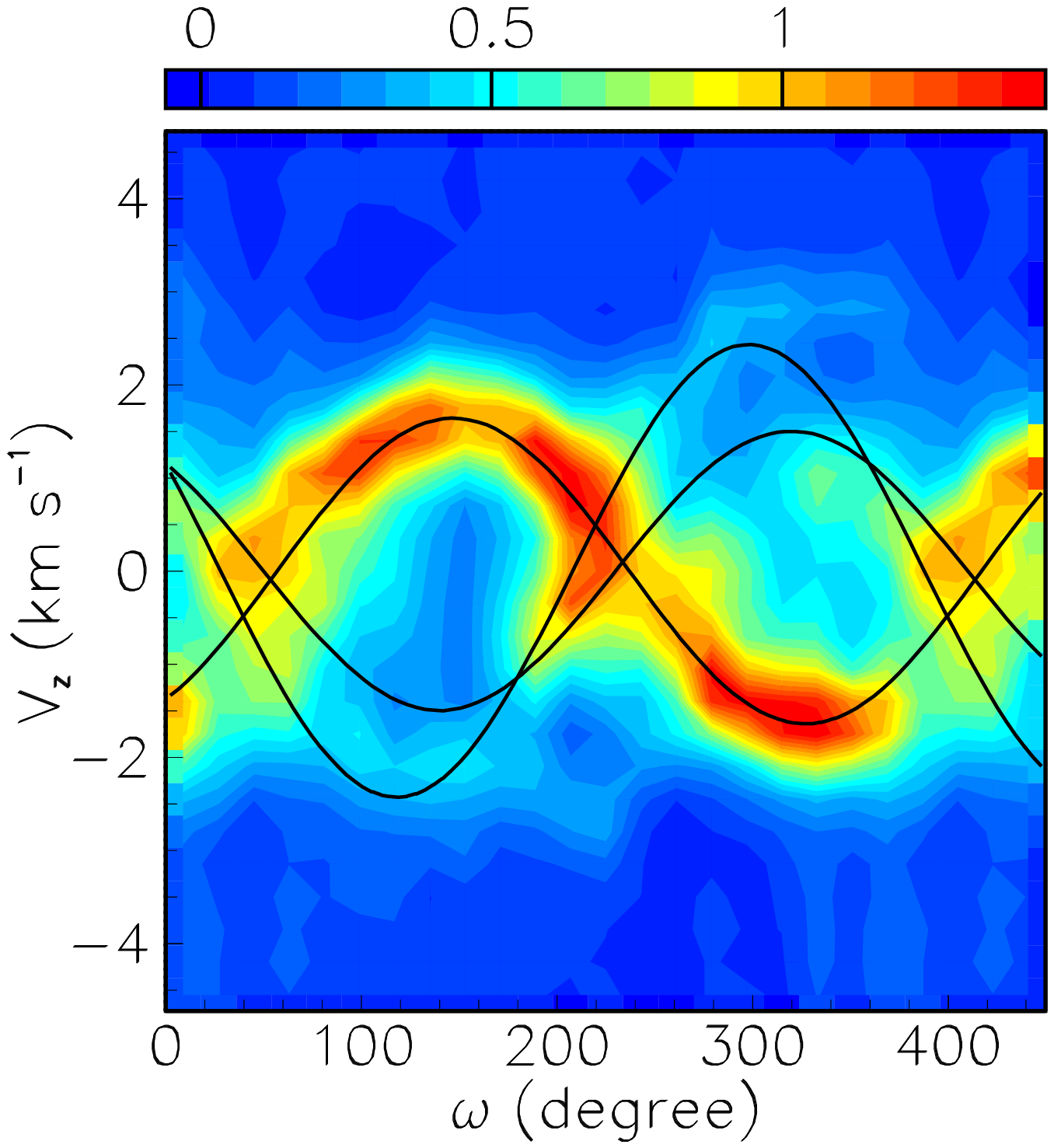}
    \caption{$^{12}$CO(2-1) line emission away from the star. Left: map of the mean Doppler velocity (moment 1) (\kms) in the interval $|V_z|$$<$2.5 \kms. The circle has a radius of 1 arcsec. Middle: PV map ($V_z$ vs $R$) of the flux density (Jy) integrated over position angle. Right: PV map ($V_z$ vs $\omega$) of the flux density (Jy) integrated over 1$<$$R$$<$3.5 arcsec and 18\dego\ in omega. The lines show the sine waves described in the text.}
 \label{fig14}
\end{figure*}

In order to study its morpho-kinematics, we use the CO(2-1) emission, which is expected to probe the CSE uniformly over large distances from the star. The left panel of Figure \ref{fig13} illustrates very clearly the transition away from the rotation regime, dominated by the disc of gas and dust, when $R$ exceeds $\sim$0.4 arcsec. However, the three rightmost panels of Figure \ref{fig13} do not show any evidence for enhanced outflow in the north/south direction. The left panel of Figure \ref{fig14} shows the map of the mean Doppler velocity of the $^{12}$CO(2-1) emission. It displays a clear asymmetry consistent with earlier observations at 11.7 and 17.9 $\umu$m by \citet{Jura2002} using the Keck I telescope equipped with the Long Wavelength Spectrometer (LWS) with a PSF of $\sim$0.5 arcsec FWHM.  It suggests the presence of a pair of back-to-back outflows in the north-west/south-east direction, clearly different from the north/south direction expected for an outflow collimated by the central gas-and-dust disc.  The central and right panels of Figure \ref{fig14} display PV maps of the CO(2-1) emission, both in $V_z$ vs $R$ and in $V_z$ vs $\omega$. The former shows that Doppler velocities are confined within $\pm$ $\sim$2.5 \kms, independently from the value of $R$, suggesting that the terminal velocity has already been reached shortly beyond 1 arcsec, where the escape velocity from a 0.66 solar mass star is $\sim$4.3 \kms. The latter shows a clear sine wave pattern, suggesting isotropically enhanced expansion in a plane inclined in the north-west/south-east direction with respect to the plane of the sky, explaining both the approximate isotropy of the intensity map (the rms fluctuations of the dependence on position angle $\omega$ of the intensity integrated over $|V_z|$$<$2.5 \kms\ are 17\% for both rings 1$<$$R$$<$2 and 2$<$$R$$<$3 arcsec) and the asymmetry of the moment 1 map. It also shows the presence of emission in phase opposition to the sine wave. Such wind morphology is reminiscent of that observed in RS Cnc \citep{Winters2021}: an equatorial enhancement together with broad outflows along its axis; it is very different from the broad pair of back-to-back outflows oriented north/south near the plane of the sky predicted by \citet{Kervella2015} and \citet{Chen2016}: these would appear as blobs of emission in the northern and southern quadrants (near $\omega$ $=$0\dego\ and 180\dego) but would not populate the eastern and western quadrants: it deserves a detailed study.

\begin{figure*}
  \centering
  \includegraphics[height=3.795cm,trim=.5cm 1.cm 2.4cm .5cm,clip]{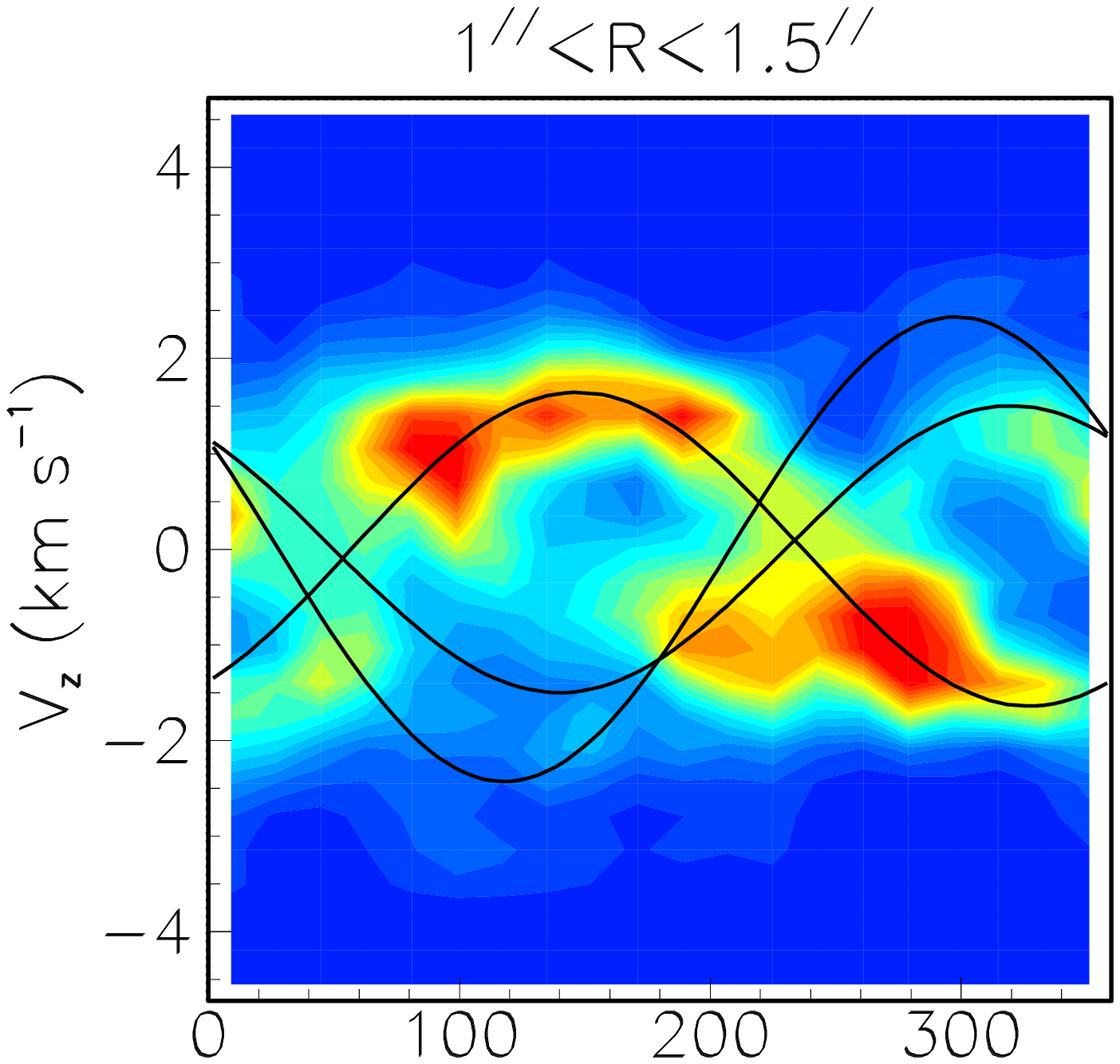}
  \includegraphics[height=3.795cm,trim=1.cm 1.cm 2.4cm .5cm,clip]{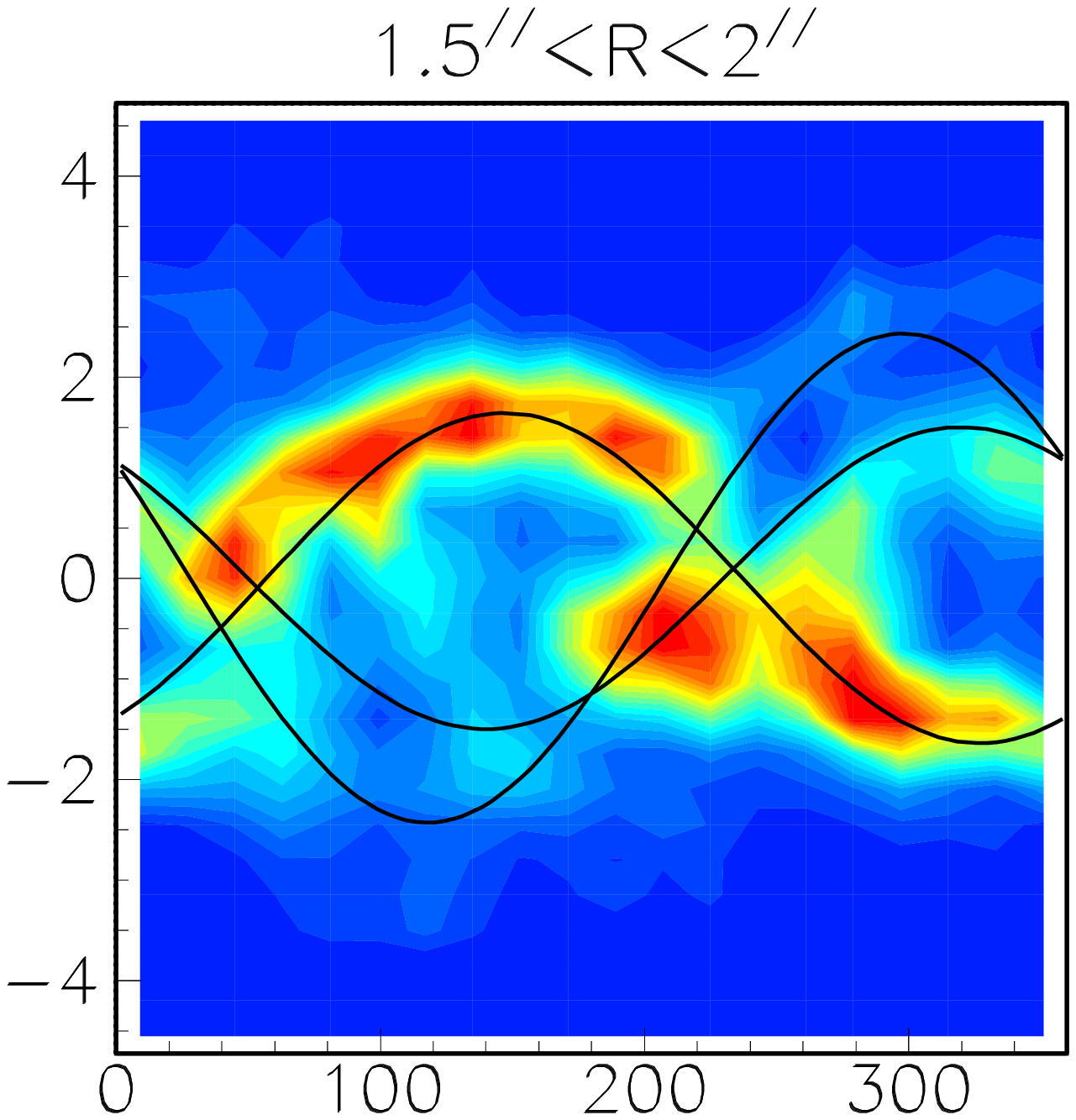}
  \includegraphics[height=3.795cm,trim=1.cm 1.cm 2.4cm .5cm,clip]{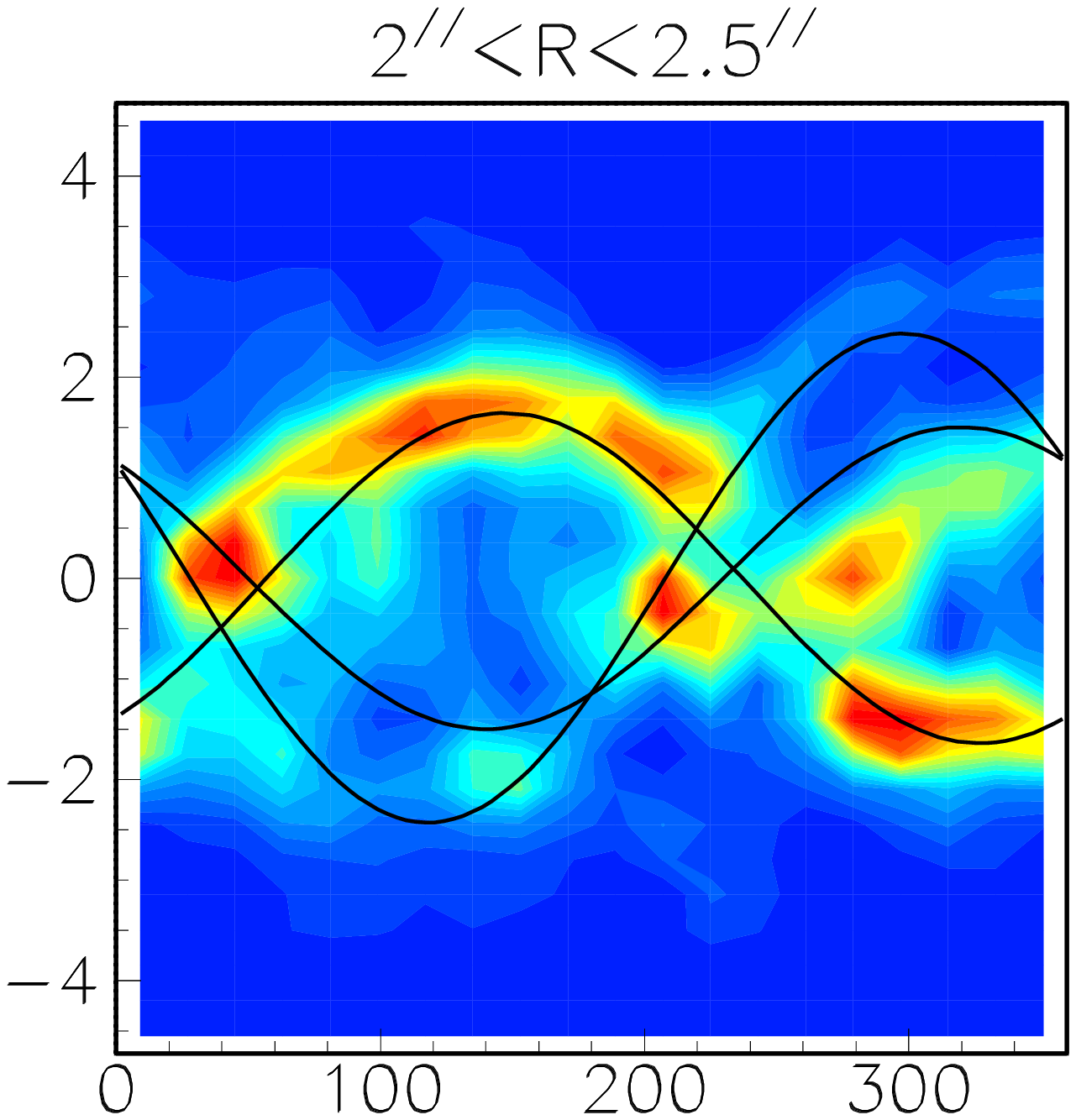}
  \includegraphics[height=3.795cm,trim=1.cm 1.cm 2.4cm .5cm,clip]{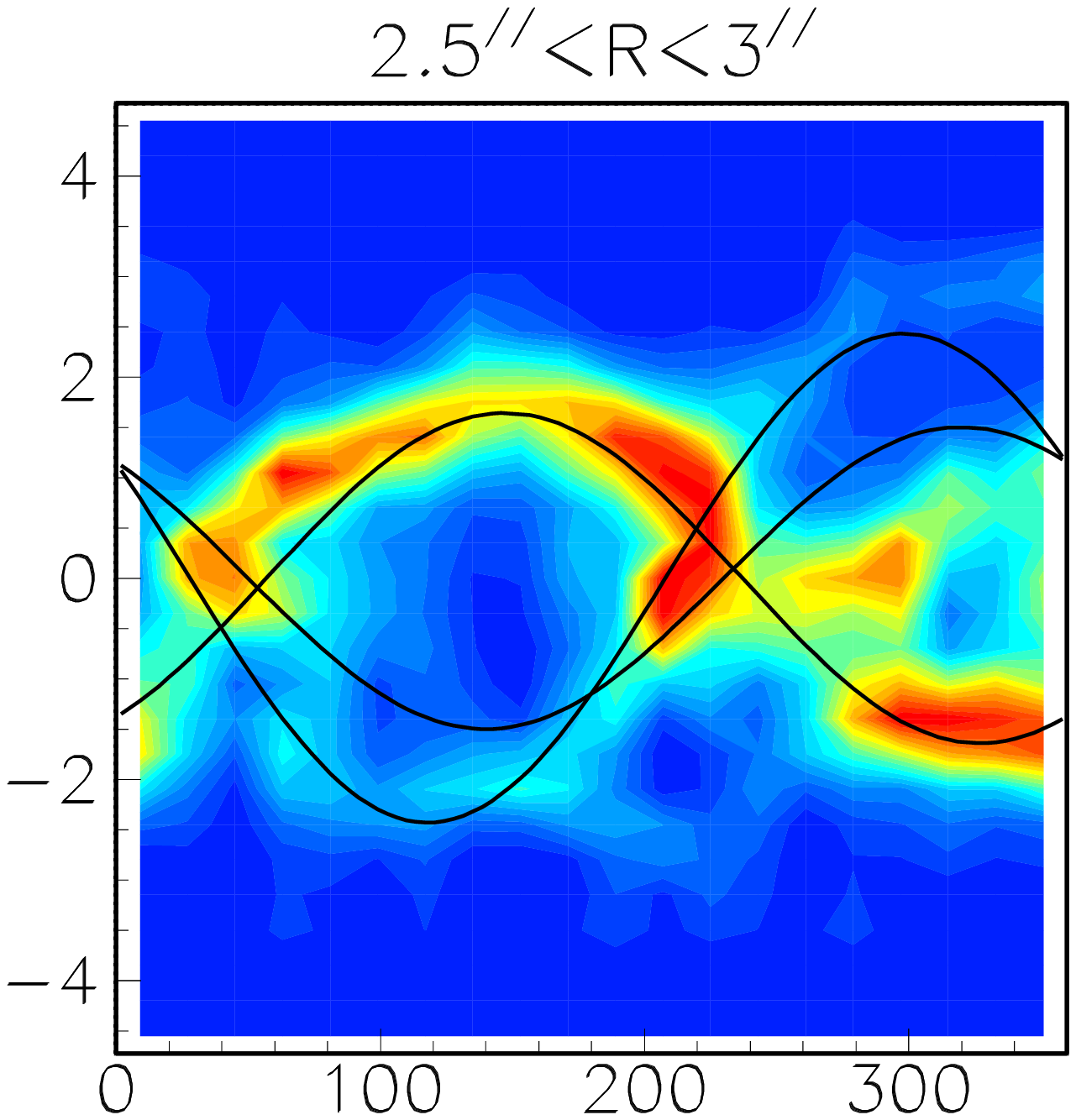}
  \includegraphics[height=3.795cm,trim=1.cm 1.cm 0.5cm .5cm,clip]{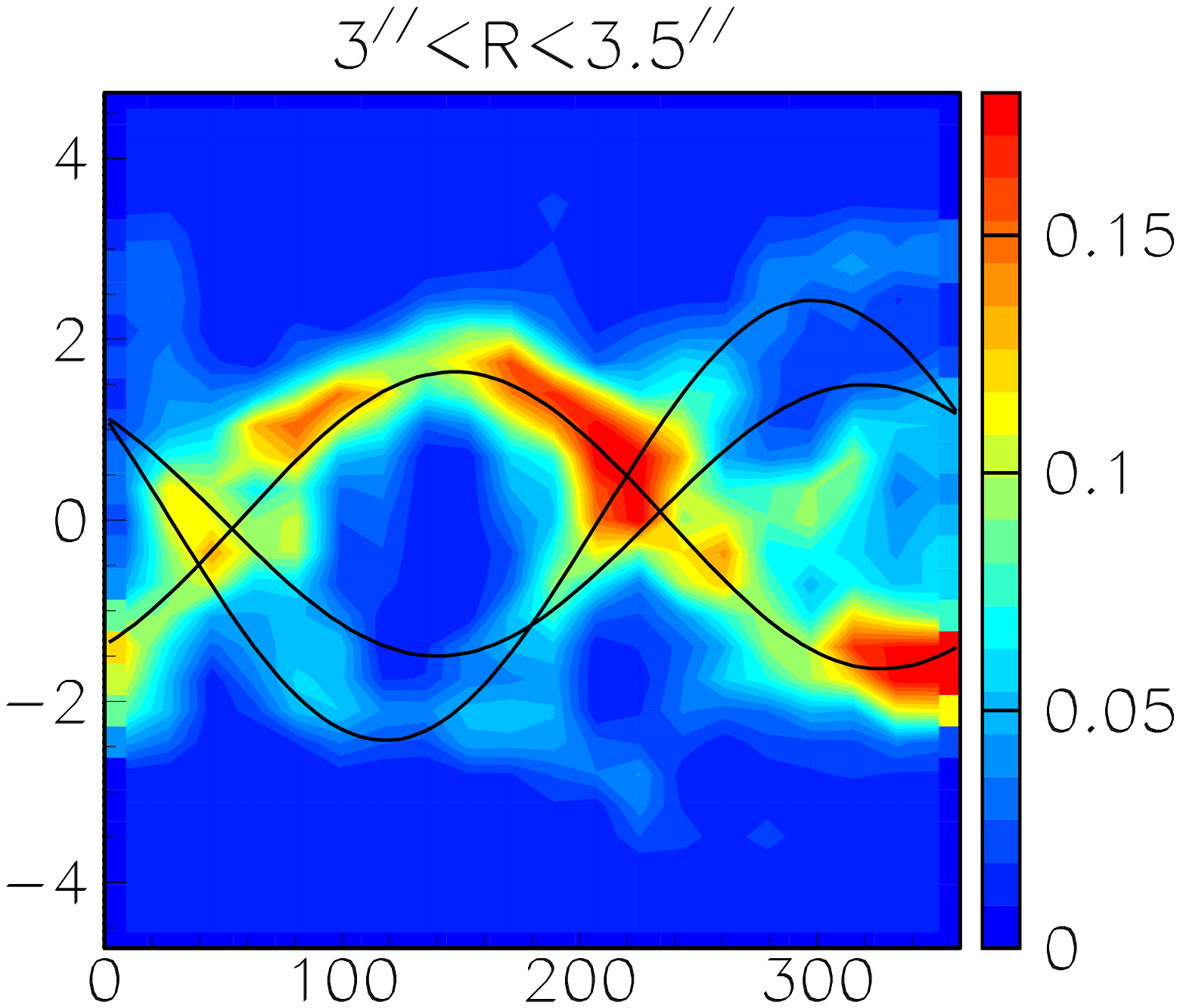}\\
  \includegraphics[height=3.39cm,trim=.5cm 1.2cm 2.4cm 2.cm,clip]{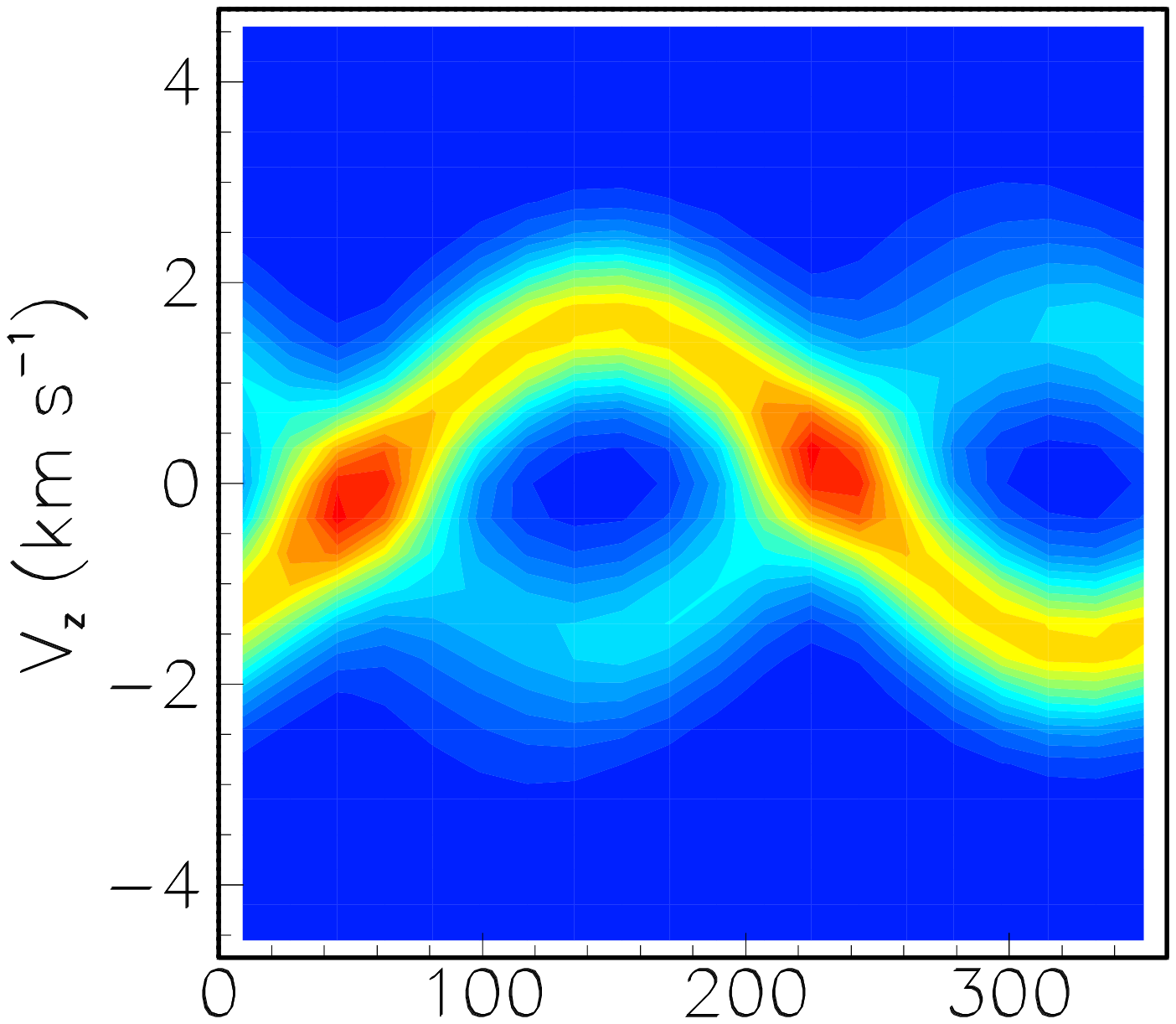}
  \includegraphics[height=3.39cm,trim=1.cm 1.2cm 2.4cm 2.cm,clip]{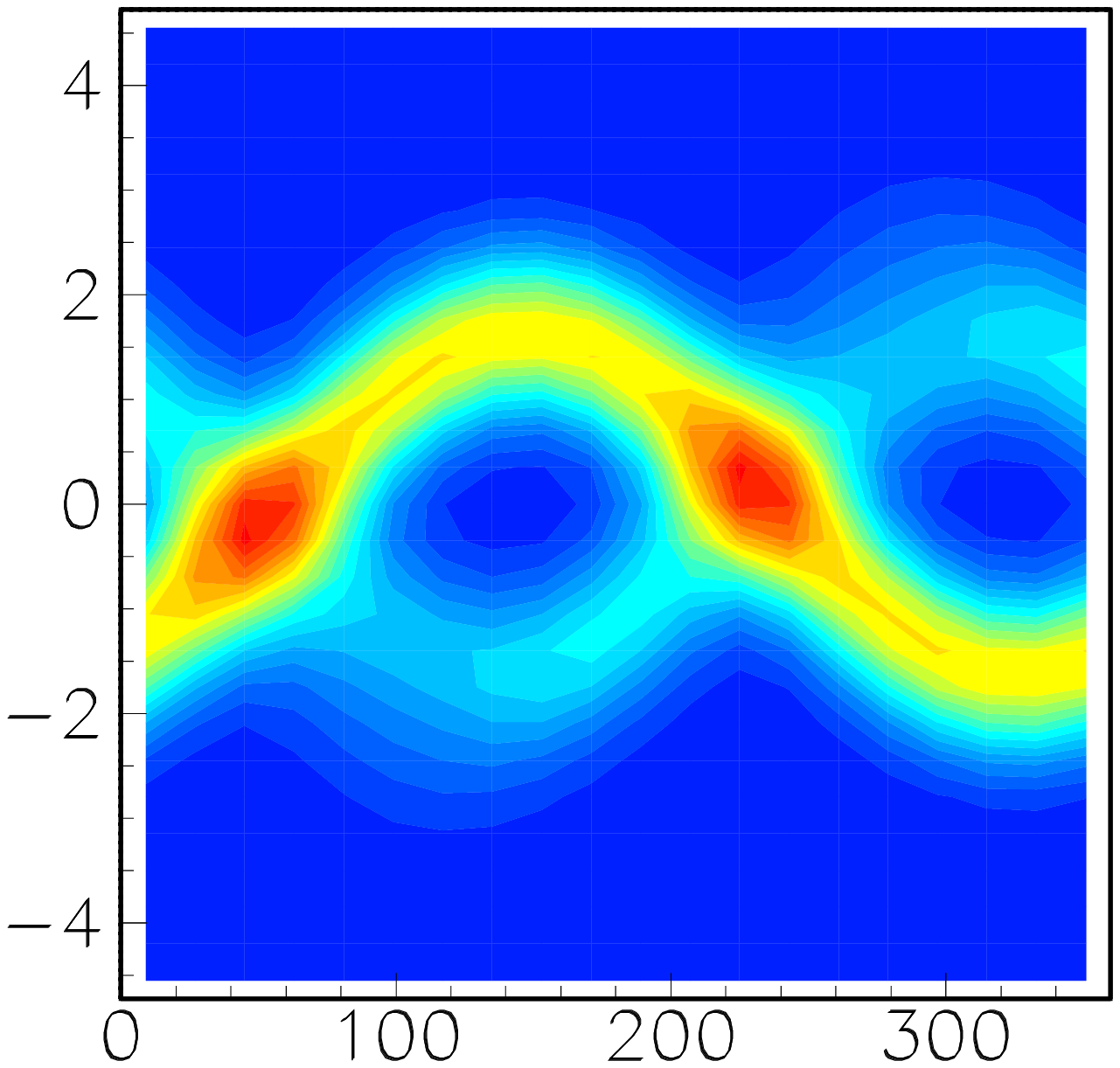}
  \includegraphics[height=3.39cm,trim=1.cm 1.2cm 2.4cm 2.cm,clip]{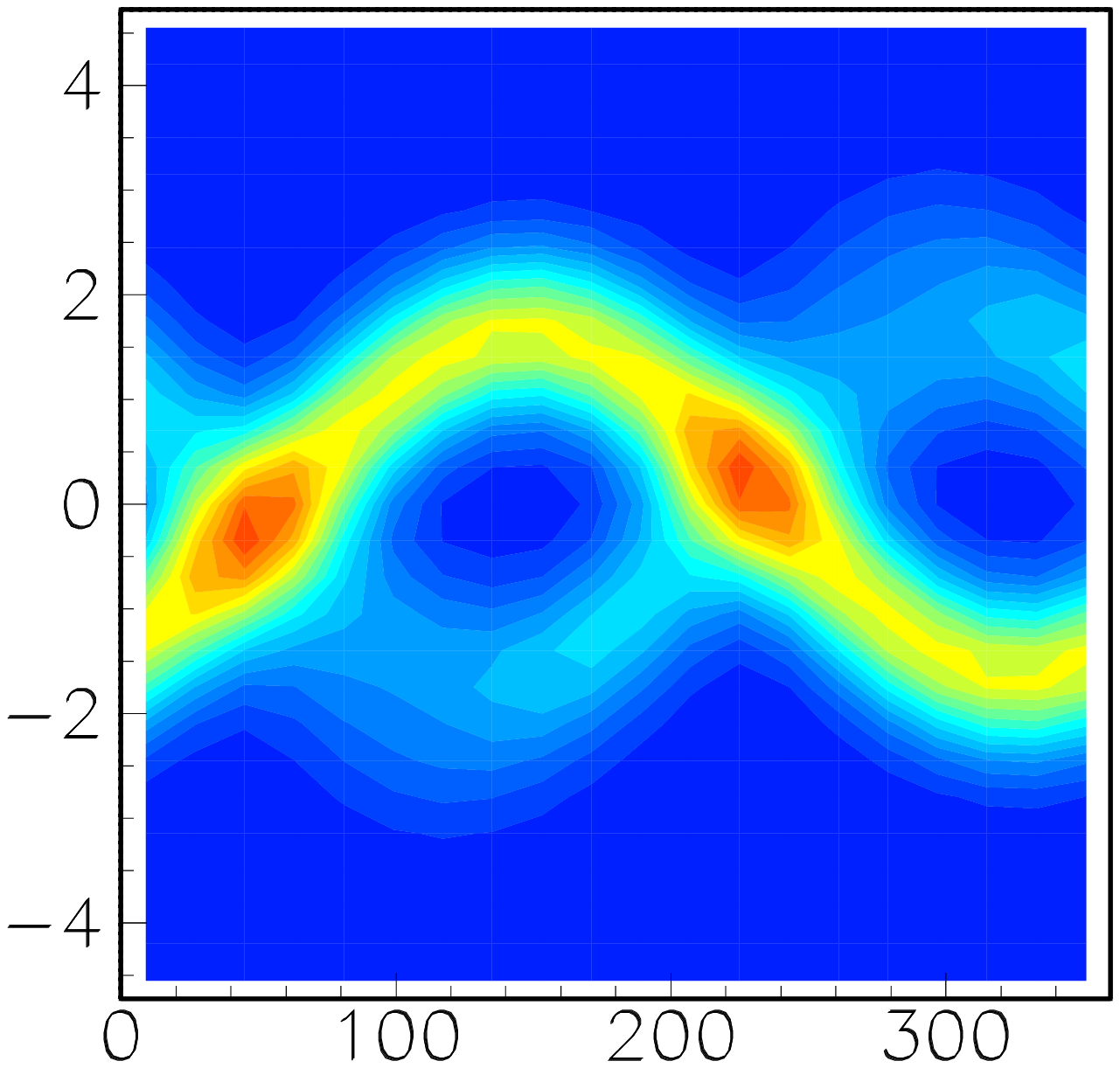}
  \includegraphics[height=3.39cm,trim=1.cm 1.2cm 2.4cm 2.cm,clip]{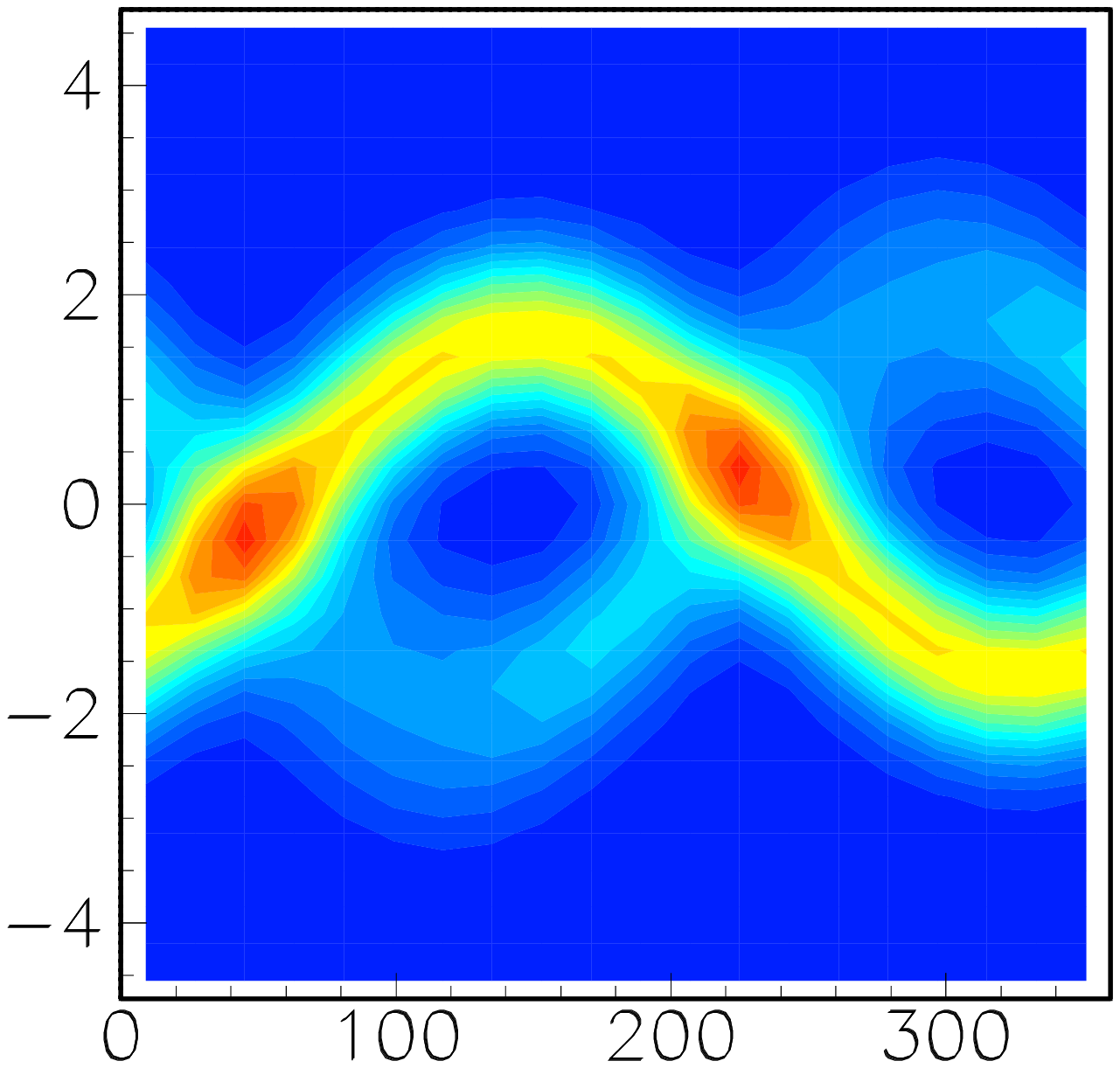}
  \includegraphics[height=3.39cm,trim=1.cm 1.2cm 0.5cm 2.cm,clip]{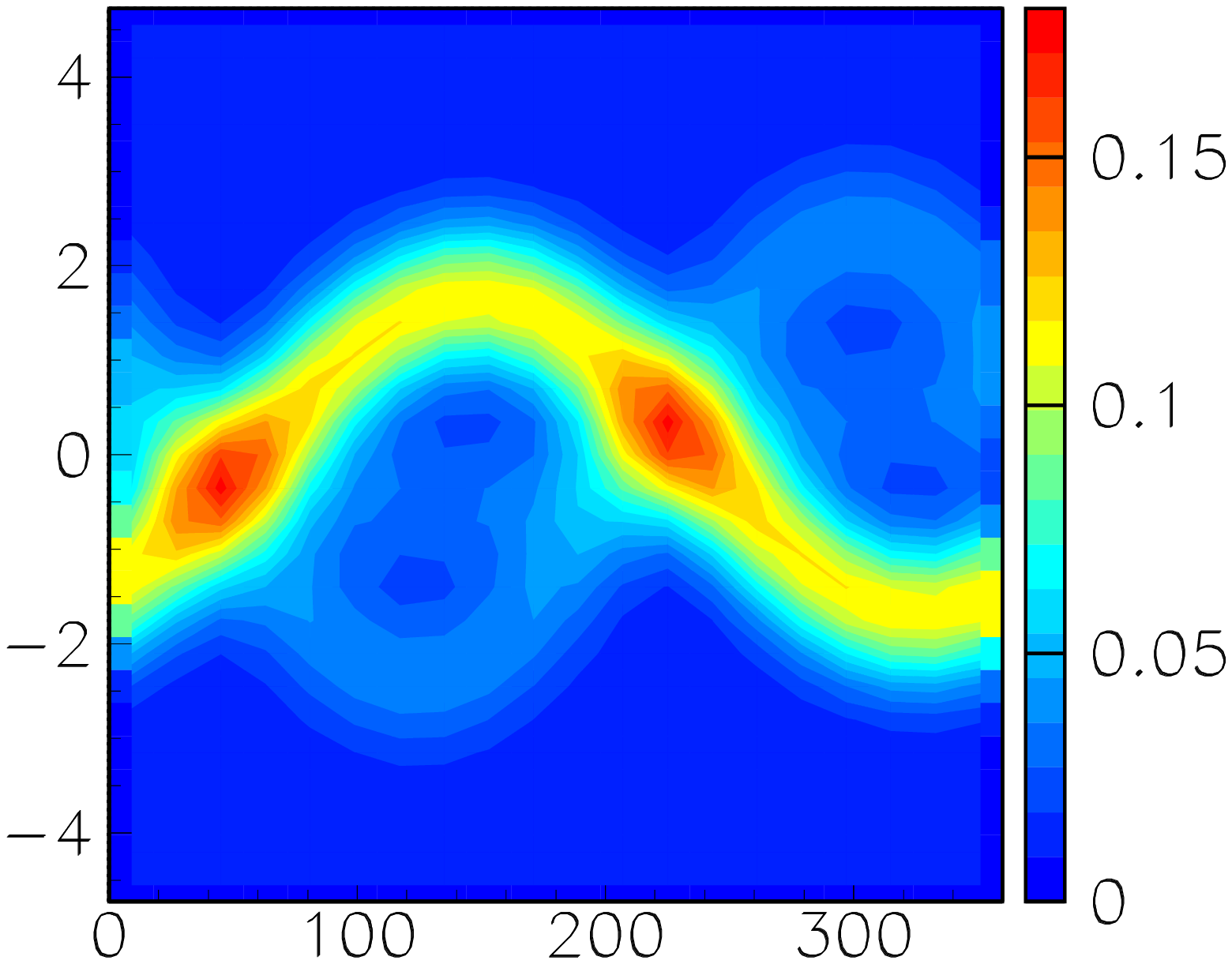}\\
  \includegraphics[height=3.42cm,trim=0.5cm 1.cm 2.4cm 2.cm,clip]{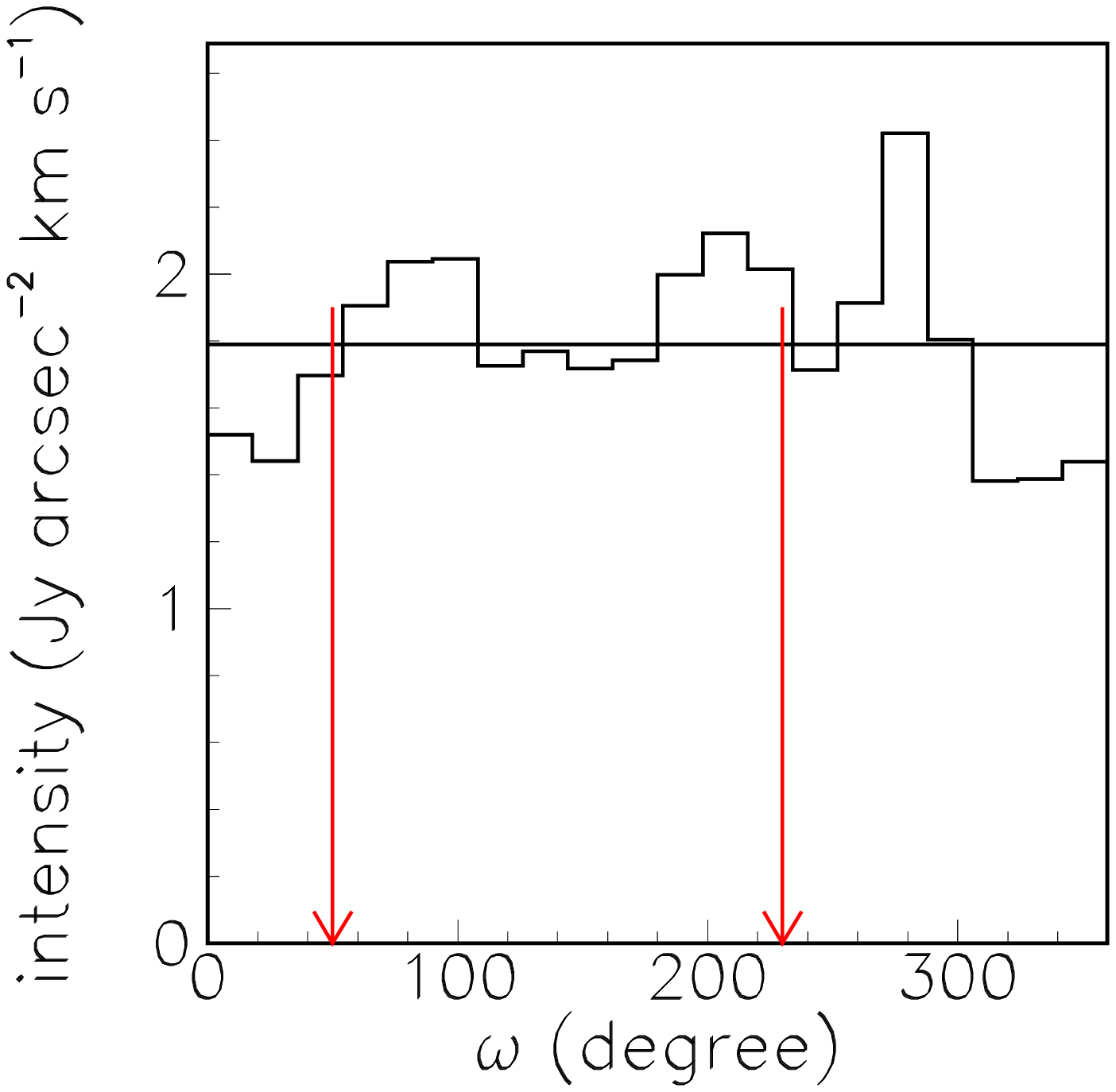}
  \includegraphics[height=3.42cm,trim=1.cm 1.cm 2.4cm 2.cm,clip]{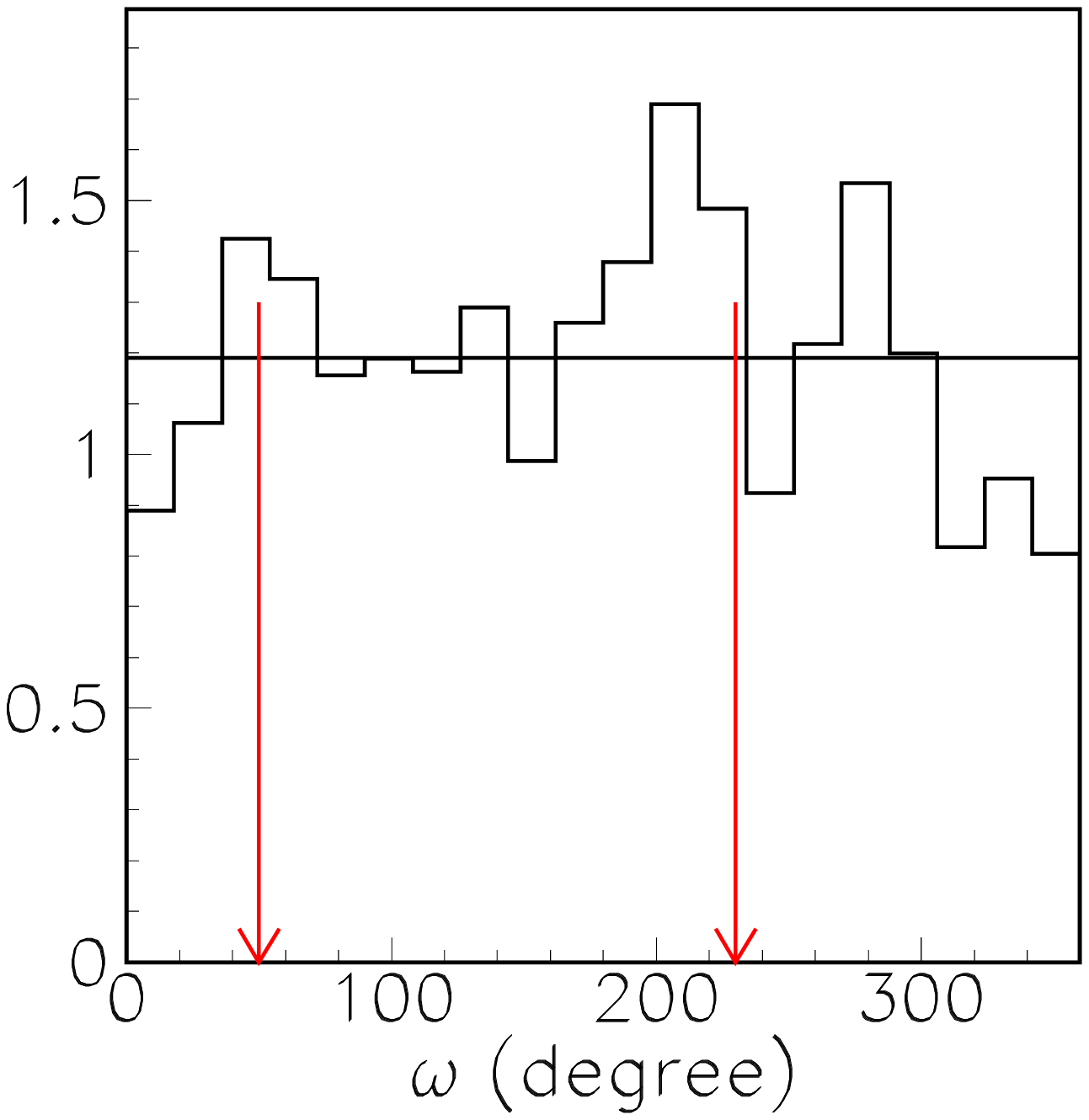}
  \includegraphics[height=3.42cm,trim=1.cm 1.cm 2.4cm 2.cm,clip]{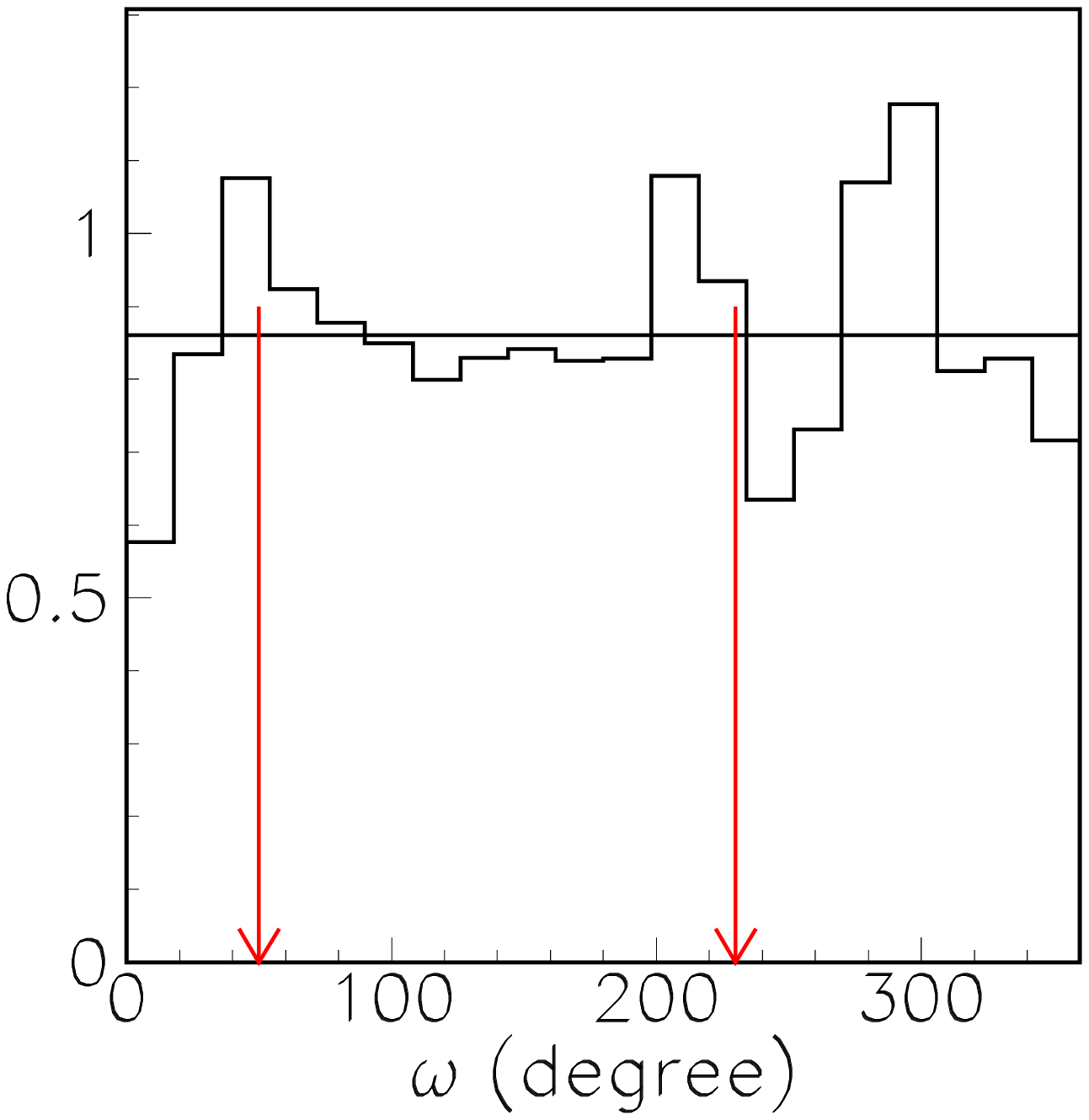}
  \includegraphics[height=3.42cm,trim=1.cm 1.cm 2.4cm 2.cm,clip]{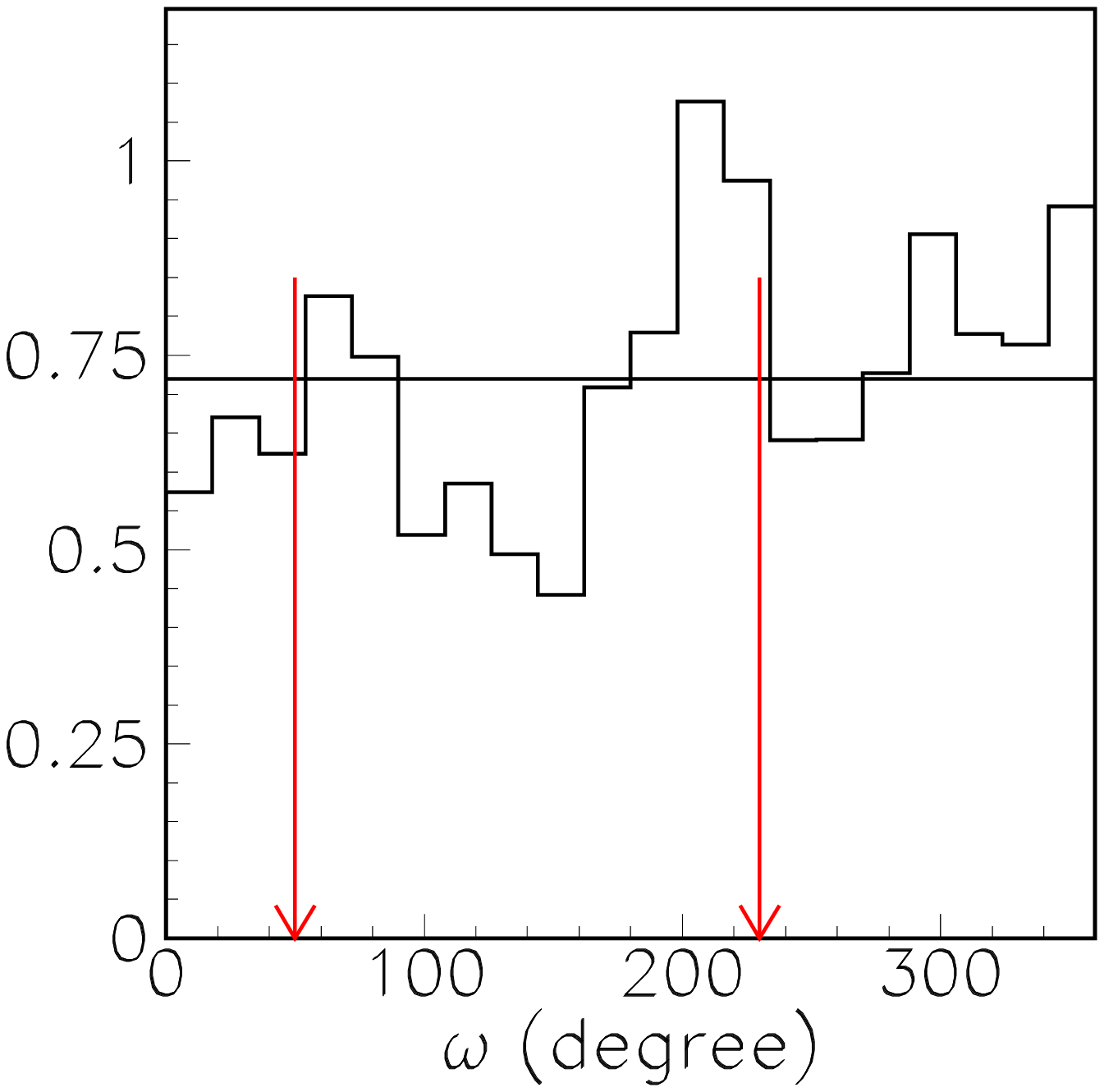}
  \includegraphics[height=3.42cm,trim=1.cm 1.cm -0.5cm 2.cm,clip]{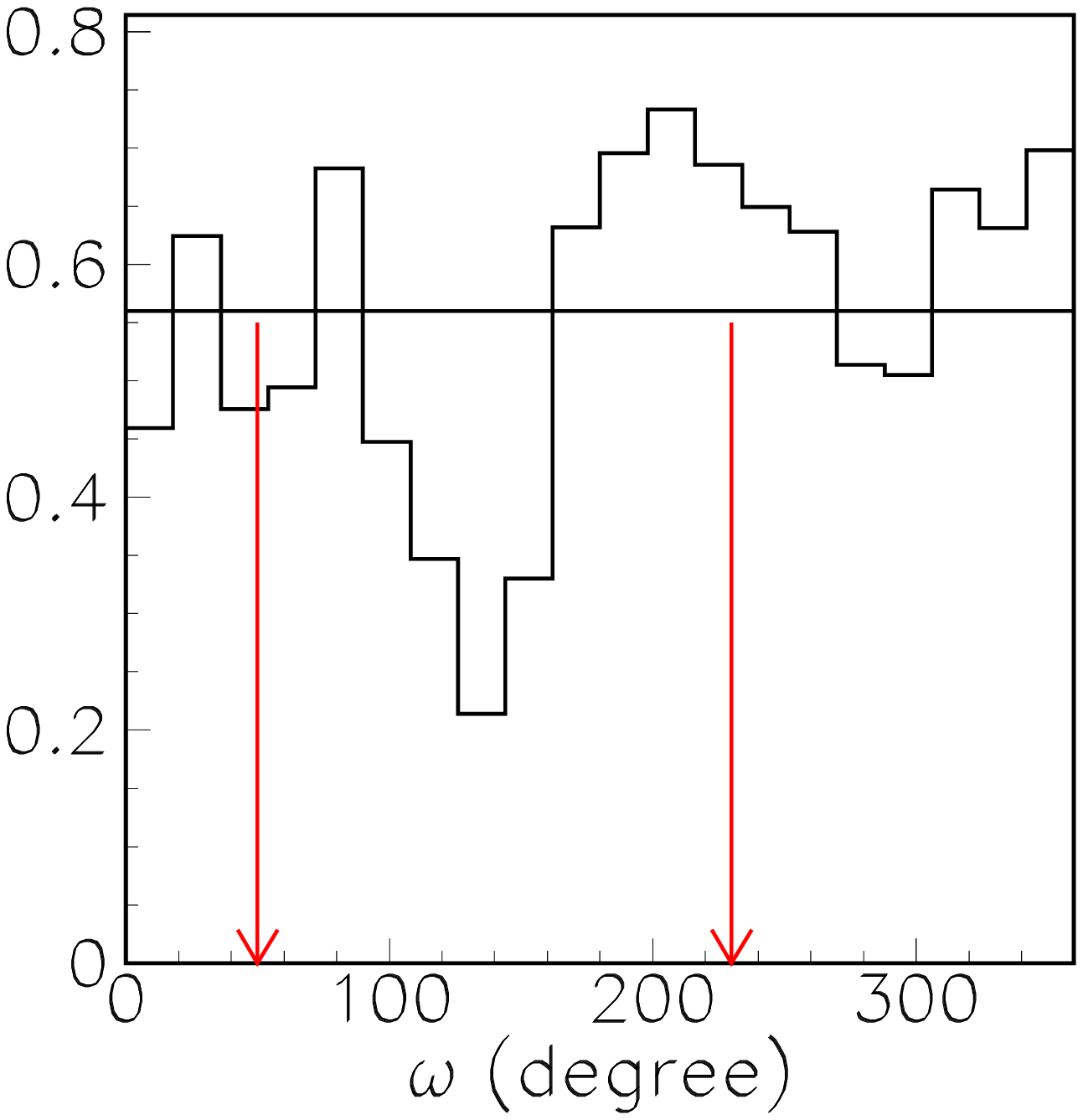}\\
  
    \caption{$^{12}$CO(2-1) line emission away from the star. Upper and middle rows: PV maps ($V_z$ vs $\omega$) of the data (up) and of the model (middle). The model sine waves are shown on the data maps. Lower panels: dependence on position angle $\omega$ of the intensity integrated over $|V_z|$$<$2.5 \kms; the red arrows indicate the north-east and south-west directions.The horizontal lines show the mean values. In all panels, from left to right, 1$<$$R$$<$1.5 arcsec,  1.5$<$$R$$<$2 arcsec, 2$<$$R$$<$2.5 arcsec, 2.5$<$$R$$<$3 arcsec and 3$<$$R$$<$3.5 arcsec. The colour scales are in units of Jy.}
 \label{fig15}
\end{figure*}

\begin{figure*}
  \centering
  \includegraphics[height=6cm,trim=.5cm 1.cm 2.4cm 1.5cm,clip]{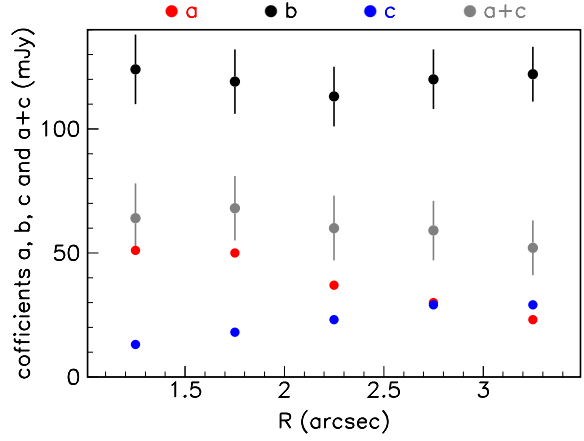}
    \caption{Dependence on $R$ of the best fit values of coefficients $a$ (red), $b$ (black), $c$ (blue) and $a+c$ (grey). Uncertainties are estimated corresponding to a 3\% increase of the best fit chi-square.}
 \label{fig16}
\end{figure*}

Figure \ref{fig15} illustrates the radial evolution of the $^{12}$CO(2-1) emission in successive intervals of $R$, each 0.5 arcsec wide, starting from $R$$=$1 arcsec. Close inspection of the $V_z$ vs $\omega$ maps may suggest the presence of two weak sine waves in the emission in phase opposition to the dominant sine wave. In order to obtain a simple parameterization of these features, we write the dominant sine wave as $V_z$ [\kms]$=$1.64sin($\omega$$-$57\dego)and the weaker sine waves as 1.50sin($\omega$$-$230\dego) and 2.43sin($\omega$$-$208\dego), respectively. Each sine wave is smeared with Gaussians having $\sigma$'s of 0.55 \kms\ in amplitude and 15\dego in phase. All these parameters have been estimated by eye to approximately match the data. In each interval of $R$, the observed dependence of the flux density on $V_z$ and $\omega$ is described as the sum of these three smeared sine waves, multiplied by coefficients $a$, $b$ and $c$ respectively. The result of the best fits is listed in Table \ref{tab3}, which lists also the mean values of the flux density in each interval of $R$ and the rms deviations with respect to the mean. The contribution of the dominant sine wave is constant over the whole $R$ range. Each of the weaker waves contributes approximately one quarter of the dominant wave; the contribution of one of them increases over the first half of the $R$ range, that of the other decreases. While this provides a convenient parameterization of what is observed, its precise form, in particular the description of the emission in phase opposition to the dominant sine wave in terms of two weaker sine waves, is somewhat arbitrary Figure \ref{fig16} illustrates the dependence on $R$ of the $a$, $b$ and $c$ coefficients. Both the intensity of the dominant sine wave ($b$) and that of the flow in phase opposition ($a+c$) are approximately constant over the whole $R$ range, the former being twice the latter. It seems therefore reasonable to be satisfied with a global description of the emission in phase opposition to the dominant sine wave, leaving the possible identification of a finer structure for a future analysis including new observations. What can be stated with confidence is the dominance of a sine wave with amplitude 1.6$\pm$0.1 \kms\ and a phase of 54\dego$\pm$6\dego, as obtained from a fit to the right panel of Figure \ref{fig14}, together with a flow of about half the intensity in phase opposition.

  Such a sine wave describes a disc, or equatorial enhancement, in radial expansion. To describe its morpho-kinematics in some detail, as done by \citet{Homan2017} for the central gas-and-dust disc or by \citet{Winters2021} for RS Cnc, is well beyond what the present data can offer. We must content ourselves with evaluations of the mean radial expansion velocity, $V_{exp}$, and of the orientation of the disc axis. The position angle of the projection of the disc axis on the plane of the sky is simply 54\dego+90\dego$=$144\dego$\pm$6\dego. However, its inclination $i$ with respect to the lines of sight, equal to that of the mid-plane of the disc with respect to the plane of the sky, is unknown. Yet, the disc cannot be too close to edge on, otherwise it would produce a spike at $-V_{exp}$  in the spectra of Figure \ref{fig5}; and it cannot be close to the plane of the sky, otherwise the sine wave would not appear so clearly. To see that better, we use the simple picture of a flared disc having a Gaussian distribution of flaring angles with an rms value of $\sigma_{fl}$ and we approximate the smearing in $V_z$ as a constant $\sigma_{Vz}$$=$0.55 \kms, independent from $\omega$. The amplitude of the sine wave, 1.6$\pm$0.1 \kms, is the product of $V_{exp}$ by the sine of $i$: $V_{exp}$$=$$1.6/\sin i$ \kms. The smearing in $V_z$ is dominated by flaring in the north-east/south-west direction ($\omega$$=$54\dego/234\dego) where $V_z$$=$0: $\sigma_{Vz}$$=$$V_{exp}\sin\sigma_{fl}\cos i$. In the north-west/south-east direction ($\omega$$=$144\dego/324\dego), where $V_z$$=$$V_{exp}\sin i$, it is instead dominated by the dispersion in the sine wave amplitude and flaring does not contribute. Table \ref{tab4} lists the values of $V_{exp}$$=$$1.6/\sin i$ and of the rms flaring angle, $\sigma_{fl}$$=$$\sin^{-1}(0.55/[V_{exp}\cos i])$ for different values of $i$.  

  \begin{table*}
    \caption{ Disc parameters as a function of the angle $i$ between the disc mid-plane and the plane of the sky.}
    \label{tab4}
    \begin{tabular}{ccccccc}
      \hline
    $i$ (degree)&
    20&
    30&
    40&
    50&
    60&
    70\\
    $V_{exp}$ (\kms)&
    4.7&
    3.2&
    2.5&
    2.1&
    1.8&
    1.7\\
    $\sigma_{fl}$ (degree)&
    7&
    11&
    17&
    24&
    37&
    71\\
    \hline
    \end{tabular}
  \end{table*}

  An approximate upper limit to $i$ is obtained by requiring that 90\dego$-$$i$ exceed $\sigma_{fl}$, which means $i<\sim$60\dego.  Indeed, for such an inclination, the disc projects on the plane of the sky as an ellipse having an axis ratio of 2, incompatible with Figure \ref{fig12}. It is more difficult to obtain a lower limit to $i$. When the median plane of the disc gets close to the plane of the sky, the expansion velocity increases and the flaring decreases in order to produce the observed sine wave.  Strictly speaking, one cannot exclude a very thin disc very close to the plane of the sky without making additional assumptions. Yet, too thin a disc would be highly improbable in the CSE of an AGB star and we retain as a reasonable range of acceptable values of $i$ the interval [20\dego,60\dego], which we write as 40\dego$\pm$20\dego, it being understood that this estimate rests on sensible but somewhat arbitrary arguments.

\begin{figure*}
  \centering
  \includegraphics[height=5cm,trim=.5cm 1.cm 1.5cm 2cm,clip]{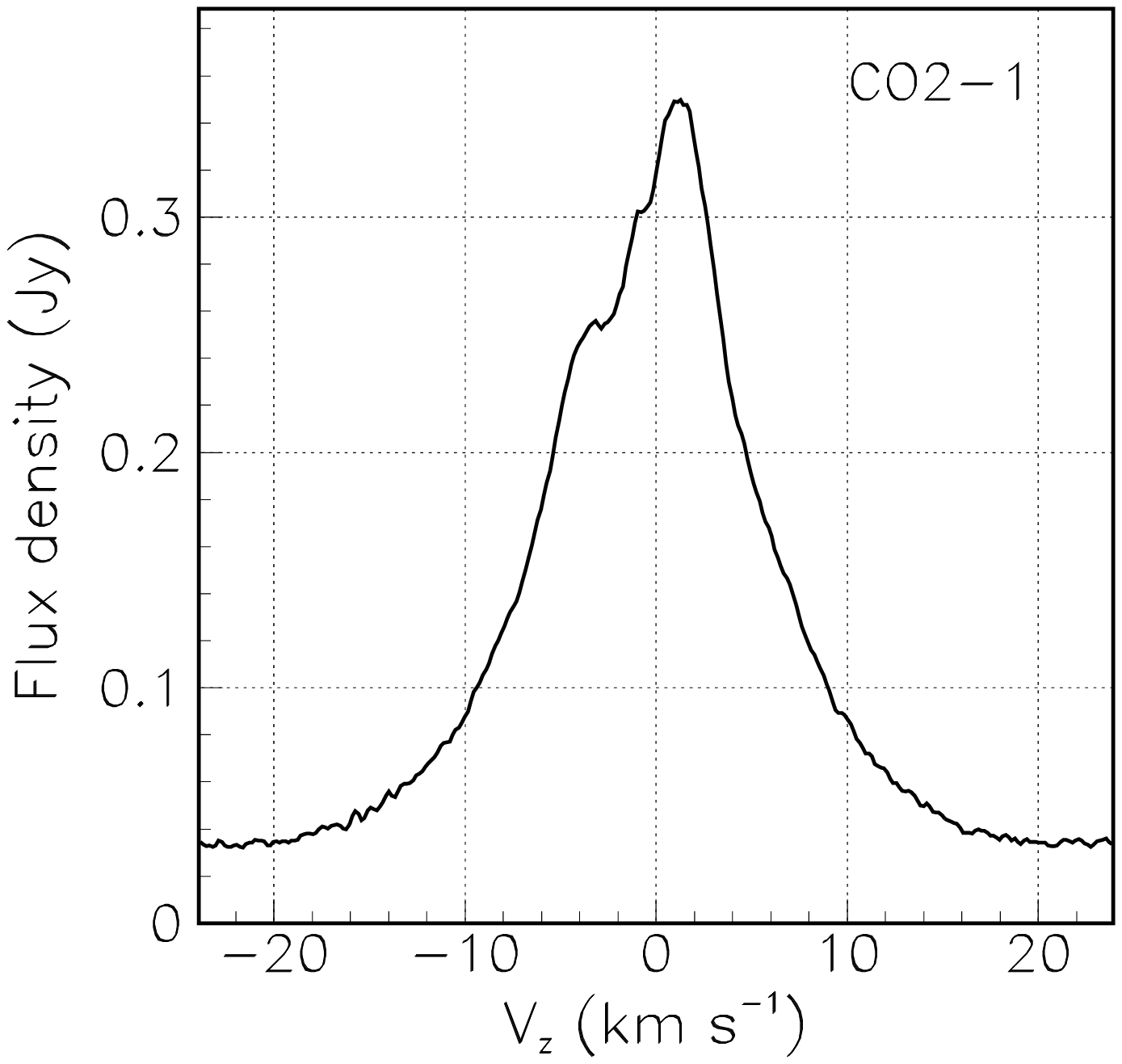}
  \includegraphics[height=5cm,trim=1.2cm 1.cm 1.5cm 2cm,clip]{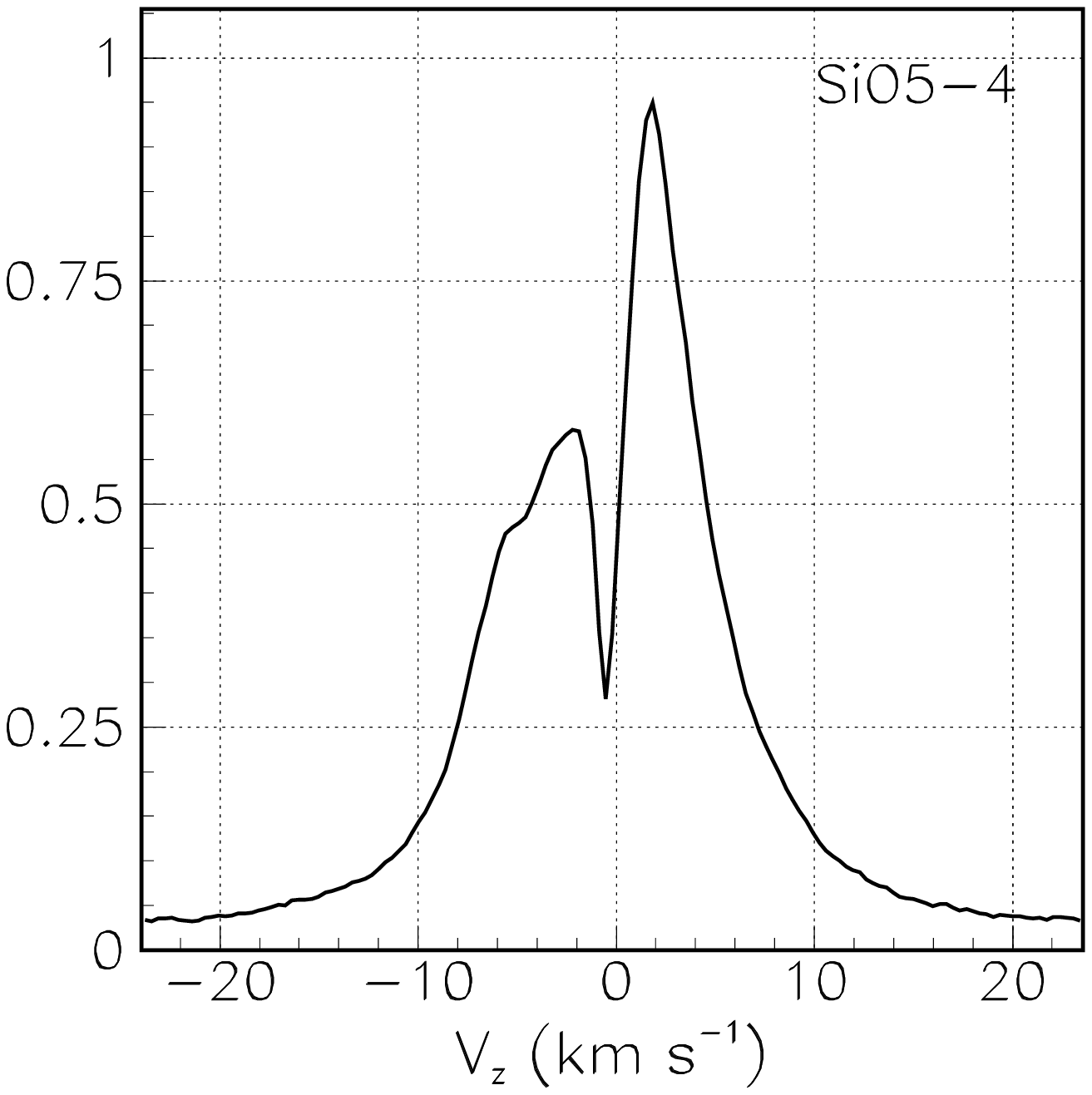}
  \includegraphics[height=5cm,trim=1.2cm 1.cm 1.5cm 2cm,clip]{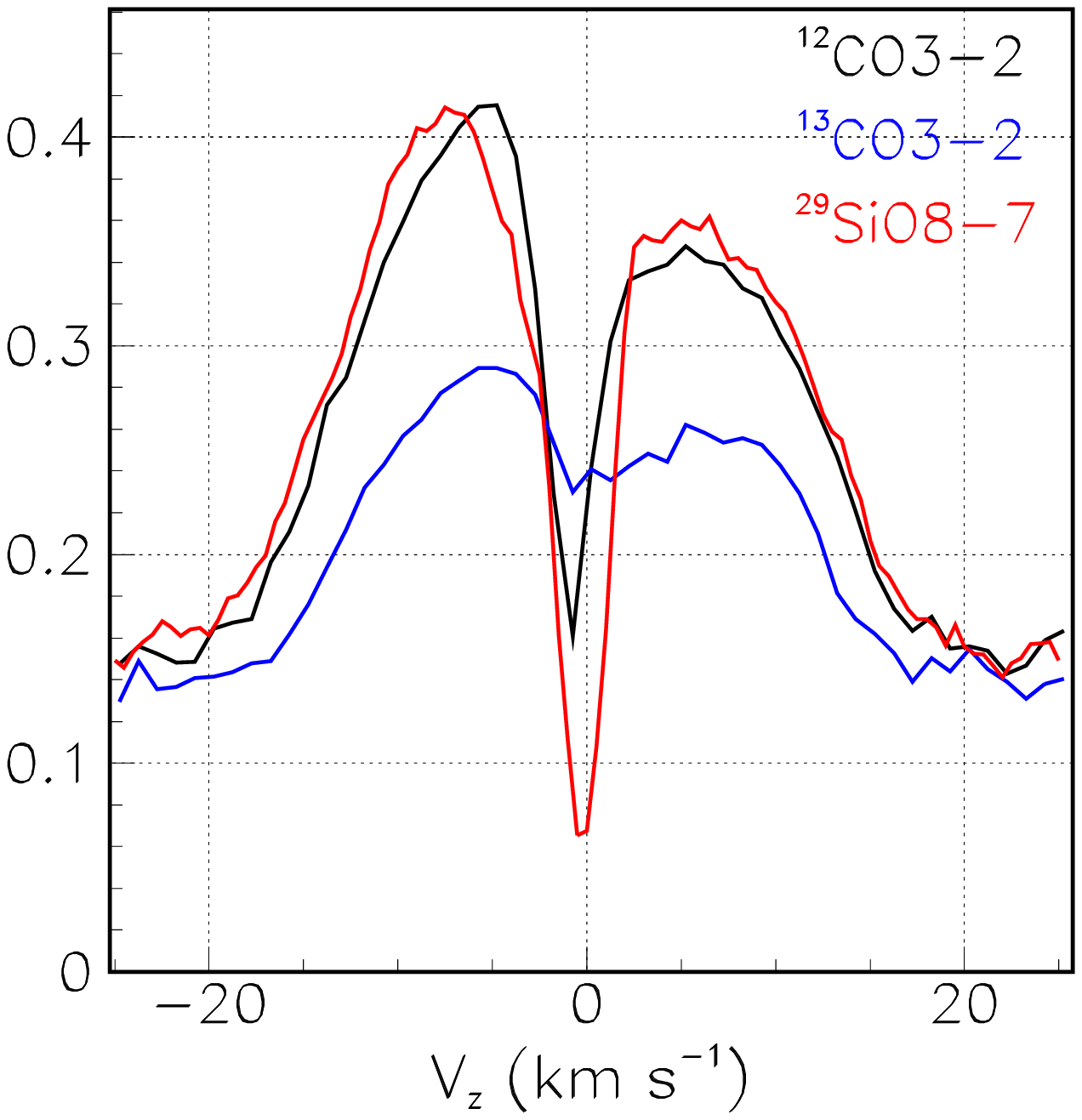}
    \caption{Emission of the $^{12}$CO(2-1) (left) and $^{28}$SiO(5-4) (middle) over the stellar disc ($R$$<$0.2 arcsec). Spectra of $^{13}$CO(3-2), $^{12}$CO(3-2) and $^{29}$SiO(8-7) over the circle $R$$<$0.06 arcsec are shown for comparison in the right panel.}
 \label{fig17}
\end{figure*}

  Figure \ref{fig17} displays the spectra of the $^{12}$CO(2-1) and $^{28}$SiO(5-4) line emissions inside a circle of 0.2 arcsec radius, namely well over the stellar disc, the beam size preventing higher resolution inspection. Both spectra show the slightly negative absorption peak that was interpreted as revealing the small expansion of the gas and dust disc, estimated in Section 3 at the approximate level of 0.8$\pm$0.5 \kms. The peak is deeper in the SiO spectrum than in the CO spectrum as expected from the shorter range of distances being probed. In addition the spectra show hints of other absorption peaks, the CO spectrum at $\sim$$-$3 \kms\ and the SiO spectrum at $\sim$$-$5 \kms. If real, these are probably unrelated to the disc wind.

\begin{table*}
  \centering
  \caption{Best fits of the $^{12}$CO(2-1) emission as the sum of three sine waves, smeared with 0.55 \kms\ in $V_z$ and 15\dego\ in $\omega$, of the form $V_z$ [\kms]$=$1.50sin($\omega$$-$230\dego), 1.64sin($\omega$$-$57\dego) and 2.43sin($\omega$$-$208\dego), respectively.  The last two columns list the mean and rms values of the intensity integrated over $|V_z|$$<$2.5 \kms. }
  \label{tab3}
  \begin{tabular}{ccccccc}
    \hline
    $R$ range &$a$&$b$&$c$&$a+c$&$<$$F$$>$&Rms{$F$}\\
    (arcsec)&\multicolumn{4}{c}{(mJy)}&\multicolumn{2}{c}{(Jy arcsec$^{-2}$ \kms)}\\
    \hline
1.0 to 1.5&
51&
124&
35&
86&
1.79&
0.38\\
1.5 to 2.0&
50&
119&
30&
80&
1.19&
0.31\\
2.0 to 2.5&
37&
113&
29&
66&
0.86&
0.20\\
2.5 to 3.0&
30&
120&
29&
59&
0.72&
0.21\\
3.0 to 3.5&
23&
122&
29&
52&
0.56&
0.19\\
\hline
  \end{tabular}
  \end{table*}

\subsection{Mass loss rate}
In order to obtain an estimate of the mass loss rate, we measure the flow of $^{12}$CO(2-1) emission across a sphere of 2 arcsec radius, where the terminal velocity can be assumed to have been reached and the emission to have occurred before the 1994 dimming episode. It has a value of 1.0 Jy arcsec$^{-2}$ \kms\ (Figure \ref{fig12} left). From Table \ref{tab1}, we see that this is 20 times the brightness produced by a column density of 1 CO molecule cm$^{-3}$ arcsec, at a temperature of 50 K. As a solar mass is 1.2$\times$10$^{57}$ proton masses and 1 arcsec$=$0.96$\times$10$^{15}$ cm, assuming an abundance ratio CO/H=10$^{-4}$, the measured brightness corresponds to a column density of 20$\times$0.963/1.2$\times$10$^{-8}$$=$1.5$\times$10$^{-7}$ \msun arcsec$^{-2}$. Taking the mean sine wave amplitude as 1.7$\pm$0.5 \kms\ and the inclination angle as 40\dego$\pm$20\dego, namely an expansion velocity 2$\pm$1 times larger than the absolute value of the Doppler velocity, we obtain a velocity projected on the plane of the sky of 1.7$\pm$1.0 \kms, namely 0.36$\pm$0.21 au yr$^{-1}$, or $\sim$6$\pm$4 mas yr$^{-1}$. In one year, the wind covers therefore an area of 2$\upi$$R$$\times$0.006=0.08$\pm$0.05 arcsec$^2$, meaning a mass loss rate of 0.08$\times$1.5$\times$10$^{-7}$=(1.2$\pm$0.7)$\times$10$^{-8}$ \msun yr$^{-1}$. This is consistent with the values obtained indirectly from single dish observations, 1.4 to 2.0 10$^{-8}$ \msun\ yr$^{-1}$ \citep{Danilovich2015, Olofsson2002}. While crude, the inclination angle of the wind mid-plane being unknown, it has the value of being the first direct measurement of the gas flow escaping the star gravity. It also shows that the wind described in the present article accounts well for earlier estimates of the mass loss rate, strengthening the case for the absence of a significant collimated north-south outflow near the plane of the sky.

\subsection{A second search for north-south bipolar outflows at short distance from the star}
\begin{figure*}
  \centering
  \includegraphics[height=6cm,trim=1cm 1.5cm 0cm 1cm,clip]{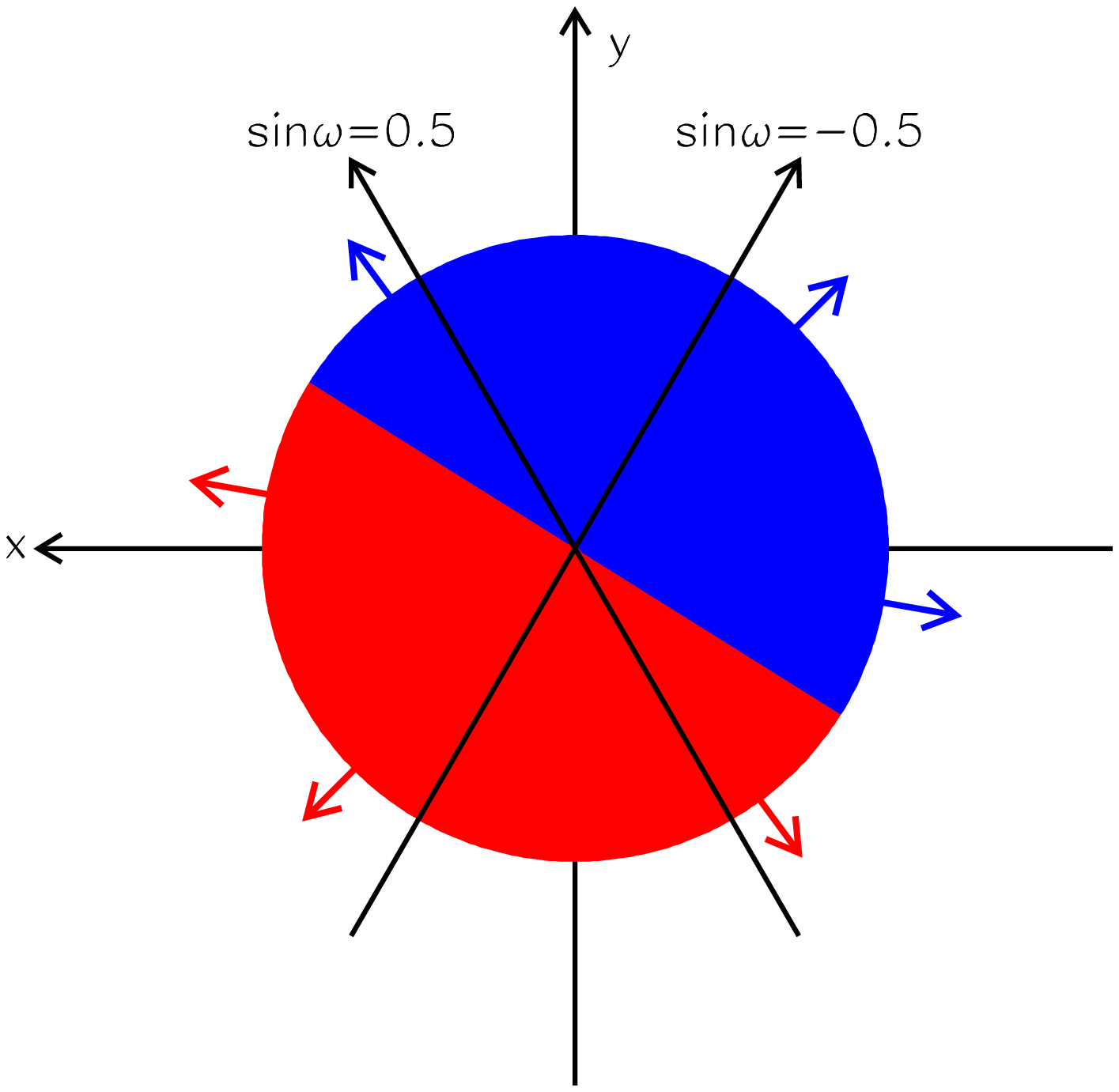}
  \includegraphics[height=6cm,trim=0cm 1.5cm 0cm 1cm,clip]{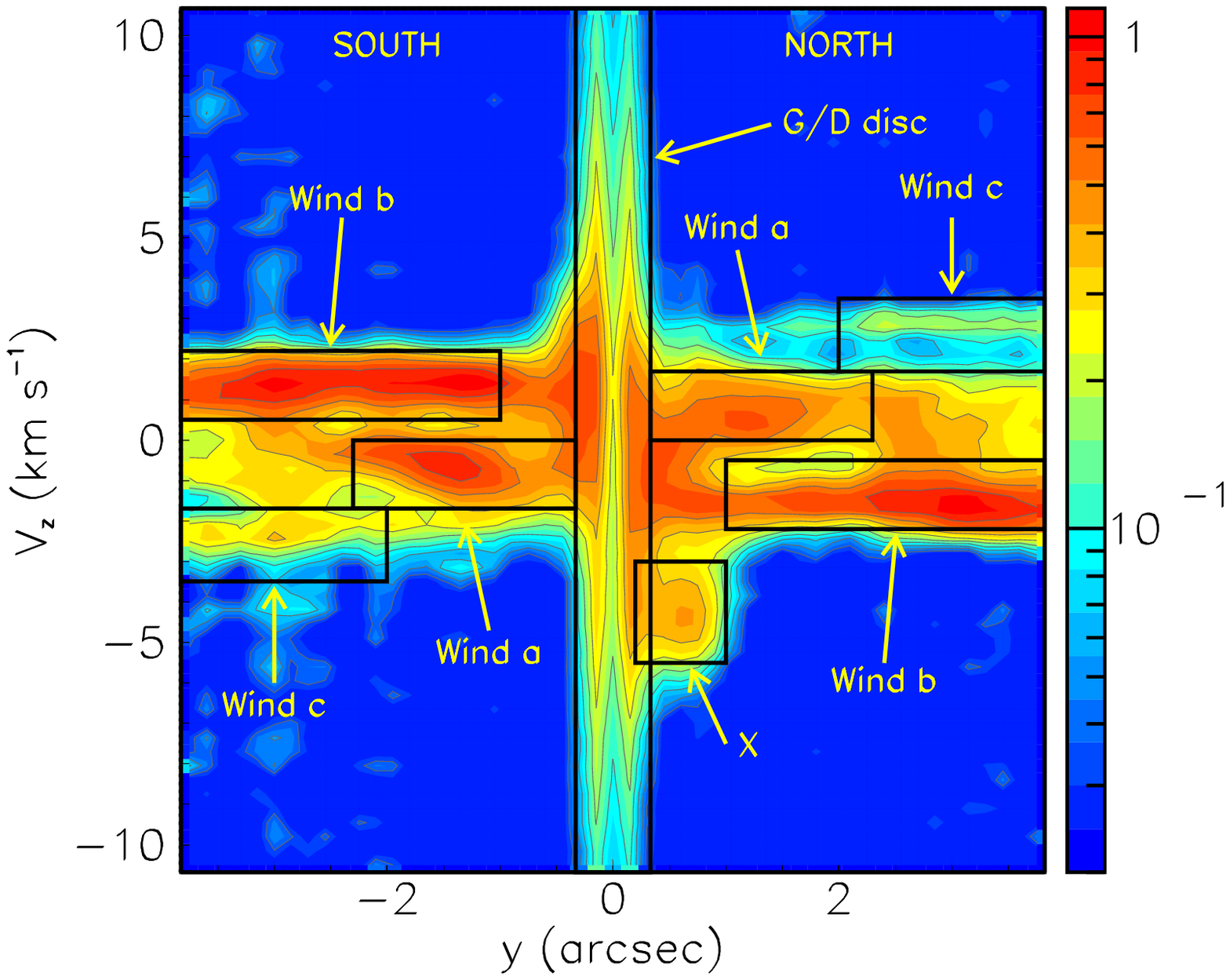}
  \caption{Searching for north-south outflows in the $^{12}$CO(2-1) emission. The left panel shows a schematic view of the expanding dominant $b$ wind, inclined in the north-west/south-east direction with respect to the plane of the sky, and of the wedge, defined as $|\sin\omega|$$<$0.5, within which the PV map shown in the right panel is integrated.  The PV map is in the $V_z$ vs $y$ plane and indicates the various components that contribute to the emission. The central gas and dust disc is indicated as G/D disc and the blob presumably associated with the loop identified by \citet{Kervella2014} is indicated as $X$. The colour scale is in unit of Jy arcsec$^{-1}$.}
  \label{fig18}
\end{figure*}

Having obtained a global view of the morpho-kinematics of the wind beyond $\sim$1 arcsec from the centre of the star, we now have a second look at the question of the expected presence of broad bipolar outflows oriented north/south. We use again the CO(2-1) emission and explore the wedge defined as $|\sin\omega|$$<$0.5, meaning $\pm$30\dego\ from the $y$ axis. The PV map, $V_z$ vs $y$, of the flux density integrated within the wedge is shown in Figure \ref{fig18}. It displays a number of remarkable features that can be easily identified: the emission of the gas and dust disc, essentially unresolved, is seen at very small values of $|y|$ and extends to very large Doppler velocities in absolute value. The emission of the expanding wind discs is seen at Doppler velocities confined to |$V_z|$$<$3 \kms\ and the three components $a$, $b$ and $c$ listed in Table \ref{tab3} are well separated: the dominant wind $b$ is red-shifted in the south direction and blue-shifted in the north direction and covers the whole y range; winds $a$ and $c$ are instead blue-shifted in the south direction and red-shifted in the north direction, wind $a$ covering shorter distances and wind $c$ larger distances from the star. The emission of the pair of back-to back outflows expected by \citet{Kervella2016} and \citet{Chen2016} would instead populate preferentially the negative $y$ hemisphere for $V_z$$>$0, the gas-and-dust disc being inclined toward us in this hemisphere; by how much depends on the velocity of the flow, claimed by these authors to exceed $\sim$20 \kms, meaning a mean $|V_z|$$>$2-3 \kms; however, the very large inclination angle of the disc, 82-84\dego, would probably make the effect difficult to detect.

\begin{figure*}
  \centering
  \includegraphics[height=6cm,trim=0cm 1.cm 1cm .5cm,clip]{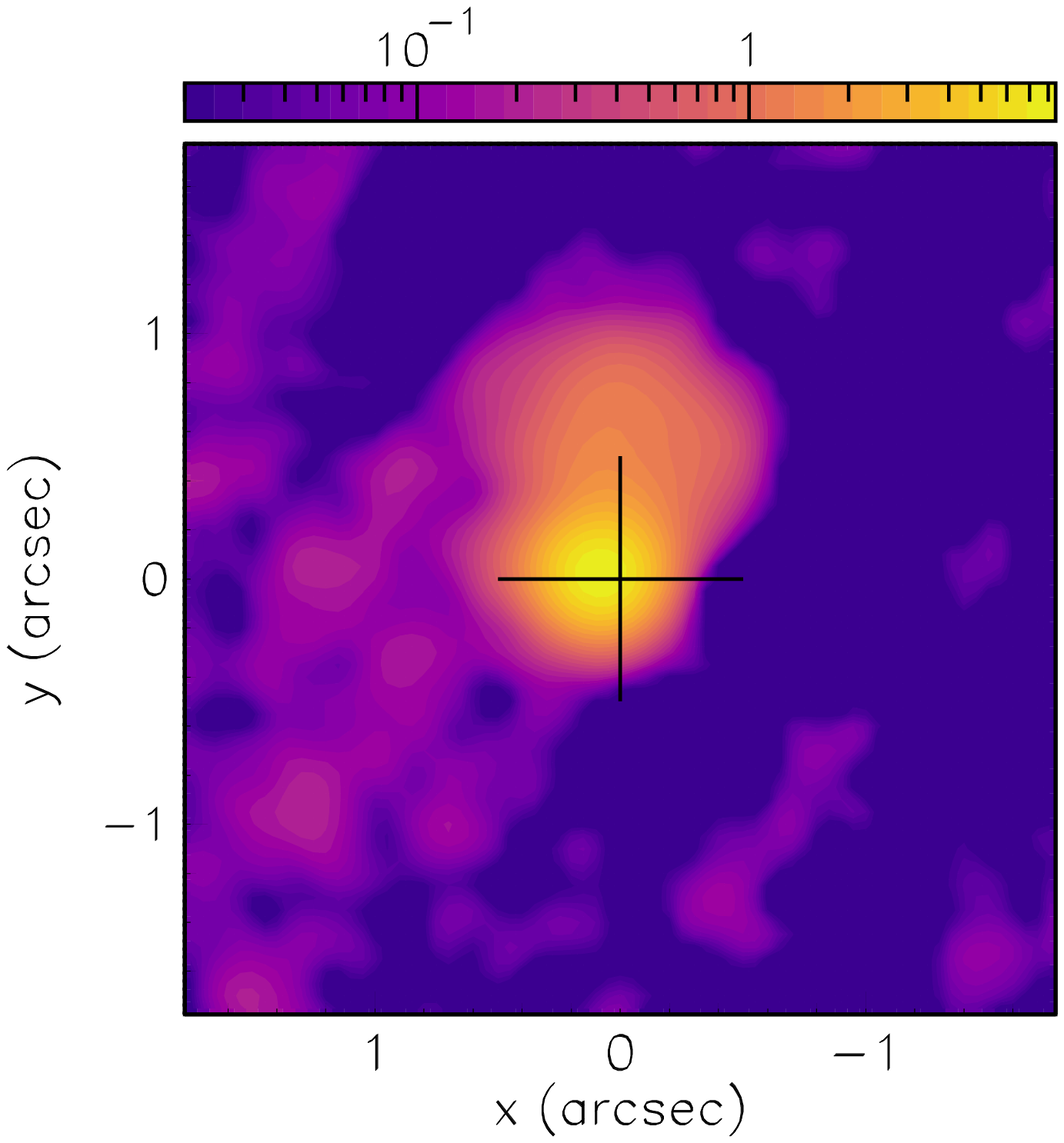}
   \includegraphics[height=6cm,trim=0cm 1.cm 1cm .5cm,clip]{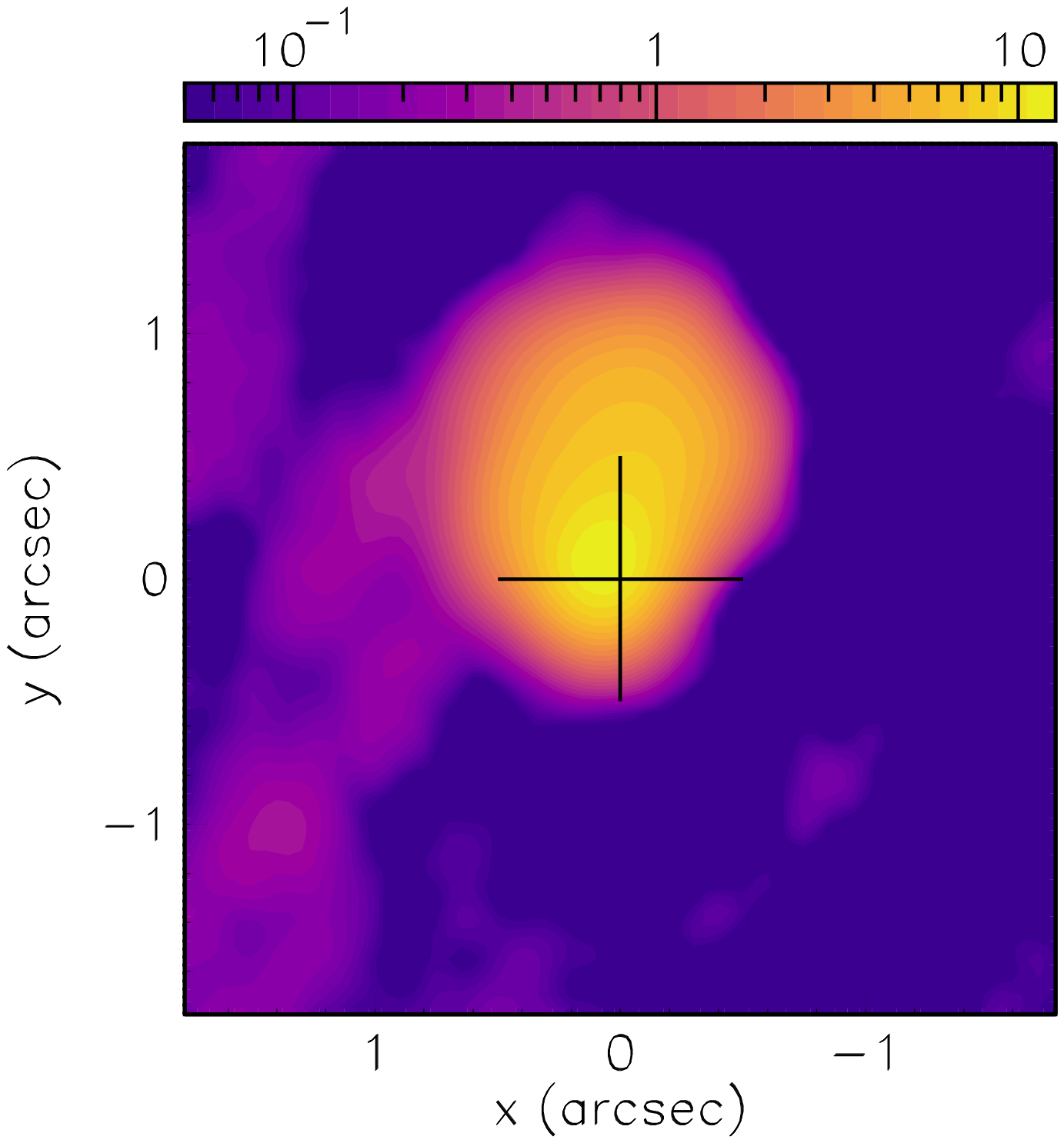}
    \caption{Intensity maps of $^{12}$CO(2-1) (left) and $^{28}$SiO(5-4) (right) emissions integrated between  $-$5 and  $-$3 \kms. The colour scales are in units of Jy arcsec$^{-2}$ \kms.}
 \label{fig19}
\end{figure*}

An unexpected feature is a blob of blue-shifted emission visible between $\sim$$-$3 and $\sim$$-$5 \kms\ in the north direction, within less than 1 arcsec from the centre of the star. This is the region of Doppler velocity where hints of absorption have been revealed in Figure \ref{fig17}. We show in Figure \ref{fig19} the intensity maps of the $^{12}$CO(2-1) and $^{28}$SiO(5-4) emission integrated between $-$3 and $-$5 \kms. The emission is confined near the star and elongated in the north direction. The expanding disc winds are not expected to contribute such relatively large Doppler velocities in this region and it is therefore probable that the blob is associated with the northern ``loop'' identified by \citet{Kervella2014} in the NACO images at 3.74 and 4.05 $\umu$m. These authors suggest an interpretation in terms of a possible interaction of a hidden companion with the dusty wind from the central star. We note that the decrease with $R$ of the absorption band observed in the $V_z$ vs $R$ map of the $^{29}$SiO(8-7) emission(lower-left panel of Figure \ref{fig6}) is probably associated with the presence of this loop.

\subsection{Interpretation}
The disc morphology of the wind up to some 200 au from the centre of the star revealed by the present analysis of the CO(2-1) emission is reminiscent of a similar morphology observed in RS Cnc \citep{Winters2021} and EP Aqr \citep[][and references therein]{Tuan-Anh2019, Homan2020}, two oxygen-rich AGB stars having mass loss rates at the 10$^{-7}$ \msun yr$^{-1}$ scale. The formation of such expanding discs, or equatorial enhancements, is not understood. In the case of EP Aqr, \citet{Homan2020} invoked the possible presence of a companion. In the present case of L$_2$ Pup, the presence of another disc within some 20 au from the centre of the star makes it even more difficult to propose a sensible interpretation. The two discs have different orientations and expansion velocities, the smaller disc being even possibly gravitationally bound to the star. Invoking the role of companions in their formation requires therefore the presence of two distinct companions. Moreover, the relation of the smaller disc with the 1994 dimming episode is not understood and its configuration in the period preceding it, during which the larger disc wind was blowing, is unknown. As remarked by \citet{Homan2017} a latitude dependent mass loss is unlikely to produce a disc that is as thin as observed for the dust disc. The same comment applies to the wind observed at large distances. This brings up an obvious question: how much of the disc morphology is already present at launching stage, and how much of it is the result of subsequent interaction? Major contributors to the former are expected to be pulsations and/or magnetic fields, and to the latter interaction with a companion. Unfortunately, very little is known of the role played by magnetic fields in the evolution of AGB stars \citep{Vlemmings2019}.

\citet{Jura2002} and \citet{Bedding2005} have reviewed our present understanding of the mechanisms governing the pulsations of semi-regular variables. Both underline the peculiarity of L$_2$ Pup, with a light curve dominated by two completely different and independent mechanisms: gradual dimming caused by circumstellar dust and seemingly random amplitude and phase changes of pulsations. The former invoke non radial pulsations that can at least in part account for the observed asymmetric mass loss, time variations of the infrared fluxes, and time variations of the position angle of the optical polarization. The latter prefer to blame these on stochastic excitation, presumably from convection. Moreover, \citet{Jura2002} note that there appears to be enough kinetic energy in the pulsations to drive the outflow, which lends support to models where the main driving mechanism of the circumstellar outflow is the stellar pulsation and where the dust plays only a secondary role \citep{Winters2000}. Qualitatively at least, there is no doubt that pulsations are major actors in the generation of the nascent wind and their irregularity must somehow impact its further evolution. Yet, it seems difficult to conceive a scenario where pulsations alone would produce relatively thin discs as observed both in the wind at large distances and in the gas and dust disc near the star.

Therefore most authors assume that the small central gas and dust disc was formed by a companion funnelling part of the spherically symmetric wind of the AGB star in its gravitational field and expelling it as a compact stream along the orbital plane. \citet{Homan2017} note that the disc could be made of both AGB wind material and evaporated planet material. If a similar mechanism were to be invoked for the formation of the wind at large distances, it would require the presence of at least another planetary companion funnelling part of the AGB wind over the past century. To the extent that planets can be expected to have orbital planes of similar inclinations, it would favour a large inclination of the larger disc. However, it seems difficult to conceive a simple scenario having the ability to generate, within a single century, at least two different winds displaying different expansion velocities and different inclinations with respect to the plane of the sky.

Our failure to detect bipolar outflows in the north/south direction deserves a comment. At short distances from the star it can be blamed in part on the inadequacy of the available observations: insufficient uv coverage for the first set and insufficient angular resolution for the second set.  At larger distances from the star, however, the wind takes the form of expanding discs that are completely unrelated to the morphology of the central gas and dust disc expected to act as a collimator. This suggests that the gas and dust disc did not exist in its present form before the 1994 dimming event.     

\section{Summary and conclusion}

The recent literature \citep{Kervella2014,Kervella2015,Kervella2016, Homan2017, Lykou2015, Ohnaka2015, Bedding2005} focuses on the morpho-kinematics of the gas and dust disc surrounding L$_2$ Pup: its likely relation with the dimming of the light curve that occurred at the end of the past century and the possible presence of a planetary companion at a distance of 2.4 au from the star. In contrast, the earlier literature underlines peculiar features apparently displayed by the star: a very low mass loss rate, at the level of 10$^{-8}$ \msun yr$^{-1}$, and a very low gas expansion velocity, at the level of 2 to 3 \kms. These were pointed out in most publications \citep[e.g.][]{Olofsson2002, Gonzalez2003, Jura2002, Winters2002} and the latter authors suggested that these peculiar features might be understood in terms of a wind driven exclusively by pulsations, with dust playing no dynamical role. Our analysis of the $^{12}$CO(2-1) and $^{28}$SiO(5-4) emissions  supports such a low value of the mass loss rate in the pre-1994 period. The observed morpho-kinematics of the wind, in the form of an inclined disc, or equatorial enhancement, expanding radially and isotropically, also suggests the importance of pulsations in generating the nascent wind but the observed confinement of the SiO emission supports the condensation of SiO molecules onto dust grains over a distance of some 80 au from the star, implying therefore that over this distance the dust transfers some of its radial momentum to the gas. Moreover, one does not see how pulsations alone could produce winds confined to relatively thin discs. It is therefore natural to consider the possibility for the wind to have been shaped by the gravity of a planetary companion, as generally accepted for the central gas and dust disc. However, if such were the case, one would have to conceive a scenario explaining how  two different wind configurations can be present within a short time period, at century scale.

In the post-1994 period, the recent VLT and ALMA observations have considerably improved our knowledge of the morpho-kinematics of the CSE and suggest that part of the L$_2$ Pup wind flows along the disc axis near the plane of the sky while the remaining part feeds the disc. However, the published literature does not directly address the properties of the wind and, to our knowledge, there exists no direct observation of gas escaping the gravity of L$_2$ Pup. In such a context, we have provided new information on the present state of the star, in particular evidence for large velocities in its vicinity, both from the shape of the absorption spectra measured over the stellar disc and from the presence of large Doppler velocity wings in the line profiles measured on line of sights crossing the star near its centre. This suggests that the standard mechanism of wind formation at stake in most oxygen-rich AGB stars also applies to L$_2$ Pup in its present state. However, the radial extension of the inner layer where shocks are expected to play a role has been found to be small: it was estimated at the scale of only 4 au, typically three times smaller than observed in stars such as Mira Ceti and R Dor. Moreover, it has not been possible to reliably reveal the presence of a gas outflow escaping the star gravity near the sky plane along the disc axis: both the complexity of the mechanisms at stake and the insufficient $uv$ coverage at short spacing have prevented it.

In summary, the main contributions of the present study can be summarized as follows:

Beyond $\sim$1 arcsec from the centre of the star, the wind of L$_2$ Pup is dominated by a disc, or equatorial enhancement, expanding isotropically with a radial velocity not exceeding some 5 \kms, inclined in the north-west/south-east direction with respect to the sky plane by $\sim$40\dego$\pm$20\dego. In addition, outflows of lower density are observed on both sides of the disc, covering large solid angles about the disc axis, contributing about half the flux of the disc. This is at strong variance with the expectation of a pair of back-to-back outflows collimated by the central gas-and-dust disc. Such outflows would not populate the eastern and western quadrants while our analysis shows that these are populated in continuity with the northern and southern quadrants. It raises new questions on the recent evolution of L$_2$ Pup: did the central gas-and-dust exist before the 1994 dimming event? if it did, why did it not act as a collimator? what has been shaping the wind observed before then? if it were a companion, it would have to be different from the planetary companion associated with the shaping of the central gas-and-dust disc. These new questions will require new observations in order to be answered.

In addition to the dominant wind, which has its north-western half inclined toward us, there are two additional winds inclined the other way.

Below $\sim$1 arcsec from the centre of the star, we failed to find any evidence for the presence of collimated outflows oriented north/south. However, difficulties inherent to the observations, uv coverage and angular resolution, and the complexity of the dynamics at stake in this region prevent us from ascertaining that there are no such outflows.

A blob of northern blue-shifted emission at Doppler velocities between $-$3 and $-$5 \kms\ has been identified and tentatively interpreted as the ``loop'' observed in L band by \citet{Kervella2014}.

Evidence for the apparent normality of the wind launching mechanism has been obtained from the presence of large velocities within $\sim$4 au from the centre of the star, the line width increasing from $\sim$5 \kms\ FWHM beyond this distance to $\sim$20 \kms\ near the star. The effect of its explicit inclusion in the models used by \citet{Homan2017} and \citet{Kervella2016} would probably be to produce smaller rotation velocities. Moreover the confinement of SiO emission within $\sim$80 au from the centre of the star suggests that acceleration by dust grains plays a significant role.

The mass loss rate has been evaluated at the level of (1.2$\pm$0.7)$\times$10$^{-8}$ \msun yr$^{-1}$.

\section*{Acknowledgements}

We thank Dr Ward Homan for useful discussions; his detailed analysis of the CO(3-2) data inspired the present work. We also thank Pr Albert Zijlstra for having shared with us his understanding of the dynamics governing the CSE of L$_2$ Pup. We thank the anonymous referee for his/her very careful reading of the manuscript and for the pertinence of his/her comments, which helped with improving the quality of the presentation. This paper uses ALMA data ADS/JAO.ALMA\#2015.1.00141.S. ALMA is a partnership of ESO (representing its member states), NSF (USA) and NINS (Japan), together with NRC (Canada), MOST and ASIAA (Taiwan), and KASI (Republic of Korea), in cooperation with the Republic of Chile. The Joint ALMA Observatory is operated by ESO, AUI/NRAO and NAOJ. We are deeply indebted to the ALMA partnership, whose open access policy means invaluable support and encouragement for Vietnamese astrophysics. Financial support from the World Laboratory, the Odon Vallet Foundation and VNSC is gratefully acknowledged. This research is funded by the Vietnam National Foundation for Science and Technology Development (NAFOSTED) under grant number 103.99-2019.368.

\section*{Data Availability}
The raw data are available on the ALMA archive ADS/JAO.ALMA\#2015.1.00141.S. The calibrated and imaged data underlying this article will be shared on reasonable request to the corresponding author.




\appendix
\section{}
\begin{figure*}
  \centering
  \includegraphics[height=12cm]{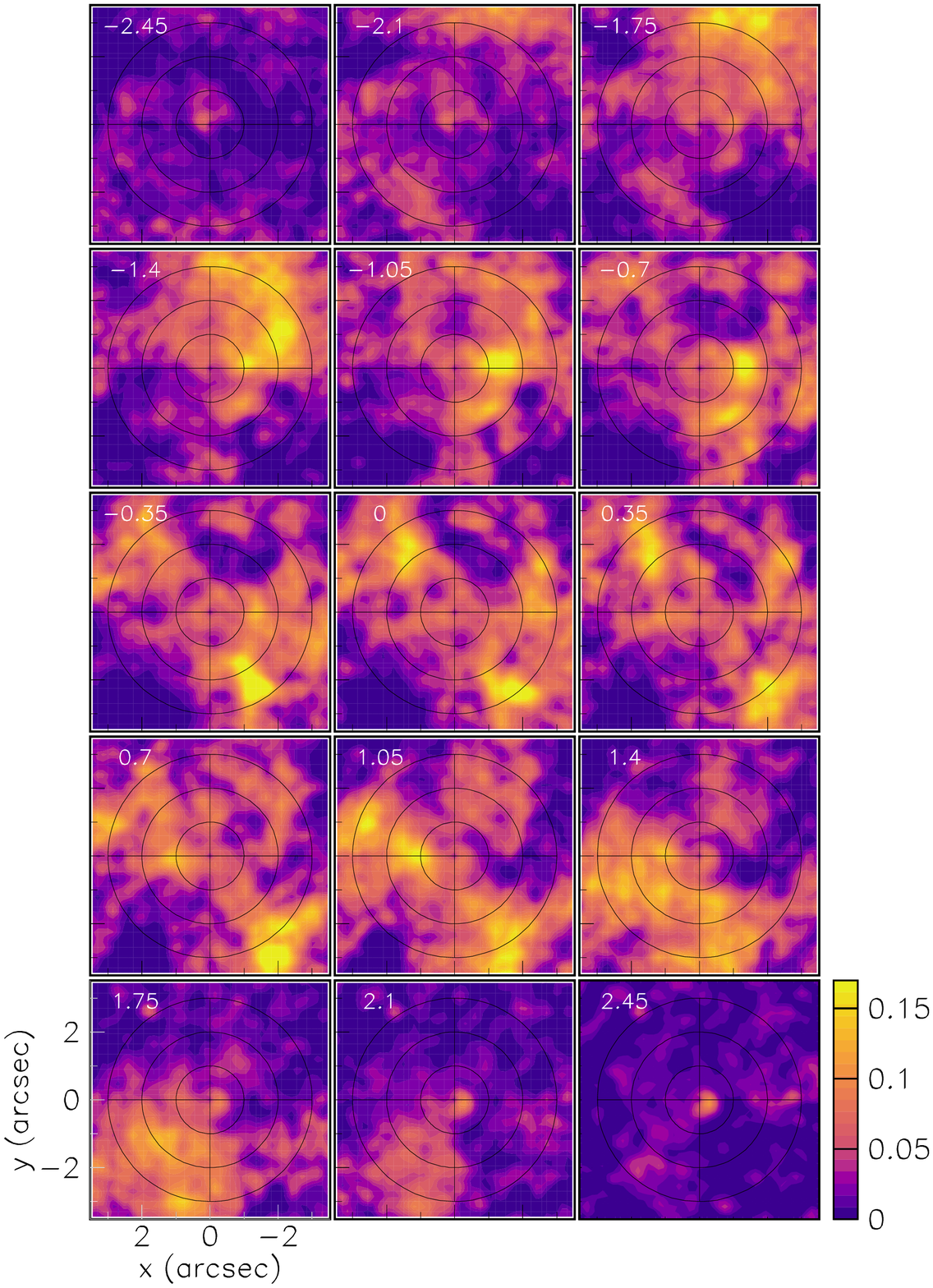}
  \includegraphics[height=12cm]{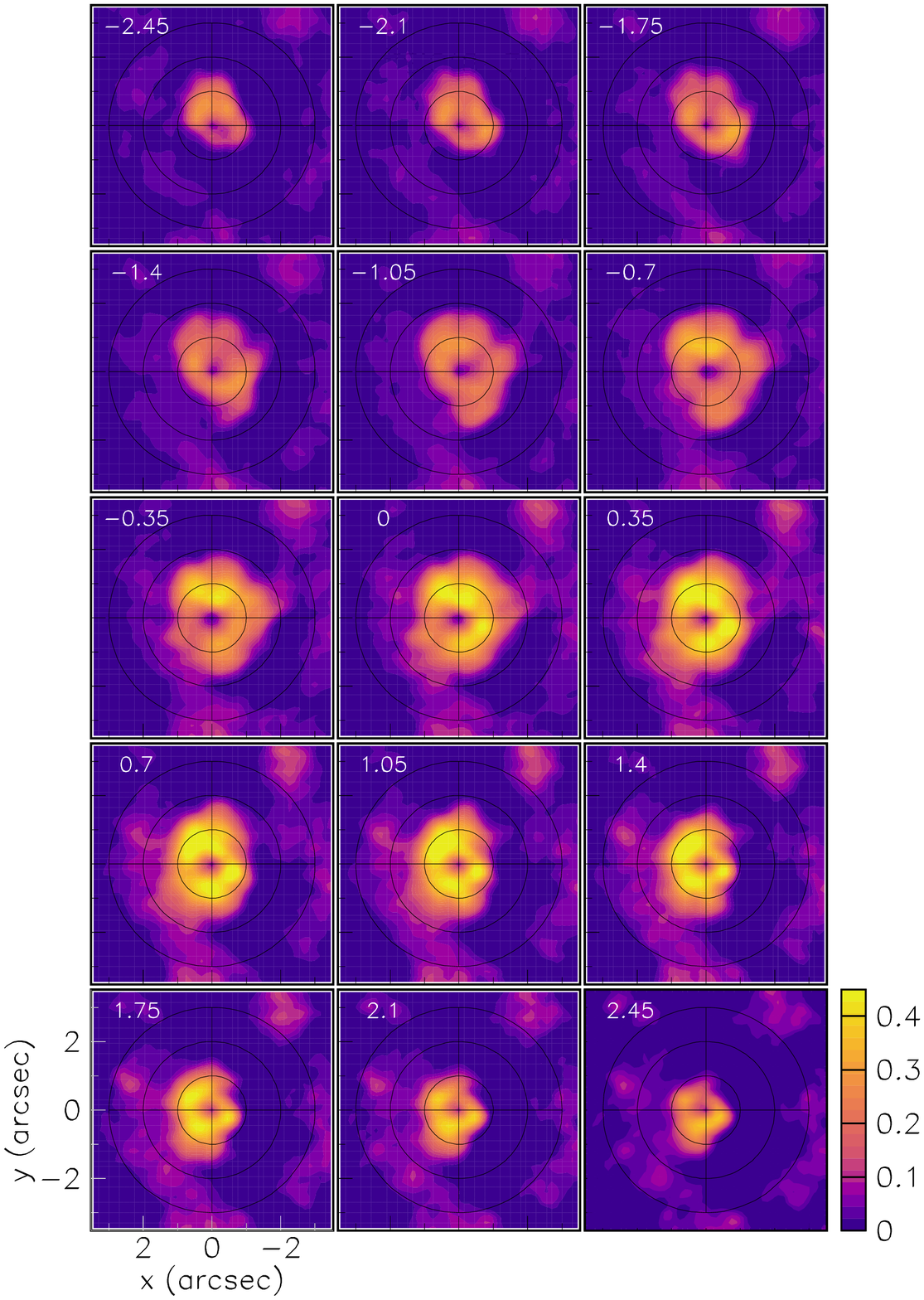}
  \caption{$^{12}$CO(2-1) (left) and $^{28}$SiO(5-4) (right) channel maps of the brightness multiplied by $R$ (Jy beam$^{-1}$ arcsec). The values of the Doppler velocity (\kms) associated with each channel are indicated in the upper-left corner of each pannel. Circles of 1, 2 and 3 arcsec in radius are shown. Note that the channel maps are shown only for $|V_z|$$<$2.5 \kms.}
\end{figure*}


\bsp	
\label{lastpage}
\end{document}